\documentclass[12pt]{article}
\usepackage[active]{srcltx}

\usepackage{amsbsy}
\usepackage{amsfonts}
\usepackage{amssymb}
\usepackage{amsmath}
\usepackage{graphicx}
\usepackage{subfigure}
\usepackage{cancel}
\usepackage[usenames, dvipsnames]{color}

\makeatletter
\renewcommand\section{\@startsection {section}{1}{\z@}%
                                   {-3.5ex \@plus -1ex \@minus -.2ex}%
                                   {2.3ex \@plus.2ex}%
                                   {\normalfont\large\bfseries}}
\renewcommand\subsection{\@startsection{subsection}{2}{\z@}%
                                   {-3.25ex\@plus -1ex \@minus -.2ex}%
                                   {1.5ex \@plus .2ex}%
                                   {\normalfont\normalsize\bfseries}}
\makeatother

\newcommand{\rf}[1]{(\ref{#1})}

\def\be{\begin{eqnarray}}
\def\ee{\end{eqnarray}}
\def\Tr{{\rm Tr}}

\def\NeqFour{{\cal N}=4}

\def\cO{{\cal O}}
\def\cJ{{\cal J}}
\def\Tr{{\rm Tr}}

\def\kk{{a}}

\def\draftnote#1{{\color{red} #1}}

\begin{document}

\vspace{ -3cm}
\thispagestyle{empty}
\vspace{-1cm}

\rightline{}

\begin{center}
\vspace{1cm}
{\Large\bf


Correlation functions of local composite operators from generalized unitarity

\vspace{1.2cm}

   }

\vspace{.2cm}
 {Oluf~Tang~Engelund\footnote{ote5003@psu.edu}
and
   Radu~Roiban\footnote{radu@phys.psu.edu}}
\\

\vskip 0.6cm

{\em

Department of Physics, The Pennsylvania  State University,\\
University Park, PA 16802 , USA

 }

\vspace{.2cm}

\end{center}

\begin{abstract}

We describe the use of generalized unitarity for the construction of correlation
functions of local gauge-invariant operators in general quantum field theories and
illustrate this method with several calculations in $\NeqFour$ super-Yang-Mills
theory involving BPS and non-BPS operators.
Form factors of gauge-invariant operators and their multi-operator generalization
play an important role in our construction.
We discuss various symmetries of the momentum space presentation of correlation
functions, which is natural in this framework and give examples involving non-BPS
and any number of BPS operators. We also discuss the calculation of correlators describing
the energy flow in scattering processes as well as the construction of the effective action
of a background gravitational field.

\end{abstract}

\newpage

\section{Introduction and summary}

Correlation functions of gauge-invariant operators are natural observables in both conformal
and non-conformal field theories. In the early days of the AdS/CFT correspondence correlation
functions of BPS operators played an instrumental role in establishing and testing it.
More recently, correlation functions have been shown to exhibit fascinating relations
to other quantities: limits in which the gauge invariant operators are null separated are
related to expectation values of null polygonal Wilson loops with or without local operators
\cite{Alday:2010zy, Alday:2011ga, Engelund:2011fg, Adamo:2011cd}
and, in $\NeqFour$
super-Yang-Mills theory in the planar limit, the same limit is related to
scattering amplitudes \cite{Eden:2010zz}.
%
Moreover, since any curve may be approximated to arbitrary precision by a null polygon,
it is in principle possible (though perhaps difficult in practice) that special limits of correlation functions
with arbitrarily many operators can be used to construct the expectation value of Wilson loops of
arbitrary shape.

Correlation functions of gauge-invariant operators also have a natural place in non-conformal
theories, such as QCD, where {\it e.g.} suitable correlation functions capture certain inclusive
properties of the final state of scattering processes. Among them are the energy correlators, originally
introduced in \cite{energy_correlators}.
As discussed in \cite{Hofman:2008ar}, they have a counterpart in conformal field theories, where
they describe the final state produced by the time evolution of some localized excitation and may be
thought of as particular analytic continuations of regular correlation functions.

Even though in conformal field theories their axioms guarantee that higher-point correlation functions
are determined by the two- and three-point functions, an explicit evaluation along this line is not
straightforward. It is therefore interesting to devise methods to directly evaluate them -- at weak and
at strong coupling -- both for generic positions of operators or directly in special limits.

The classic approach to the calculation of correlation functions makes use of Feynman diagrams
in momentum space, position space or superspace; in $\NeqFour$ sYM theory three-point functions
of scalar operators of dimension $\Delta\le 5$ have been evaluated systematically using such
methods in \cite{Georgiou:2012zj}.
An efficient reorganization of this approach is the Lagrangian insertion formalism \cite{Intriligator:1998ig,
Howe:1999hz, Eden:2000mv}.
In this theory, and in general in all theories in which the planar dilatation operator
defines an integrable Hamiltonian, the calculation of three-point (as well as that of higher-point) functions
benefits from use of integrable model techniques \cite{Okuyama:2004bd, Roiban:2004va, Alday:2005nd,
Escobedo:2010xs, Caetano:2011eb}.
More recently it was proposed \cite{Ananth:2012tf} and illustrated for the correlation function of four BPS
operators in $\NeqFour$ sYM theory that light-cone superspace can be efficiently used for this purpose.
The correlation functions of chiral stress tensor multiplets in $\NeqFour$ sYM theory enjoy special
properties at the integrand level \cite{Eden:2011we} -- a permutation symmetry which becomes manifest
when the integrand is constructed in the Lagrangian insertion formalism.
This was used to devise an efficient method for the construction
of four-point correlation functions of stress tensor multiplets \cite{Eden:2012tu} which led to the
evaluation of the five-loop correction to the anomalous dimension of the Konishi multiplet \cite{Eden:2012fe}.
The result confirms  the integrability-based predictions \cite{5loopKonishi, 5loopKonishiTBA}.

In conformal field theories that have a string theory dual, certain classes of correlation functions of local
operators may be evaluated at strong coupling  using semiclassical expansion on the string theory side
\cite{Janik:2010gc, Buchbinder:2010vw, Janik:2011bd, Buchbinder:2011jr, Zarembo:2010rr, Costa:2010rz,
Roiban:2010fe, Buchbinder:2010ek}. It was moreover argued
\cite{Raju:2010by, Raju:2011mp} that AdS supergravity scattering amplitudes with suitable boundary
conditions follow an on-shell recursion relation of the same type as the gauge theory on-shell recursion
relations.

Independently, increasingly efficient perturbative computational techniques -- generalized unitarity,
on-shell recursion relations, {\it etc.} -- led to remarkable progress in our understanding of  scattering amplitudes
of $\NeqFour$ sYM theory and to the discovery of new and powerful symmetries -- dual superconformal
symmetry \cite{Drummond:2006rz, Bern:2006ew}, color/kinematics duality \cite{Bern:2008qj} -- which
severely constrain the scattering matrix.
As we will explain in the next section, correlation functions of local gauge invariant operators may be interpreted as
special scattering amplitudes of sources in the theory obtained by adding to the action the operators
and their sources.
One might naturally expect that the modern techniques developed for S-matrix calculations can also be
used to calculate correlation functions; as we shall see, this is indeed the case. While correlation functions
are naturally functions of the positions of the operators, an approach that mirrors the calculation of scattering
amplitudes will yield their Fourier-transformed expressions, {\it i.e.} the momentum space correlation functions.
%
Since position space correlation functions are conformally invariant the momentum space correlators should
also have this property (albeit non-manifestly) and thus should be annihilated by the Fourier-transform of the
conformal group generators.

The relation between correlation functions and scattering amplitudes implies that the conformal symmetry
of the former becomes -- in the null separation limit -- the dual conformal symmetry of the latter~\cite{Eden:2010zz}.
Momentum space expressions for correlation functions may also be used to search for additional symmetries, which
are hidden in the position space expressions.\footnote{The relation to amplitudes might suggest a relation
between momentum space symmetries of correlation functions and position space symmetries of amplitudes.
Such a relation may however be obscured by the null limit, which is not very transparent in momentum space.
While such a relation can exist only in the null limit, it is not clear why momentum space correlation functions
cannot have additional symmetries.}

Apart from constructing position-space correlation functions, momentum space correlators can also be used
to construct quantum effective actions in background fields. As we shall briefly discuss in \S~\ref{Tcorrelators},
coupling an action with
external fields is essentially equivalent to deforming the action by various operators, with the background fields
acting as sources. Compared to the deformations relevant for the calculation of correlation functions, the only
difference is that, depending on the specifics of background fields, sources may appear nonlinearly. Nevertheless,
scattering amplitudes of background fields, perhaps expanded for small momenta/weakly varying external fields,
yield the terms relevant for the construction of their off-shell effective action.

Recently, the color/kinematics duality \cite{Bern:2008qj} has emerged as an important property of color-dressed
amplitudes in certain gauge theories with only adjoint fields and antisymmetric structure constant couplings.
It states that, given some amplitude, there exists a presentation of its integrand such that, if the color factors
of some integrals obey the Jacobi identity,  then the numerator factors  of those integrals also obey a
Jacobi-like identity.
There is by now substantial evidence in favor of the duality,
both at tree level~\cite{OtherTreeBCJ,virtuousTrees,Square,
Oconnell, ExplicitForms,Sondergaard:2009za} and at loop level \cite{BCJLoop}.
A consequence of this duality is the
existence of nontrivial relations between the color-ordered partial
tree amplitudes of gauge theory~\cite{Bern:2008qj}, which have been proven both
from field theory~\cite{Feng} and string theory~\cite{Bjerrum1} perspectives.
We will explore whether a duality of this type exists for correlation functions.
A natural expectation is that, if present, it would be most easily visible in
momentum space correlation functions.

The structure of the operators makes it unlikely that color/kinematics duality
always holds for all internal edges of the integrals that appear. Indeed, color/kinematics duality acts naturally on
integrals associated to graphs constructed from 3-point vertices.  Due to the operator insertions however, correlation
functions naturally have multi-point vertices. One may attempt to define the duality by resolving the
higher-point vertices into sequences of three-point vertices in color space. In general however, these vertices
carry both antisymmetric structure constants as well as the symmetric ones, $d_{abc}$, and their
generalizations.
It is therefore not clear whether for a generic correlation function Jacobi relations can exist for all
internal edges. A notable exception\footnote{We thank H.~Johansson
for providing this example, discussions on this point and for sharing his insight
into \cite{BjerrumBohr:2011xe}.} which appears consistent with color/kinematics duality is
provided by a four-gluon operator with the color structure given by $d_{abcd}$.
If a Jacobi identity is not present and color and kinematic factors are no longer linked,
it is still possible that numerator factors of various integrals are nonetheless
nontrivially related to each other\footnote{This is reminiscent of the $\beta$-deformed $\NeqFour$
sYM theory  \cite{Jin:2012mk}, where tree-level kinematic factors are related despite absence of a relation
for the corresponding color factors. The importance of such relations remains an open question.}.
Regardless of these details, we will argue that color/kinematics duality should hold for all internal
edges not directly connected to one of the operators up to contact terms collapsing at least one of these
latter edges.


In this paper we discuss the use of generalized unitarity for the construction of momentum space
correlation functions of BPS and non-BPS operators. While we will be mainly concerned with
describing the details and subtleties of this approach, we also illustrate this method by recovering
the known example of the four-point BPS operators and also by constructing infinite classes
of new ones: such as the $n-$point function of BPS operators and the $n-$point function of BPS
operators and some number of twist-2 non-BPS operators at the next-to-leading order.
%
In our examples we will focus on the $\NeqFour$ sYM theory; as for scattering amplitudes however, the
method we describe here may be applied with suitable care to all quantum field theories. Essential ingredients
in this construction are the (super-)form-factors of local gauge-invariant operators as well as their generalizations
involving  several operators. Unitarity and generalized unitarity have already been applied to the construction of
higher-loop form factors in \cite{vanNeerven:1985ja, Gehrmann:2011xn, Bork:2012tt, Brandhuber:2012vm}.

In \S~\ref{general_unitarity} we will describe the similarities  between scattering amplitudes and correlation
functions and the use of generalized unitarity for the construction of the latter for general gauge-invariant
operators, the relevance of generalized form factors and the need for regularization and renormalization. We
will also identify the components of correlation functions which may exhibit color/kinematics duality as well
as may contain some hidden consequences of dual conformal invariance.
In \S~\ref{formfactors} we collect the known expression of the MHV super-form factor of the chiral stress tensor
multiplet and list (while relegating the details to appendices) the generalized (two-chiral stress tensor multiplet)
form factor as well as the MHV super-form factor of scalar non-BPS operators and of the general twist-2 non-BPS
operators. We will use them in the examples we discuss in \S~\ref{examples}.

We construct the leading order and the next-to-leading order correlation function of four BPS operators
in momentum space, Fourier-transform it to position space and recover the known results \cite{Eden:1998hh,
Eden:1999kh}. While the calculations are mainly carried out in components, in \S~\ref{susy_methods} we
illustrate the use of manifestly supersymmetric methods for the construction of correlation functions.
We also derive the (connected part of the) next-to-leading order correlation function of $n$ BPS operators
and check that its null limit reproduces the $n$-point MHV amplitude, as originally shown in \cite{Eden:2010zz}.
In \S~\ref{BPSNBPS}
we compute (the connected part of) the correlation function of one and two general non-BPS twist-2
operators and an arbitrary number of BPS operators. Similarly to the discussion of correlators of
BPS operators, we verify explicitly that their null limits reproduce the $n$-point MHV amplitude.
We also discuss the structure of the correlator of  $m$ twist-2 and $n$ BPS operators in a split configuration.
For the three-point function~\footnote{While writing up this paper we received \cite{Plefka:2012rd} in which
the three-point function of one twist-2 non-BPS operator and two BPS operators was computed through a
different method.} we discuss the appearance of the anomalous dimension of the non-BPS
operator, its renormalization, as well as the vanishing of the three-point function in limits in which the
twist-2 operator becomes a conformal descendant of a BPS operator.

In \S~\ref{Tcorrelators} we include a general discussion of correlation functions involving stress tensors,
which may be used to determine the effective action in non-dynamical gravitational background.
We also discuss the energy correlators, which capture the energy flow in the time evolution of some
composite state. After recalling their definition, we illustrate this
non-standard type of correlation functions by evaluating several examples to first nontrivial order:
the one- and two-point energy correlator in the state created by a BPS scalar operator as well as the
one- and two-point energy correlator in the state created by a twist-2 non-BPS operator.
Section \S~\ref{outlook} contains our conclusions and a summary of our results. Several appendices
contain detailed derivation of some of the formulae used in the main body of the paper.

\section{Generalized unitarity and correlation functions \label{general_unitarity}}

Generalized unitarity has been developed as a very efficient tool for the calculation of on-shell
scattering amplitudes in quantum field theories.
From a modern perspective it is interpreted as a specific organization of the Feynman graphs
contributing to an amplitude  which exposes the vast simplifications occurring when off-shell
Green's functions are amputated and placed on shell:\footnote{In the unitarity construction of scattering
amplitudes based on the optical theorem the imaginary part of an amplitude has this property manifestly;
the real part is constructed through a dispersion integral.}
each generalized cut is, on the one hand, a product of tree-level amplitudes and on the other
it is the subset of Feynman graphs that contain the cut propagators.
From this perspective it appears natural that anything that is expressible in terms of Feynman
graphs and is gauge invariant may be constructed through generalized unitarity-type
methods.\footnote{It is presumably possible to generalize this statement to gauge-variant quantities
at the expense of having non-vanishing contributions from ghosts and unphysical degrees of freedom.}
An example in this direction are form factors; particular form factors have recently been considered at various loop
orders in \cite{vanNeerven:1985ja, Gehrmann:2011xn, Brandhuber:2011tv, Bork:2010wf,
Brandhuber:2012vm, Bork:2012tt} in $\NeqFour$ sYM theory as well as in QCD \cite{Gehrmann:2011aa, Johansson:2012zv}.

While this interpretation of generalized unitarity makes clear its applicability, it is not difficult to
give form factors the interpretation of scattering amplitudes.  In a similar spirit correlation functions,
which are the focus of our paper, can be given the same interpretation.~\footnote{From the perspective
of the optical theorem, this implies that correlation functions should also be expressible in terms
of dispersion integrals.}

\subsection{Correlation functions as scattering amplitudes}

Let us start with the path integral expression of an $n$-point correlation function in some quantum field theory
defined by the Euclidian action $S_E[\Phi]$,
\be
\langle {\cal O}_1(x_1)\dots {\cal O}_n(x_n)\rangle = \int[D\Phi]\; {\cal O}_1(x_1)\dots {\cal O}_n(x_n)
e^{-S_E[\Phi]} \ ,
\ee
where $\Phi$ generically denote the fields of the theory and the local operators ${\cal O}_i$ are
gauge-invariant combinations of them. All correlation functions
may be packaged into a generating functional, by introducing local sources for all possible/desired local
operators:
\be
Z[\cJ_1,\cJ_2,\dots]=\int[D\Phi]\;
e^{-S_E[\Phi]-\int d^dx\sum_i\,\cJ_i(x)\cO_i(x)} \ .
\ee
Correlation functions of specific operators are then extracted by differentiating with respect to the relevant
sources and setting all sources to zero:
\be
\langle {\cal O}_1(x_1)\dots {\cal O}_n(x_n)\rangle =
\frac{\delta^n}{\delta \cJ_1(x_1)\dots \delta \cJ_n(x_n)}Z[\cJ_1,\dots]\Big|_{\cJ_i\rightarrow 0} \ .
\ee
The sources $\cJ_i$ may be interpreted as non-dynamical fields. If
$S_E$ is the action of a conformal field theory with a closed string theory dual \cite{Maldacena:1997re},
then the sources may be
interpreted as the boundary values of the fields describing the closed string states
\cite{Gubser:1998bc, Witten:1998qj}. In particular, for BPS operators, they are just the boundary
values of the fields of the relevant supergravity theory.

The construction above is formally identical to that of scattering amplitudes; the only difference is that, for
scattering amplitudes,  $\cJ$ are  sources for the fundamental fields.\footnote{Also, scattering amplitudes
are constructed in Lorentzian signature with the time-ordered operators. For correlators in Euclidian signature
one may formally introduce a radial ordering of operators. }
From this perspective, if we interpret the
sources $\cJ$ as fields, we may also interpret the correlation functions of operators as scattering amplitudes
of their associated source-fields $\cJ$ in the theory with the modified action
\be
S_E^{\text{mod}}[\Phi,\cJ]=S_E[\Phi]+\int d^dx\sum_i\,\cJ_i(x)\cO_i(x) \ .
\label{deformedSE}
\ee
We may therefore view the scattering amplitudes of source-fields $\cJ_i$ as the correlation
functions of the corresponding operators ${\cal O}_i$. If the field theory has a string
theory dual we may interpret this as the scattering amplitude of closed strings with specified
boundary conditions.
Similarly, we may interpret the scattering amplitudes of one source-field $\cJ$ and any
number of fundamental fields $\Phi$ as the form factor of the operator corresponding
to that source,
\be
\langle \cO|\Phi_1\cdots\Phi_m\rangle \ ,
\ee
as also mentioned in \cite{Bork:2010wf}. From the perspective of a string theory dual we may interpret
this as the scattering amplitude of a closed string into open strings. Amplitudes with several sources
and fundamental fields have a similar interpretation in the context of gauge/string duality; from a field
theory perspective we will refer to them as "generalized form factors" or as "multi-operator form factors".
\footnote{It is important to stress that these are not necessarily related to form factors of multi-trace operators.}

The gauge invariance of the operators ${\cal O}_i$ and of the corresponding source fields $\cJ_i$
guarantees that this action is BRST invariant. As in the absence of source fields, it is possible to use
a sequence of Ward identities \cite{Veltman:1963th} to show that scattering amplitudes with external
unphysical fields or ghosts do not contribute to unitarity cuts. \footnote{The required Ward identities do not hold
if at least one of the operators -- and hence the deformed action \rf{deformedSE} -- is gauge-variant.}

As described above, while $\cJ$ are interpreted as fields, they are nevertheless non-propagating and thus
they may appear only as external  lines of an amplitude and do not have a well-defined on-shell condition.
To make an even closer analogy with scattering amplitudes one may formally promote them to propagating
fields by assigning them massive quadratic terms with
different masses for each source; they may then appear both as internal and external lines.
\footnote{
An example in this direction is the interaction term $H\Tr[F^2]$ generated at one-loop level through
a top-quark loop in the Standard Model (and it is responsible for the gluon-fusion production
of the Higgs boson). The two-loop amplitude $H\rightarrow ggg$ in this theory, {\it i.e.} the two-loop form
factor of $\Tr[F^2]$ with three external states, was discussed recently in \cite{Gehrmann:2011aa, Duhr:2012fh,
Johansson:2012zv}.
By formally taking the Higgs mass
to infinity one forces the Higgs boson to appear only as an external state. The resulting amplitudes are
either generalized (in the sense described below) form factors of $\Tr[F^2]$ or correlation functions of
these operators.}
Restricting the
integrands of the resulting amplitudes to terms that have no poles as the mass of the sources is varied
guarantees that all contributions with internal sources are projected out. Last, the remaining mass dependence
is solely associated to the norm of the external momenta and should participate in the Fourier-transform to
position space.

\subsection{On the presentation, structure and symmetries of correlation functions
\label{presentation_and_symmetries}}

Correlation functions of operators with definite dimension transform covariantly under position space
conformal transformations and are strongly constrained by them.
However, as for scattering amplitudes, the generalized unitarity method yields scattering amplitudes
for source fields carrying definite momenta and thus gives the momentum space form of correlation functions
\be
\langle {\widetilde \cO_1}(q_1)\dots {\widetilde \cO_n}(q_n) \rangle
\ee
where ${\widetilde \cO_i}(q_i)$ are the Fourier-transform of the usual position space operators
\be
{\widetilde \cO}(q)=\int \frac{d^dq}{(2\pi)^d}\,e^{iq\cdot x}\,\cO(x) \ .
\ee
Position-space conformal invariance will be hidden in momentum space; nevertheless, since momentum space
correlation functions should not have holomorphic anomalies\footnote{Holomorphic anomalies are likely to
appear when the operator's momenta are restricted to be null. Since position space correlation functions should
be conformally invariant, we expect that the constrained momenta should not play an important role in the
Fourier-transform.} \cite{Cachazo:2004by} (unlike scattering amplitudes
of fundamental fields) one may test whether they are annihilated by  the momentum space form
of conformal generators -- in particular the special conformal generator
\be
\label{scgenerators}
K^\mu&=&\sum_{i=1}^nK^\mu_{i}
\\
K_i^\mu &=& -2(\Delta_i-3)\frac{\partial}{\partial q_\mu^i}
-\left(q^\mu_i \frac{\partial^2}{\partial q^i_\nu\partial q_i^\nu}
  -2 q^i_\nu\frac{\partial}{\partial q_\nu^i}\frac{\partial}{\partial q_\mu^i}\right)
  +2\frac{\partial}{\partial q^i_\nu}(\Sigma^i)_\nu{}^\mu
\nonumber
\ee
where the sum runs over all operators, $\Delta$ is the conformal dimension and $\Sigma^i$ represents
$SL(2,C)$ on the $i-$th operator.
Conformal symmetry should emerge as a manifest symmetry upon inverse Fourier-transform to position
space, if the operators are chosen to have definite dimension.

The unitarity method provides an efficient framework for systematically constructing and verifying the
expression for any multi-loop amplitude in a massless field theory. This method, along with various refinements, has
already been described in some detail elsewhere \cite{UnitarityMethod, BDDPR,
GeneralizedUnitarity, BCFGeneralized, FiveLoop, CachazoSkinner}; see also
\cite{Alday:2008yw, Beisert:2010jr, JPhysA_Volume} for recent reviews.
Here we will discuss the additional information needed for the construction of correlation functions of local
operators.
A color-dressed generalized unitarity cut is a sum over products of color-dressed amplitudes ${\cal A}_{(i)}$,
\be
i^c \sum_{\text{states}}{\cal A}_{(1)}{\cal A}_{(2)}\dots {\cal A}_{(m)}
\ee
where the $c$ cut lines are placed on shell and we included a factor of $i$ for each cut propagator; each
cut line appears twice, leaving one amplitude factor and entering another. While not necessary, the amplitude
factors ${\cal A}_{(i)}$ can be chosen to be at tree level. In our case of cuts of scattering amplitudes of source
fields, the factors ${\cal A}_{(i)}$ are either amplitudes of fundamental fields or (generalized) form factors.
The cut construction of an amplitude formally proceeds by matching all cuts onto an ansatz in terms of
Feynman integrals.

A systematic strategy, which is designed to keep under control the size of the ansatz,  is the maximal cut
method in which one begins by first constraining the ansatz to reproduce all the $(3L+(n-3))$-particle
cuts ({\it i.e.} the maximal cuts, in which the maximal number of propagators for an $n$-point amplitude are cut)
and then systematically
proceeds to relax the cut condition on one propagator (next-to-maximal cuts), two propagators (next-to-next-to-
maximal cuts) and so on. At each step one may reduce the expression of the cuts to cuts of master
integrals by generalizing the methods of refs.~\cite{Kosower:2011ty, Johansson:2012zv} beyond two-loop order
and to all massive external legs.
The advantage of this method is that, for any one cut, only a small part of the ansatz is relevant.
For the resulting expression to be correct it must reproduce all generalized cuts. Since some cuts are special cases
of others, it suffices to verify that it reproduces a spanning set of cuts -- {\it i.e.} a set of cuts that guarantee that
all the other ones are satisfied. For simple (low-order) cases one may construct amplitudes by analyzing directly the
relevant spanning set of cuts.

\begin{figure}[ht]
\begin{center}
\includegraphics[height=50mm]{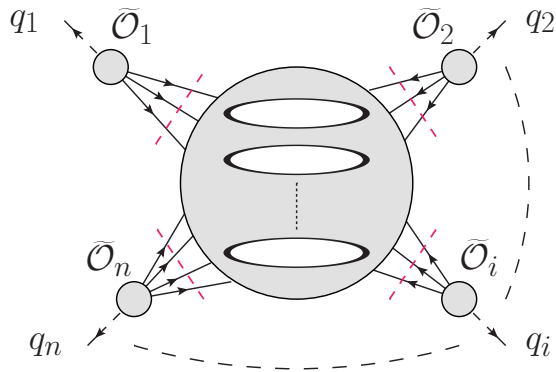}
\caption{Generalized unitarity guarantees the presence of multi-loop amplitudes in the
correlation function calculation. Additional contributions involves at least one form factor
with more fields than the corresponding operator.\label{use_form_factors} }
\end{center}
\end{figure}

It is therefore clear that an essential ingredient in the construction of the (momentum space form of) correlation
functions are the form factors and generalized form factors of the corresponding operators. While it is obvious
that there exists a very close relation between off-shell Green's functions of fundamental fields and correlation
functions of local operators, the generalized unitarity makes it clear that a similar though slightly weaker relation
exists between scattering amplitudes of fundamental fields and correlation functions.
Let us consider the generalized cut shown in fig.~\ref{use_form_factors}: the external lines are attached
to blobs representing color-dressed form factors while Feynman diagrammatics guarantees that
the blob with no external lines is some multi-loop scattering amplitude of fundamental fields.
This structure is independent of the number of external legs of each form factor, with the highest-loop amplitude
appearing together with the form factors with the smallest number of external legs. Other cuts are necessary
in order to identify the terms in which propagators exposed in fig.~\ref{use_form_factors} are collapsed. It is
moreover necessary to consider cuts involving generalized form factors.
The terms captured only by these cuts are those in which there is no (sequence of) propagator(s) between
two operators; upon Fourier-transform to position space, such terms lead to contact terms which we know are
relevant only in the OPE limit; it is possible that at least some contact terms may be inferred from the
requirement that the position space Euclidean correlator satisfies all relevant Ward identities
(see {\it e.g.} \cite{Petkou:1999fv} for a discussion of 3-point functions and the determination of contact terms
and \cite{Osborn:1993cr} for the contact terms required by conformal invariance).

This structure of correlation functions points to the fact that, in theories whose color-dressed
scattering amplitudes of fundamental fields exhibit larger symmetries, such as $\NeqFour$
sYM theory, momentum space correlation functions should be at least partly constrained by
these symmetries.
Among the remarkable properties of $\NeqFour$ sYM theory are the dual conformal invariance and
color/kinematics duality of its scattering amplitudes. They may have some consequences on correlation
functions as well.
Indeed, the cut in fig.~\ref{use_form_factors} in which the form factors have the smallest number of external lines
implies that there should exist a presentation of correlation functions such that the numerator factors
of the contributing Feynman integrals obey Jacobi identities for all internal lines not attached to an
operator, up to contact terms that collapse one of these lines.
Depending on the structure of the operators, in the complete correlation function the internal legs directly
attached to any one of them may not necessarily obey a Jacobi identity and a case by case study appears
necessary.
For example, it was shown \cite{Broedel:2012rc} that scattering amplitudes with an insertion of
$\Tr[F^3]$ operator -- {\it i.e.} the zero-momentum form factors of this operator -- exhibit color/kinematics
duality; the arguments above imply that we should expect that the correlation functions of several
such operators should also exhibit color/kinematics duality in the zero-momentum limit. The fate of
color/kinematics duality for the correlation function of $\Tr[F^3]$ operators with generic momenta
is an interesting question, albeit one which we will not explore here.
%
%
In general, color Jacobi identities exist only in the special case in which the color factors of operators
contain at most one factor of the symmetric structure constants or their higher-rank generalizations
and thus the correlation functions of only such operators may be expected to obey color/kinematics
duality for all internal lines; the possible appearance of contact terms in Jacobi transformations on internal
lines requires nevertheless a case by case study to ascertain whether color/kinematics duality is in fact realized.
This structure is relatively similar to the properties of amplitudes in the $\beta-$deformed $\NeqFour$
sYM theory~\cite{Jin:2012mk}.

Similarly, if excising the operators leads to a planar amplitude\footnote{It is important to note that, since the
source fields $\cJ$ are  color-singlets,
planar integrals are only a subset of the leading terms of the larger $N_c$ expansion. Indeed, if the
non-planarity of a integral arises solely because external legs (or sources) are attached to internal lines, then that
diagram yields, in fact, leading-color contributions. This has already appeared in the calculation of higher-loop
form factors \cite{Gehrmann:2011xn, Brandhuber:2012vm}.}, then the terms in the correlation
function which contribute to this cut may exhibit some trace of the dual conformal invariance of that
amplitude. In the complete correlator the non-planar terms (if present) as well as the presence of the operator
insertions will break the usual dual conformal invariance.

Regularization and renormalization are important issues which need to be carefully considered.
Since all external lines of momentum space correlation functions (or, equivalently, of source field
amplitudes) are effectively massive, no infrared divergences can appear and thus infrared regularization
is not necessary.
Ultraviolet divergences have two possible origins: divergences due to the structure
of the undeformed Lagrangian and divergences due to the presence of the deformation by the gauge-invariant
operators.
The former may be eliminated by standard renormalization of the Lagrangian and lead to $\beta$-functions
for the various coupling constants. In $\NeqFour$ sYM theory they are absent. The later are eliminated by
renormalizing the deformations and are solely related to the anomalous dimensions of
operators. In particular, if one used renormalized operators, with the renormalization
factors given in dimensional regularization by the usual expression in terms of the anomalous dimension of
the operator
\be
Z_\gamma = \exp\Big(\frac{1}{\epsilon}\int_0^1\frac{dt}{2t} \, \gamma(t\,g^2)\Big) \ ,
\label{Zfactor}
\ee
($g$ is the gauge coupling constant), the correlation functions would be finite in the ultraviolet as well. In
general the anomalous dimension $\gamma$ requires a separate calculation. In $\NeqFour$ sYM in the
planar limit they are provided by integrability.

One therefore has several options for accounting for the required renormalization: (1) one uses from the
outset dimensional regularization and renormalized operators or (2)
one carries out the calculation in four dimensions, regularizes dimensionally (if necessary) the integrals
appearing in the final result and searches for potentially missing $\mu$-type integrals\footnote{We recall
here that $\mu$-type integrals are integrals whose integrand vanishes identically when evaluated in
four dimensions.} such that multiplication by the appropriate $Z$-factors \rf{Zfactor} renders it finite
as $\epsilon\rightarrow 0$ or (3) one uses another regulator, such as a higher-derivative regulator, for both the
calculation of the amplitude and the calculation of the renormalization factors.

While the strategy outlined here applies to generic quantum field theories, in the following we will restrict
ourselves to $\NeqFour$ sYM theory. In this case the sources break maximal supersymmetry to the subalgebra
that leaves invariant  the deformation \rf{deformedSE}.
BPS operators preserve some amount of supersymmetry, their anomalous
dimension vanishes identically and their $Z$-factors equal unity. In their case no regularization
is necessary.
For the calculation of correlation functions that include non-BPS operators the deformed action \rf{deformedSE}
is formally nonsupersymmetric\footnote{The same is true, in fact, for the action deformed by several
BPS operators
each of which preserves a different non-overlapping or partly overlapping subsets of supercharges.}.
Nevertheless, since
supersymmetry breaking is confined to  the operator insertions, we expect that most of the features of
maximally supersymmetric calculations will continue to exist here as well.
Since to leading order (tree-level in position space) all correlation functions are rational functions and at
the next-to-leading order only one-loop bubble integrals are divergent, a systematic regularization procedure
starts being necessary only at the next-to-next-to-leading.

Quite generally, operators with definite anomalous dimension are (complicated) linear combinations of simpler
single-term operators with coefficients given \cite{Staudacher:2004tk},
in the planar limit and for operators with at least one large charge,
in terms of the solution to the Bethe equations \cite{Beisert:2005fw}.
The form factors of the former operators are linear combinations of the form factors of the latter operators
with the same coefficients\footnote{Since the $L$-loop mixing coefficients are proportional to $\lambda^{L-1}$,
the $L$-loop form factor requires use of the $(L+1)$-loop eigenvectors of the dilatation operator.}.
To compute correlation functions of such operators we evaluate the correlation
functions of the generic terms in each of them and then sum them with the appropriate integrability-determined
coefficients. This strategy applies both to 3-point as well as to higher-point functions.

As noted in \cite{Brandhuber:2011tv}, scattering amplitudes may be interpreted as the (generalized)
form factors of the
zero-momentum Lagrangian. Similarly, form factors with more external legs than fields in the operator
may be interpreted ({\it e.g.} through the MHV vertex expansion) as generalized form factors of the operator and
a suitable number of zero-momentum on-shell Lagrangians. In particular, scattering amplitudes with more than
four external fields can be interpreted as generalized form factors of several on-shell Lagrangians.
It then follows that the generalized unitarity-based construction of the integrand  of the
next$^k$-to-leading order momentum space correlation functions  of some operators is the same
as the construction of the leading order correlation function of those operators and $k$ on-shell
Lagrangians\footnote{At each order beyond the leading order one needs to add one additional
internal line and hence one additional operator with at least three fields.}.
We therefore see a formal parallel between the generalized unitarity calculation and the Lagrangian insertion
method of \cite{Intriligator:1998ig, Howe:1999hz, Eden:2000mv}.

The various monomials that enter the expression of operators with definite anomalous dimension may not
always have the same number of fields. A simple example is the stress tensor, whose expression contains a
bilinear in the field strength which has two, three and four fields. This is, in fact, the generic structure of
terms  on the higher levels of supersymmetry multiplets with the increase in the number of fields being due
to non-linear terms in supersymmetry transformations.
Clearly, these terms contribute to form
factors with different number of external legs; in the standard organization of the Lagrangian, in which
the coupling constant appears as an overall factor, these terms contribute to different loop orders.~\footnote{From
a position space perspective one might be tempted to interpret all of them as contributing to the same
order -- {\it e.g.} at tree level -- as they do not involve any integration.}
It has been suggested in \cite{CaronHuot:2010ek, Eden:2011yp, Eden:2011ku} their contribution cancels against
contact terms arising from terms proportional to the equations of motion from Lagrangian insertions and,
at least for the purpose of constructing the null limit of correlation functions, they may be set them aside.
Their close relation to contact terms (which in a momentum space framework should appear as Feynman
integrals with cancelled propagators) suggests that a shortcut to determining the contribution of nonlinear
terms in supersymmetry transformations to correlation functions may be imposing the
Ward identities of the various symmetries on (the Euclidian) correlation functions found in the absence of
these terms \cite{Petkou:1999fv, Osborn:1993cr, Eden:2000mv}.

\section{Some single-operator and multi-operator form factors \label{formfactors}}

We have argued in the previous section that tree-level form factors and generalized form factors are
essential for the construction of correlation functions. They may be constructed in several ways, mirroring
the various methods for the construction of tree-level scattering amplitudes. As in that case it is useful to
assemble them into super-form factors; they are labeled by the
coordinates of two superspaces: the usual on-shell $\NeqFour$ superspace, with Grassmann coordinates $\eta^A$
and an off-shell superspace for the multiplet of operators, with Grassmann coordinates $\gamma$.
A convenient one is the harmonic superspace \cite{Harmonic_superspace};
the on-shell superspace fields\footnote{\label{superfieldfootnote}
Specific on-shell external fields may be extracted in the usual way, by specifying their
position in the on-shell  $\NeqFour$ multiplet
\begin{eqnarray}
\Phi(\eta)=
   g_-
+\eta_A\psi^A
+\frac{1}{2!}\eta_{A}\eta_{B}\phi^{AB}
+\frac{1}{3!}\eta_{A}\eta_{B}\eta_{C}\epsilon^{ABCD}\psi_{D}
+\frac{1}{4!}\eta_{A}\eta_{B}\eta_{C}\eta_{D} \epsilon^{ABCD}g_+~~.
\label{eq:superfield}
\nonumber
\end{eqnarray}
} may also be written in harmonic
superspace form \cite{Eden:2011yp, Eden:2011ku}.
Whenever necessary, we will follow the notation there as well as in \cite{Brandhuber:2011tv,
Bork:2011cj} and collect some details in Appendix~\ref{harmonic_superspace}.

A possible approach to the construction of (super-)form factors is to use the MHV vertex rules. This strategy was first applied to the
construction of form factors of a particular gluon operator in \cite{Dixon:2004za} and more recently
in \cite{Brandhuber:2010ad}. This method requires independent information on MHV form factors;
they are defined to be those that have the minimal number of fermionic coordinates
$\eta$.
This definition mirrors that of MHV amplitudes.

BCFW-like recursion relations may also be used if a suitable shift can be found. In this approach
only the tree-level form factor with the minimal number of fundamental fields is necessary. As in
the case of scattering amplitudes of fundamental fields, different BCFW shifts yield different presentations
of the same form factors.

In this section we collect several examples of form factors and generalized form factors with arbitrary
number of external states  which will be useful in the examples of correlation functions we will discuss
in the next section.

\subsection{The tree-level MHV form factors and generalized form factors of the chiral stress tensor multiplet}

The tree-level MHV super-form factor ${\cal F}^{\text{MHV}}_{\cal T}$
of the chiral part of the stress tensor multiplet ${\cal T}$ was found in
ref.~\cite{Brandhuber:2011tv, Bork:2011cj} through a combination of symmetry constraints
and explicit calculations:
\begin{eqnarray}
{\cal F}^{\text{MHV}}_{\cal T}&=&\prod_{i=1}^n \frac{1}{\langle i, i+1\rangle}
\;
\delta^{(4)}(q-\sum_i\lambda_i{\tilde\lambda}_i)
\delta^{(4)}(\gamma_+-\sum_i\lambda_i\eta_{+;i})
\delta^{(4)}(\sum_i\lambda_i\eta_{-;i})
\label{MHVff}
\ee
The two $\delta-$functions may be combined into a single one by multiplying their arguments
with the suitable harmonic variables:
\be
{\cal F}^{\text{MHV}}_{\cal T}&=&\prod_{i=1}^n \frac{1}{\langle i, i+1\rangle}
\;
\delta^{(4)}(q-\sum_i\lambda_i{\tilde\lambda}_i)
\delta^{(8)}(\sum_{i=1}^n\eta_{Ai}\lambda_i^\alpha-1^+_{Aa}\gamma^{a\alpha}_{\tilde{1}+}) \ .
\label{modMHVTff}
\end{eqnarray}
The form factor of the chiral primary operator is extracted as the coefficient of $(\gamma_+)^4$
while that of the on-shell Lagrangian as the coefficient of $(\gamma_+)^0$.

The simplicity of the MHV form factor resembles that of MHV amplitudes. As discussed in \cite{Brandhuber:2011tv},
the conjugate super-form factor may be obtained by conjugation and Grassmann Fourier transform
of the super-form factor of the chiral primary operator:
\be
{\cal F}^{\text{N}^\text{max}\text{MHV}}_{\cal T}&=&
\delta^4(q-\sum_i\lambda_i{\tilde\lambda}_i)
\delta^{(4)}(\gamma_+-\sum_i\lambda_i\eta_{+;i})
\prod_{i=1}^n \frac{1}{[ i, i+1]}
\;
\nonumber\\
&& \qquad\qquad
\int \prod_i d^4{\tilde \eta}_i\,e^{i\eta_{Ai}{\tilde \eta}_{i}^{A}}
\delta^{(4)}(\sum_i{\tilde \lambda}_i {\tilde \eta}_{i}^{A} u_{+,A}^{a})
\ee
An alternative expression, in terms of a single auxiliary Grassmann integral, is
\be
{\cal F}^{\text{N}^\text{max}\text{MHV}}_{\cal T}=
\delta^4(q-\sum_i\lambda_i{\tilde\lambda}_i)
\delta^{(4)}(\gamma_+-\sum_i\lambda_i\eta_{+;i})
\prod_{i=1}^n \frac{1}{[ i, i+1]}
\int  d^4\omega_{{\dot \alpha}}^a
\,\delta^{(2)}(\omega_{\dot\alpha}{\tilde\lambda}_i^{\dot\alpha}-\eta_{-;i})
\ .
\ee
%

One might wonder whether there exist generalized MHV form factors -- {\it i.e.} MHV form factors
with several operator insertions.  The supersymmetry Ward identities discussed in \cite{Brandhuber:2011tv}
may be easily extended to such cases: for the chiral stress tensor multiplet, a possible solution is obtained
by simply replacing $\gamma_+$ by the sum of the $\gamma_+$ coordinates of all the operators,
$\sum_i \gamma_{+;i}$:
\begin{eqnarray}
{\cal F}^{\text{MHV}}_{{\cal T}_1,\dots {\cal T}_k}={\rm F} \;
\delta^{(4)}(q-\sum_i\lambda_i{\tilde\lambda}_i)
\delta^{(4)}(\sum_{j=1}^k\gamma_{\tilde{j}+}-\sum_i\lambda_i\eta_{+;i})
\delta^{(4)}(\sum_i\lambda_i\eta_{-;i}) \ ,
\label{gen_MHVff}
\ee
with some coefficient ${\rm F}$\footnote{The supersymmetry Ward identities discussed
in \cite{Brandhuber:2011tv} suggest  that ${\rm F}=\prod_{m=1}^n \langle m, m+1\rangle^{-1}$.}.
This expression however contains at most four $\gamma_+$ coordinates
and therefore will not contain {\it e.g.} the generalized form factor of two chiral primary operators (CPO-s)
(which would require eight $\gamma_+$ coordinates)\footnote{It contains however the generalized form
factor of the CPO and any number of chiral Lagrangians.}

We may alternatively consider a generalization of eq.~\rf{modMHVTff} in which  we replace
$1^+_{Aa}\gamma^{a\alpha}_{\tilde{1}+}$ by $\sum_{k=1}^n k^+_{Aa}\gamma^{a\alpha}_{\tilde{k}+}$.
However, extracting  various components -- such as the two-CPO component -- out of this expression one
quickly finds that they cannot be generated by Feynman diagrams; for example, the two-CPO form factor
will not have any scalar external lines.

Last, one could potentially imagine an MHV form factor with different harmonic variables for each operator
that would not allow for a representation in terms of a single eight-dimensional Grassmann $\delta-$function
as it would only satisfy the supersymmetric Ward identities up to equations of motion. However,
simple counting suggests that the generalized form factor of two CPO-s cannot
appear in an MHV generalized form factor because the relevant form factor should have at least
twelve Grassmann $\delta-$functions (eight to saturate the eight $\gamma$ integrals isolating the
two chiral primaries and  four isolating two external scalar fields).

\subsection{Generalized tree-level NMHV form factor of two BPS operators \label{NMHV_2T}}

Generalized form factors, as well as the N$^k$MHV form factors of $(k+1)$ BPS operators,
are an important ingredient in the construction of higher-loop
corrections to correlation functions of CPO-s; in particular, they are necessary to identify
potentially degenerate contributions in which propagators between operators are canceled.
The evaluation of this form factor turns out to be most efficient through the MHV vertex expansion
\cite{CSW, Dixon:2004za}. We should sum over all graphs with two vertices
in which either vertex is a super-form factor, fig.~\ref{2T_nmhv}(a), or one vertex is a generalized two-operator
MHV form factor and the other one is a regular MHV amplitude, fig.~\ref{2T_nmhv}(b):
%
\be
\mathcal{F}^{NMHV}=\mathcal{F}^{NMHV}_{(a)}+\mathcal{F}^{NMHV}_{(b)} \ .
\ee
Since, as we argued in the previous section, an MHV generalized form factor for the lowest component
of the chiral stress tensor multiplet does not exist, the diagrams of the second type have vanishing contributions
\be
\mathcal{F}^{NMHV}_{(b)} = 0 \ .
\ee
For higher components of the chiral stress tensor multiplet this contribution may be nonvanishing; we will
however not be interested in it here.


\begin{figure}[ht]
\begin{center}
\includegraphics[height=33mm]{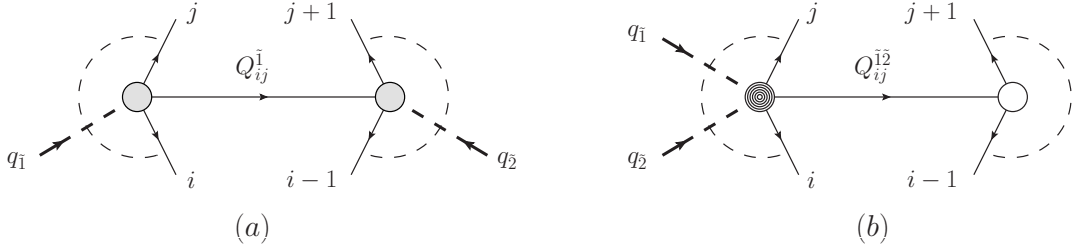}
\caption{The two generic terms in the MHV vertex expansion of the generalized super-form factor with
two insertions of the chiral stress tensor. Grey blobs represent regular single-operator MHV super-form factors,
the hashed blob represents an MHV generalized super-form factor and the white blob represents a regular MHV scattering super-amplitude. The momenta carried by the two operators are $q_{\tilde 1}$ and
$q_{\tilde 2}$.\label{2T_nmhv}}
\end{center}
\end{figure}

Relegating the details to Appendix~\ref{evaluation_of_formfactors}, we list here the final result for
$\mathcal{F}^{NMHV}_{(a)}$:
\be
\mathcal{F}^{NMHV}_{(a)}
&=&{}
\delta^8(1^+_a\gamma_{\tilde{1}+}^a+2^+_a\gamma_{\tilde{2}+}^a-\sum_{m=1}^n\eta_m\lambda_m)
\delta^4(q_{\tilde{1}}+q_{\tilde{2}}-p_1\cdots-p_n)\nonumber\\
&&\qquad
\times\; \frac{1}{3!\prod_{m=1}^n\langle m\,m+1\rangle}
\Bigg[\sum_{w=1}^2\sum_{i=1}^n\sum_{j=i+1}^{i-2}A^{\tilde{w}}_{ij}
         +\sum_{w=1}^2\sum_{i=1}^nB^{\tilde{w}}_i\Bigg]
\label{2BPS}
         \ ,
\ee
where $A^{\tilde{w}}_{ij}$ and $B^{\tilde{w}}_{i}$ are defined as:
\be
&&A^{\tilde{w}}_{ij}
={}\epsilon^{TABC}\big(\eta_{Ti}(Q^{\tilde{w}}_{ij})^2-w^+_{Ta}
\langle\gamma^a_{\tilde{w}+}|Q^{\tilde{w}}_{ij}|i]+\sum_{r=i}^j\eta_{Tr}\langle r|Q^{\tilde{w}}_{ij}|i]\big)
(w^+_{Ab}\langle\gamma^{b}_{\tilde{w}+}i\rangle-\sum_{r=i}^j\eta_{Ar} \langle ri\rangle)\nonumber\\
&&\quad\times
\big(w^+_{Bc}\langle\gamma^{c}_{\tilde{w}+}i\rangle-\sum_{r=i}^j\eta_{Br}\langle ri\rangle\big)
\big(w^+_{Cd}\langle\gamma^{d}_{\tilde{w}+}i\rangle-\sum_{r=i}^j\eta_{Cr}\langle ri\rangle\big)
\frac{1}{(Q^{\tilde{w}}_{ij})^2(Q^{\tilde{w}}_{i+1j})^2}\frac{\langle jj+1\rangle}
{\langle ji\rangle\langle ij+1\rangle} \ ,~~~
\\
&&
B_i^{\tilde{w}}=-\frac{3w^+_{Ab'}w^+_{Bc'}\epsilon^{b'c'}\epsilon^{TABC}}{2(Q_{ii}^{\tilde{w}})^2}
\bigg(-\eta_{Ti}\eta_{Ci}\epsilon_{bc}\langle
\gamma^b_{\tilde{w}+}i\rangle\langle\gamma^c_{\tilde{w}+}i\rangle
-\tfrac{1}{4}\epsilon^{a'd'}w^+_{Ta'}w^+_{Cd'}\langle
\gamma^1_{\tilde{w}+}\gamma^1_{\tilde{w}+}\rangle
\langle\gamma^2_{\tilde{w}+}\gamma^2_{\tilde{w}+}\rangle\nonumber\\
&&\quad
+\eta_{Ti}w^+_{Cb}
\epsilon^b_{\phantom{b}c}\langle\gamma^c_{\tilde{w}+}i\rangle\delta_{ad}\langle
\gamma^a_{\tilde{w}+}\gamma^d_{\tilde{w}+}
\rangle-\eta_{Ci}w^+_{Tb}\epsilon^b_{\phantom{b}c}\langle
\gamma^c_{\tilde{w}+}i\rangle\delta_{ad}\langle\gamma^a_{\tilde{w}+}\gamma^d_{\tilde{w}+}
\rangle\bigg)\ . ~~~
\ee
It is not difficult to check that the simplest such form factor has two external scalars and it has no
momentum dependence, as one might expect based on Feynman diagram calculation.

\subsection{Tree-level MHV form factor of a non-BPS scalar operator \label{tree_su2}}

As discussed in \S~\ref{general_unitarity}, one of the building blocks of correlation functions
of non-BPS operators are their tree-level form factors and generalized form factors.
Non-BPS operators with definite scaling dimension are typically
linear combinations of simpler single-term operators with the same quantum numbers. Their
relative coefficients are functions of the coupling constant and, at least for long operators, may be
determined using the integrability of the dilatation operator of $\NeqFour$ sYM theory
\cite{Staudacher:2004tk, Beisert:2005fw}.
The form factors of non-BPS operators may be therefore interpreted -- order by order in perturbation theory --
as the sum of form factors of these single-term operators with the same coupling-dependent coefficients.
We thus need to focus on them.

The operator mixing determining the eigenvectors of the dilatation operator is, in general, rather complicated
with the only constraints arising from charge conservation and the  fact that such mixing can occur only
between operators with the same classical dimension. There exist, however, "closed sectors"~\cite{Beisert:2003jj}
in which
operator mixing involves a very restricted class of operators. An example is the so-called $SU(2)$-sector
which contains operators of any dimension constructed out of two complex scalar fields which are not conjugate
of each other.

The R-symmetry properties of the $\NeqFour$ sYM fields implies that, at tree-level, it is more
efficient to evaluate the form factor of scalar operators of arbitrary charges rather than restricting
ourselves to specific charges .
We carry out this calculation in Appendix~\ref{evaluation_of_formfactors} using a BCFW recursion relation.
Defining
\be
\label{Hfactor}
\mathrm{H}_{aAbB}&=&\eta_{Aa}\eta_{Bb}-\eta_{Ba}\eta_{Ab}+\delta_{ab}\eta_{Ba}\eta_{Ab},
\\
\label{Sigmamat}
\left(\Sigma_{a_1b_1}\right)^{\alpha}_{\phantom{\alpha}\gamma}&=&\lambda_{a_1}^{\alpha}
\lambda^{\beta}_{b_1}\varepsilon_{\beta\gamma} \ ,
\ee
the MHV super-form factor of the operator
${\cO}_{{A_1B_1}\dots{A_kB_k}}=\mathrm{Tr}(\phi_{A_1B_1}\cdots\phi_{A_kB_k})$ with $n$ external fields
is given by:
\be
\langle{\widetilde\cO}_{{A_1B_1}\dots{A_kB_k}}(q)|1\dots n \rangle=
\frac{\delta^4(q-\sum_{l=1}^np_l)}{\prod_{m=1}\langle m, m+1\rangle}
\sum_{\{a_1,b_1\cdots,a_k,b_k\}}\left(\prod_{i=1}^k\mathrm{H}_{a_iA_ib_iB_i}\right)
\mathrm{Sp}\left(\prod_{j=1}^k\Sigma_{a_jb_j}\right),
\label{scalarNBPS}
\ee
where the first sum runs over all over the sets $\{a_1,b_1\cdots,a_k,b_k\}$ with
$a_1\leq b_1<a_2\leq b_2\cdots b_{k-1}<a_k\leq b_k$ and its cyclic permutations
and $\mathrm{Sp}$ stands for the trace over the
$SL(2,C)$ indices of the product of $\Sigma$ matrices.
By restricting the pairs $(A_i, B_i)$ to only two values, {\it e.g.} $\{(1,2), (1,3)\}$ or $\{(1,2),(1,4)\}$,
one finds the $SU(2)$ sector operators and their form factors.

The tree-level form factors \rf{scalarNBPS} can be used to construct correlation functions\footnote{They
could, of course, also be used to construct higher-loop form factors for these operators, determine their
mixing and anomalous dimensions, {\it etc}.}; one first constructs
the correlation functions of the single-term operators $\cO_{{A_1B_1}\dots{A_kB_k}}$, then one takes appropriate
linear combinations with integrability-determined coefficients and finally one renormalizes the result by multiplying
with the corresponding factors $Z_{\gamma_i}$, cf. eq.~\rf{Zfactor}.

\subsection{Tree-level MHV form factor of a general twist-2 operator \label{tree_twist2}}

In the next section we will construct examples of $n$-point functions with one twist-2 spin-$S$ operator. These
operators are linear combinations of $\cO_{2,S, x}^{AB, CD}=\mathrm{Tr}(D_+^x\phi^{AB}D_+^{S-x}\phi^{CD})$
\be
{\cal O}^{AB, CD}_{2,S}=\sum_{n=0}^S \,c_{S,n}(\lambda)\cO^{AB, CD}_{2,S,n} \ ,
\label{opS}
\ee
where the coefficients $c_n(\lambda)$ are determined by requiring that these operators have definite
anomalous dimensions. At one loop and if $\phi^{AB}=\phi^{CD}=Z$, they are given \cite{Belitsky:2003ys}
in terms of the Gegenbauer polynomials\footnote{The Gegenbauer polynomials appearing for
operators discussed here are the same as the Legendre polynomials with index $S$.}
\be
\label{nonBPS_1loop}
{\cal O}_{2,S} = (n\cdot D_{1}+n\cdot D_{2})^S\,C_S^{1/2}
\left(\textstyle{\frac{n\cdot D_{2}-n\cdot D_{1}}{n\cdot D_{2}+n\cdot D_{1}}}\right)
\Tr[Z(\xi_2)Z(\xi_1)]\Big|_{\xi_1=\xi_2}
\ ,
\ee
where $n\cdot D$ are covariant derivatives in the adjoint representation in the light-like direction
specified by the vector $n$; one possible choice is $n\cdot D=D_+$. The one-loop coefficients in \rf{opS}
are $c_{S,n,0}=(-1)^n \left(\begin{smallmatrix}S\cr n\end{smallmatrix}\right)^2$; the two-loop
coefficients may be found in \cite{Belitsky:2007jp}.

We will describe in Appendix \ref{evaluation_of_formfactors} the evaluation of the form factors of the
operators ${\cal O}^{AB, CD}_{2,S}$ through  a BCFW recursion relation. It is convenient to
express the result in terms of the $\eta$-dependent combinations $\mathrm{H}$ in eq.~\rf{Hfactor}:
\be
&&\!\!\!\!\!\!\!\!
\langle {\widetilde \cO}_{2,S, x}^{AB, CD}|1\dots n \rangle
=\frac{\delta^4(q-\sum_{l=1}^np_l)}{\prod_{m=1}^n\langle m, m+1\rangle}
\sum_{\{a,b,c,d\}}\mathrm{H}_{aAbB}\mathrm{H}_{cCdD}\sum_{k=b}^{c-1}\sum_{l=d}^{a-1}
\left(\sum_{r=l+1}^k p_r^-\right)^x\left(\sum_{s=k+1}^lp_s^-\right)^{S-x}
\nonumber\\
&&\qquad \times
\left(\frac{\langle b|\sigma^-\cancel{p}_k|c\rangle}{2p_k^-}
+\frac{\langle b|\cancel{p}_{k+1}\sigma^-|c\rangle}{2p_{k+1}^-}-\langle bc\rangle\right)
\Bigg(\frac{\langle d|\sigma^-\cancel{p}_l|a\rangle}{2p_l^-}
+\frac{\langle d|\cancel{p}_{l+1}\sigma^-|a\rangle}{2p_{l+1}^-}-\langle da\rangle\Bigg) \ ,
\label{twist_2_form_factor}
\ee
where $q$ is the momentum of ${\widetilde {\cal O}}$,
the sum runs over all sets $\{a,b,c,d\}$ where $a\leq b<c\leq d$ or $d<a\leq b<c$ {\it etc}.
For small number of external particles this expression may be easily verified using Feynman graphs.
As in the case of the
scalar non-BPS operators, taking the appropriate linear coupling constant-dependent combinations of these
expressions yields the form factors of twist-2 spin-$S$ operators with definite anomalous dimensions.

\section{Examples of correlation function construction\label{examples}}

Using the form factors constructed in the previous section we shall construct examples of correlation
functions to leading order (LO) and next-to-leading order (NLO). Some of these correlators have been known for some
time and we reproduce their expressions. We will also discuss various limits and properties of our results.

\subsection{Correlators of BPS operators}

Correlation functions of four chiral stress tensor multiplets have been evaluated to high loop order
\cite{Eden:2012tu, Eden:2012fe} using a hidden symmetry of their position space integrand \cite{Eden:2011we}.
We will discuss here, to a much lower loop order, a unitarity-based approach to the same correlator. As we will
see, this construction generalizes quite easily to correlators of any number of operators.

The simple structure of the CPO-s implies that the construction of their 4-point function
through generalized unitarity is almost
identical to that in terms of Feynman diagrams. From \rf{MHVff} it follows that the (MHV or anti-MHV)
form factor with two external lines of the chiral primary operator has no momentum dependence, which is the same
as the vertex containing the source of the operator and two scalars. For more operators we need to use the
generalized form factors; nevertheless, for scalar operators, they are quire similar to the results of Feynman
diagram calculations.

R-charge conservation implies that the correlation function of four CPO-s (two-index symmetric
traceless dimension-2 operators) is determined by six functions of the positions of the operators \cite{Eden:1998hh}:
\be
\label{general4CPO}
&&
\langle \Tr[\phi^{I_1}\phi^{J_1}]\Tr[\phi^{I_2}\phi^{J_2}]
             \Tr[\phi^{I_3}\phi^{J_3}]\Tr[\phi^{I_4}\phi^{J_4}]  \rangle
\\
&&\quad
=a_1\delta_{12}^2\delta_{34}^2+a_2\delta_{13}^2\delta_{24}^2+a_3\delta_{14}^2\delta_{23}^2
+b_1\delta_{13}\delta_{14}\delta_{23}\delta_{24}
+b_2\delta_{12}\delta_{14}\delta_{32}\delta_{34}
+b_3\delta_{12}\delta_{13}\delta_{42}\delta_{43} \ .~~~
\nonumber
\ee
with
\be
\delta_{ij}^2 = \delta^{I_i}_{\{I_j}\delta^{J_i}_{J_j\} }
\qquad
\delta_{13}\delta_{14}\delta_{23}\delta_{24}=
\delta^{\{I_1}_{\{I_3|}\delta^{J_1\}}_{\{J_4}\delta^{\{I_2}_{I_4\}}\delta^{J_2\}}_{|J_3\}} \ ,
\ee
where $I_i$ and $J_i$ are $SO(6)$ indices.
As mentioned before, since the sources are singlets under the $\NeqFour$ gauge group, each term in the
momentum space correlator may be extracted from un-ordered source-field scattering amplitudes.

\begin{figure}
  \centering
  \subfigure[]{%
    \includegraphics[height=35mm]{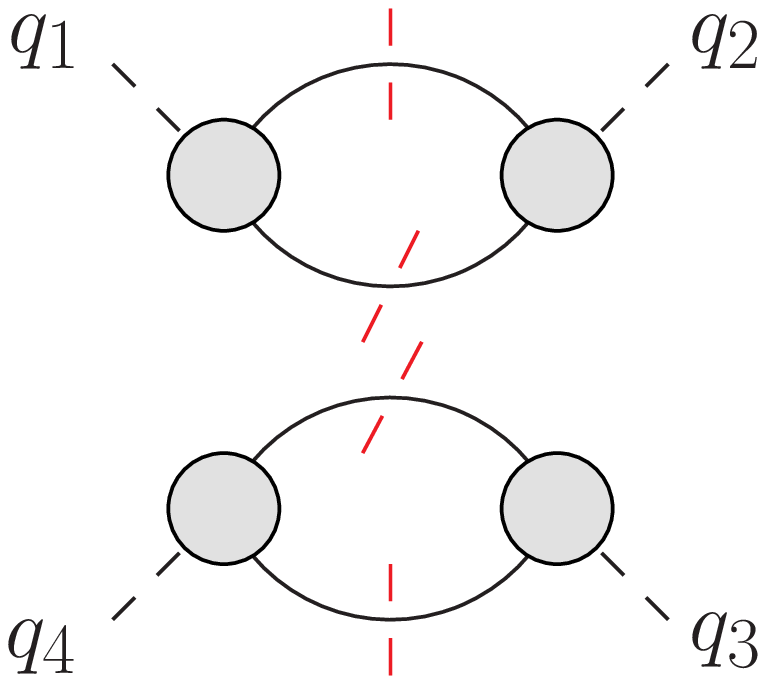}
    \label{qcut_4pt_disconnected}
  }
  ~~~~
  \subfigure[]{%
    \includegraphics[height=35mm]{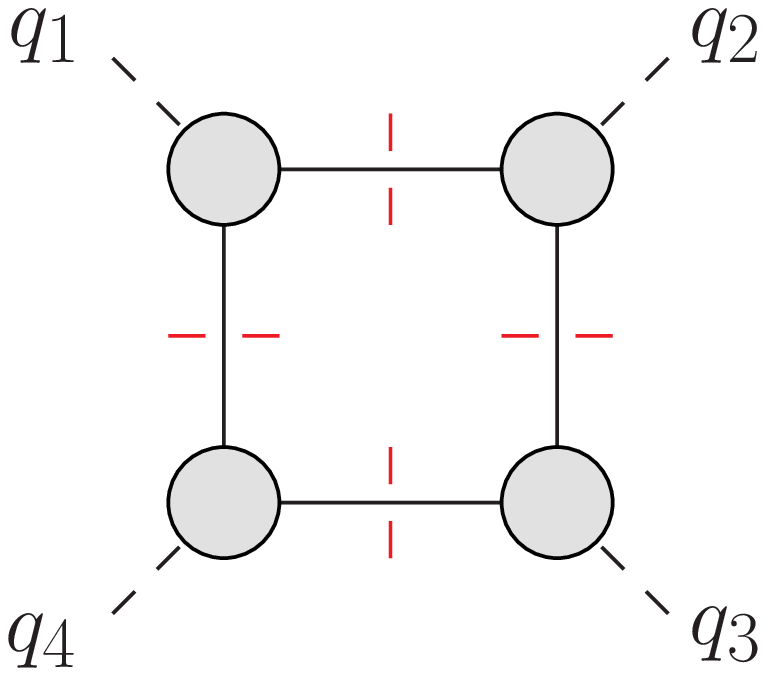}
    \label{qcut_4pt}
  }
  \caption{Disconnected and connected quadruple cuts.}
  \label{quadruplecut}
\end{figure}

\subsubsection{Leading order}

As it is well-known, at this order the contributions to the (Fourier-transform of the) coefficient functions $a_i$
come only from disconnected graphs. In our language, the generalized (quadruple) cut contributing to {\it e.g.}
$a_1$ is shown in fig.~\ref{qcut_4pt_disconnected}.
The $b_i$ coefficients receive contributions only from connected graphs.
To determine them one may consider two-particle cuts with the 2-operators form factors discussed in
\S~\ref{NMHV_2T}. A close inspection of these form factors with two external fields implies however
that there is always a propagator between the two operators; this in turn implies that the leading order correlator
is determined by its quadruple cuts, such as the one shown in fig.~\ref{qcut_4pt} and its non-cyclic
permutations. Depending on the specific choice of R-charge carried by the four operators only some of these
cuts may exist.

Choosing as representatives of the four CPO-s the operators
$\cO_1=\Tr[\phi^{12}\phi^{12}]$, $\cO_2=\Tr[\phi^{34}\phi^{34}]$,
$\cO_3=\Tr[\phi^{13}\phi^{13}]$ and $\cO_4=\Tr[\phi^{24}\phi^{24}]$
only the disconnected quadruple cut is non-zero. This correlation function determines
the coefficient $a_1$.  The relevant form factors which determine the quadruple cut are
extracted from \rf{MHVff} (by appropriately choosing the harmonic variables) and are constant;
it is therefore trivial to find that the momentum space representation of this correlator and thus of $a_3$ is
\footnote{Some UV regularization is assumed.}
\be
{\tilde a_3}&=&
\langle {\widetilde\cO}_1(q_1){\widetilde\cO}_2(q_2){\widetilde\cO}_3(q_3){\widetilde\cO}_4(q_4)\rangle
\\
&=&
\delta(q_1+q_2)\delta(q_3+q_4)
\int\frac{d^4 p}{(2\pi)^4}\frac{1}{p^2(p+q_1)^2}\int\frac{d^4 l}{(2\pi)^4}\frac{1}{l^2(l+q_3)^2} \ .
\ee
It is easy to Fourier-transform this expression to position space before the $p$ and $l$ integrals are
carried out, with the expected result. By permuting the labels of the various operators it is not difficult to find
the coefficients $a_1$ and $a_2$.

Choosing as representatives of the four CPO-s the operators
$\cO_1=\Tr[\phi^{12}\phi^{23}]$, $\cO_2=\Tr[\phi^{14}\phi^{24}]$,
$\cO_3=\Tr[\phi^{13}\phi^{13}]$ and $\cO_4=\Tr[\phi^{24}\phi^{34}]$
the only non-zero quadruple cut is the one shown in fig.~\ref{qcut_4pt}. From the R-charge assignment
it is easy to see that this correlator determines the coefficient $b_2$ in eq.~\rf{general4CPO}.
The quadruple cut may be easily evaluated to be consistent with
\be
{\tilde b_2}=
\langle {\widetilde\cO}_1(q_1){\widetilde\cO}_2(q_2)
{\widetilde\cO}_3(q_3){\widetilde\cO}_4(q_4)\rangle
= \delta^4(\sum_{i=1}^4q_i)I_{4m}[1](q_1, q_2, q_3, q_4) \ .
\label{b2t_LO}
\ee
where $I_{4m}[1](q_1, q_2, q_3, q_4)$ is the standard 4-mass scalar box integral. Previous arguments,
based on the structure of the generalized form factor with two CPO insertions, imply that this should be
the complete result. One may nevertheless check the relevant 2-particle cuts are correctly reproduced.
Thus, this is the complete momentum-space form ${\widetilde b_2}$ of $b_2$.
As in the case of the disconnected contribution, it is not difficult to Fourier-transform this
expression\footnote{To this end one treats independently the momentum of each propagator and introduces
a $\delta-$function for each three-point vertex. Using an integral representation of these four $\delta-$functions
makes all integrals trivial.}
and recover the standard position-space form of $b_2$:
\be
b_2 = \frac{1}{x_{12}^2x_{23}^2x_{34}^2x_{41}^2} \ .
\ee
This expression is annihilated by the special conformal generators; consequently, their momentum space
expression \rf{scgenerators} should annihilate the 4-mass box integral \rf{b2t_LO}.

\subsubsection{Next-to-leading order \label{NLO_BPS_4pt}}

As in the case or regular scattering amplitudes, we may either proceed systematically through the maximal cut
method\footnote{Depending on the complexity of the correlator one may also use integral reduction
strategy of \cite{Johansson:2012zv} generalized to all-massive external legs, which reduces integrals to a
basis at the level of maximal cuts.} or, in simple cases, bypass some
steps analyze directly a spanning set of color-dressed cuts.  In the case
at hand -- the correlation function of four CPO-s -- we may use the fact that we expect that
it should be UV-finite\footnote{A divergence would signal the presence of an anomalous dimension for at least
one of the operators,  which contradicts the fact that the operators are chiral primaries.}. Thus, all contributing
Feynman integrals should have at least five propagators. Ignoring disconnected and non-1PI contributions \footnote{The cut
in fig.~\ref{NLO_cuts}(a) also captures the non-1PI correction to the correlation function. The additional cut
that is needed to determine the disconnected part may be interpreted as determining the next-to-leading order
correction to the two-point function of two of the four operators.},  this implies that the contributions to the
momentum
space form of the connected part of the coefficient functions $a_i, \, b_j$ should be determined by the cuts in
fig.~\ref{NLO_cuts} as well
as the permutations of their external leg labels. As explained in \S~\ref{general_unitarity},  each blob is either
a color-dressed form factor or a color-dressed amplitude; depending on the specifics of operators, color-dressed
form-factors of single-trace operators may be -- up to the color factor -- the same as color-ordered form factors.
An example of such an operator is the CPO.

\begin{figure}[ht]
\begin{center}
\includegraphics[height=28mm]{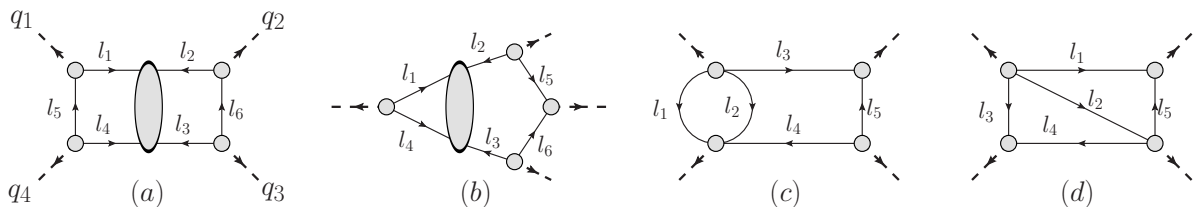}
\caption{Color-dressed cuts for connected part of the next-to-leading order 4-point correlator.
There are 2 cuts of type $(a)$, 4 cuts of type $(b)$,  4 cuts of type $(c)$ and 2 cuts of type $(d)$.
Other members of each set are related by relabeling.  Each blob with an external line is a (generalized) color-dressed
form factor; blobs with no external lines are color-dressed on-shell scattering amplitudes.  \label{NLO_cuts}}
\end{center}
\end{figure}

The color algebra is nontrivial only for the cuts in figs.~\ref{NLO_cuts}$(a)$ and \ref{NLO_cuts}$(b)$. Writing
the 4-point amplitude as
\be
{\cal A}(l_1, l_2, l_3, l_4)=M_1\,f^{a_1a_2}{}_{d}f^{a_3a_4 d}
                                        +M_2\,f^{a_1a_3}{}_{d}f^{a_4a_2 d}
                                        +M_3\,f^{a_1a_4}{}_{d}f^{a_2a_3 d}
\label{fullA4BCJform}
\ee
and using the fact that the 2-point form factors are just Kronecker delta-functions in color space (with indices
in the adjoint representation -- {\it e.g.} the color factors of the two form factors at the left of the cut
in fig.~\ref{NLO_cuts}(a) are just $\delta^{a_1 a_4}$) it follows that the cuts in figs.~\ref{NLO_cuts}$(a)$ and
\ref{NLO_cuts}$(b)$ are proportional to
\be
(M_2-M_1)f^{a_1a_4}{}_{d}f^{a_4a_2 d}\delta_{a_1 a_4}\delta_{a_3 a_2} \ .
\ee
Further using the relation between $M_i$ and color-ordered amplitudes it follows that the cuts in
figs.~\ref{NLO_cuts}$(a)$ and \ref{NLO_cuts}$(b)$ are
\be
\label{NLOexp1}
{\cal C}_{\ref{NLO_cuts}(a)}&=&
{\widetilde{\cal F}}(q_1, l_1, l_2)
{\widetilde{\cal F}}(q_2, l_6, l_2)
{\widetilde{\cal F}}(q_3, l_3, l_6)
{\widetilde{\cal F}}(q_4, l_5, l_4)
A(l_1, l_3, l_4, l_2)
\\
{\cal C}_{\ref{NLO_cuts}(b)}&=&
{\widetilde{\cal F}}(q_1, l_1, l_4)
{\widetilde{\cal F}}(q_2, l_5, l_2)
{\widetilde{\cal F}}(q_3, l_6, l_5)
{\widetilde{\cal F}}(q_4, l_3, l_6)
A(l_1, l_3, l_4, l_2)
\label{NLOexp2}
\ee
where $A(l_1, l_3, l_4, l_2)$ is the color-ordered amplitude\footnote{This becomes the corresponding
super-amplitude if one uses super-form factors.} with the relevant external legs.\footnote{The coefficients
$M_i$ in the four-point amplitude \rf{fullA4BCJform} may be chosen to obey  color/kinematics duality
\cite{Bern:2008qj}
on the unique off-shell leg of this amplitude. Thus, the corresponding cuts in figs.~\ref{NLO_cuts}$(a)$ and
\ref{NLO_cuts}$(b)$ obey it as well.}
This expression may also be justified using the photon decoupling identity.

As in the leading order case, choosing as representatives of the four CPO-s the operators
$\cO_1=\Tr[\phi^{12}\phi^{12}]$, $\cO_4=\Tr[\phi^{34}\phi^{34}]$,
$\cO_2=\Tr[\phi^{13}\phi^{13}]$ and $\cO_3=\Tr[\phi^{24}\phi^{24}]$
determines the contribution to the coefficient $a_3$.  It is not difficult to see that the cuts
in figs.~\ref{NLO_cuts}$(b)$, $(c)$ and $(d)$ vanish identically because $SO(6)$ invariance forbids
propagators between form factors. The cut in fig.~\ref{NLO_cuts}$(a)$ is
\be
\label{disconnected_4pt}
{\cal C}_{\ref{NLO_cuts}(a)}= 1
\ee
which implies that ${\widetilde a_3}$ is given by the bow tie integral (see fig.~\ref{integrals_2loop_4pt}$(a)$):
\be
{\widetilde a_3}=\int \frac{d^4 p d^4 l}{(2\pi)^8}\frac{1}{p^2(p+q_1)^2(p+q_1+q_4)^2l^2(l+q_2)^2(l+q_2+q_3)^2}
=\text{BTie}(4,1|2,3) \ ;
\ee
This may be Fourier-transformed to position space with the result
\be
a_3 = \frac{1}{x_{41}^2 x_{23}^2}\int\frac{d^4 x_0}{x_{01}^2x_{02}^2x_{03}^2x_{04}^2} \ .
\ee
Permuting the labels of the operators we may similarly obtain $a_1$ and $a_2$.

\begin{figure}[ht]
\begin{center}
\includegraphics[height=26mm]{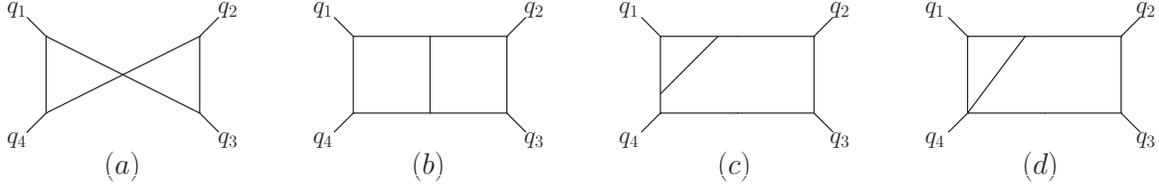}
\caption{2-loop integrals: $\text{BTie}(4,1|2,3)$, $\text{DB}(4,1|2,3)$,
$\text{TriP}(4,1|2,3)$ and $\text{TriB}(4|1|2,3)$.   \label{integrals_2loop_4pt}}
\end{center}
\end{figure}

To determine the NLO correction to $b_2$ we choose as representatives of the four CPO-s the
operators to be $\cO_1=\Tr[\phi^{12}\phi^{23}]$, $\cO_2=\Tr[\phi^{14}\phi^{24}]$
$\cO_3=\Tr[\phi^{13}\phi^{13}]$ and $\cO_4=\Tr[\phi^{24}\phi^{34}]$.
All cuts are nontrivial:
\be
\label{connected_ab}
{\cal C}_{\ref{NLO_cuts}(a)}&=& \frac{\langle l_1 l_4\rangle\langle l_2 l_3\rangle}
                                                          {\langle l_1 l_3\rangle\langle l_4 l_2\rangle}=\frac{(q_1+q_4)^2}{(l_1+l_3)^2}
~,~~\quad\qquad\qquad\qquad
{\cal C}_{\ref{NLO_cuts}(b)}=\frac{\langle l_1 l_4\rangle\langle l_2 l_3\rangle}
                                                          {\langle l_1 l_3\rangle\langle l_4 l_2\rangle}=\frac{q_1^2}{(l_1+l_3)^2}
                                                          ~~~
\\
\label{connected_cd}
{\cal C}_{\ref{NLO_cuts}(c)}&=&\frac{1}{2 l_1\cdot l_2}
\left[\frac{\langle l_1 l_3\rangle  [l_1 l_4]}
{\langle l_2 l_3\rangle [l_2 l_4]}
+\frac{\langle l_2 l_3\rangle  [l_2 l_4]}{\langle l_3 l_1\rangle [l_4 l_1]}+2\right]
~,\qquad
{\cal C}_{\ref{NLO_cuts}(d)}= \frac{\langle l_1 l_3\rangle [l_4 l_5]}
{\langle l_1 l_2\rangle\langle l_2 l_3\rangle[l_2 l_5][l_4 l_2]}  \ .
\ee
%
%
Using the methods developed for the construction of scattering amplitudes of fundamental fields ({\it e.g.}
matching onto an ansatz, or simply expressing the cuts in terms of momenta and inspecting the result) it
is not difficult to find (the integral representation of) a function that has these cuts:
\be
{\widetilde b_2}
&=&\delta^{(4)}(\sum_{i=1}^4 q_i)\Big[
    (q_1+q_2)^2\text{DB}(1,2|3,4)+(q_1+q_4)^2\text{DB}(4,1|2,3)
 \cr
&&\quad\quad\quad~
+q_1^2\text{TriP}(1|2,3,4)+q_2^2\text{TriP}(2|3,4,1)+q_3^2\text{TriP}(3|4,1,2)+q_4^2\text{TriP}(4|1,2,3)
\nonumber\\[1pt]
&&\quad\quad\quad~
-\text{TriB}(1|2|3,4)-\text{TriB}(2|1|3,4)-\text{TriB}(2|3|4,1)-\text{TriB}(3|2|4,1)
\cr
&&\quad\quad\quad~
-\text{TriB}(3|4|1,2)-\text{TriB}(4|3|1,2)-\text{TriB}(4|1|2,3)-\text{TriB}(1|4|2,3)
\Big]
\label{2loop4pt}
\ee
Here $\text{DB}(a,b|c,d)$ is the double-box integral with legs $a$ and $b$ on one one-loop box sub-integral,
$\text{TriP}(a|b,c,d)$ is the triangle-pentagon integral with external leg $a$ on the triangle point and
$\text{TriB}(a|b|c,d)$ is the triangle-box integral with external leg $a$ at the 4-point vertex and external
leg $b$ on the triangle point; these integrals are shown in figs.~\ref{integrals_2loop_4pt}$(b)$, $(c)$ and $(d)$,
respectively.

The expression for ${\widetilde b_2}$ in eq.~\rf{2loop4pt} can be Fourier-transformed to position space;
we immediately recover the expression for the integral representation of the coefficient $b_2$ found in
\cite{Eden:1998hh}. The easiest way to carry out the Fourier-transform is to (1) treat the momentum of
each propagator as independent and introduce the appropriate $\delta-$functions enforcing momentum
conservation at each vertex (2) use an integral parametrization of the $\delta-$functions to carry out
the momentum integrals and (3) at this stage the Fourier-transform becomes trivial and identifies the
integration parameters for the $\delta-$functions of vertices with external lines with the position of the
operators.
The resulting integrals (over the position of the vertices with no external lines) may be further
reduced using a strategy outlined in the Appendix~A of \cite{Beisert:2002bb} and we find the expected result
\cite{Eden:1999kh}; we collect some of the details in Appendix~\ref{Fourier}

\subsubsection{Brief comments on supersymmetric methods \label{susy_methods}}

While the calculation in the previous subsection was carried out in components, it is not difficult to
streamline it by making use of the super-form factor of the chiral stress tensor multiplet. In such an
approach it is not necessary to choose representatives of the CPOs; rather, products of the harmonic
variables appearing in each form factor play the role of the various $\delta-$functions appearing in
eq.~\rf{general4CPO}. In the following we will continue constructing component correlation
functions; in this subsection however we will illustrate the supersymmetric methods and recover the
LO and NLO correction to the four-point correlation function of chiral stress tensor multiplets.

The quadruple cut in fig.~\ref{qcut_4pt} is given by\footnote{We use the standard convention
$p\rightarrow -p ~~\mapsto  |p\rangle \rightarrow -|p\rangle,\;|p] \rightarrow |p]$.
The Grassmann coordinates $\eta[i]$ and $\eta[j]$ have different harmonic coordinates
since they correspond to different multiplets -- $\eta[x]_{-,i}^a=\eta_{A, i}\,u_x{}_A^{+a}$ --
where $x$ labels the operator to which $\eta$ belongs to.}
\be
\label{qcut_MHV_4}
\langle {\cal T}(q_1){\cal T}(q_2)
{\cal T}(q_3){\cal T}(q_4)  \rangle\Big|_{\text{4-cut}}
\!\!&=&\!\!
\delta^4(\sum q_i)
\prod_{i=1}^4\frac{1}{\langle l_i l_{i+1}\rangle\langle l_{i+1} l_i\rangle}
\int d^4\eta_{1}d^4\eta_{2}d^4\eta_{3}d^4\eta_4
\\
&&
\delta^{(4)}(\gamma_{\tilde{1}+}-l_1\eta[1]_{+;1}+l_2\eta[1]_{+;2})
\delta^{(4)}(l_1\eta[1]_{-;1}-l_2\eta[1]_{-;2})
\cr
&&
\delta^{(4)}(\gamma_{\tilde{2}+}-l_2\eta[2]_{+;2}+l_3\eta[2]_{+;3})
\delta^{(4)}(l_2\eta[2]_{-;2}-l_3\eta[2]_{-;3})
\cr
&&
\delta^{(4)}(\gamma_{\tilde{3}+}-l_3\eta[3]_{+;3}+l_4\eta[3]_{+;4})
\delta^{(4)}(l_3\eta[3]_{-;3}-l_4\eta[3]_{-;4})
\cr
&&
\delta^{(4)}(\gamma_{\tilde{4}+}-l_4\eta[4]_{+;4}+l_1\eta[4]_{+;1})
\delta^{(4)}(l_4\eta[4]_{-;4}-l_1\eta[4]_{-;1})
\nonumber
\ee
Carrying out the $\eta$ integrals (either directly or though the methods described in \cite{Bern:2009xq})
we find the cut of a 4-mass box integral multiplied by the product of harmonic variables $(12)(23)(34)(41)$
and $\delta^{(4)}(\gamma_{\tilde{1}+})
\delta^{(4)}(\gamma_{\tilde{2}+})
\delta^{(4)}(\gamma_{\tilde{3}+})
\delta^{(4)}(\gamma_{\tilde{4}+})$, implying that only the CPO components have non-vanishing
four-point function \cite{Eden:2010zz}.

Through similar manipulations and using the higher-point form factors \rf{MHVff} it
is not difficult to
compute the cuts in fig.~\ref{NLO_cuts}; up to a factor of $\delta^{(4)}(\gamma_{+1})
\delta^{(4)}(\gamma_{+2})
\delta^{(4)}(\gamma_{+3})
\delta^{(4)}(\gamma_{+4})$, cuts $(a)$ and $(d)$ are:
\be
{\rm C}_{\ref{NLO_cuts}(a)} &=& (14)(23)
                                               + (12)(23)(34)(41)\frac{(q_1+q_4)^2}{(l_1+l_3)^2}
                                               + (13)(32)(24)(41) \frac{(q_1+q_4)^2}{(l_1+l_2)^2}
\\
{\rm C}_{\ref{NLO_cuts}(d)} &=&
                                               (12)(23)(34)(41)\Big[
-\frac{-(q_1+q_2)^2+(l_2-l_5)^2+(l_2+l_3)^2}{(l_2+ l_3)^2(l_2- l_5)^2}
+\frac{q_4^2}{(l_2+ l_3)^2(l_2-l_4)^2}
\cr
&&\qquad\qquad\quad\quad
+\frac{q_2^2 }{(l_1+ l_2)^2(l_2- l_5)^2}
-\frac{-(q_1+q_4)^2+(l_1+l_2)^2+(l_2-l_4)^2}{(l_1+ l_2)^2(l_2- l_4)^2}
\Big] \ .
\ee
We recognize the various momentum-dependent factors as the component cuts found previously.
Together with the cuts obtained by permuting the labels of the external lines these expressions
can be used to recover the results of the \S~\ref{NLO_BPS_4pt}.

\subsubsection{Higher-point correlation function of BPS operators \label{higherpts_BPS}}

The calculation of the correlation function of four scalar BPS operators may be easily extended to
the correlator of any number of operators. The main difference is that, unlike the four-point correlator, the
R-symmetry constraints are more difficult to write in compact form. Similarly to eq.~\rf{general4CPO}, one is
to identify the $SO(6)$ singlets in the product of $n$ ${\bf 20}'$-dimensional representations of
$SO(6)$.
By picking suitable combinations of operators and evaluating their correlation functions we may then
extract the coefficient of each individual $SO(6)$ singlet.

\begin{figure}[ht]
\begin{center}
\includegraphics[height=35mm]{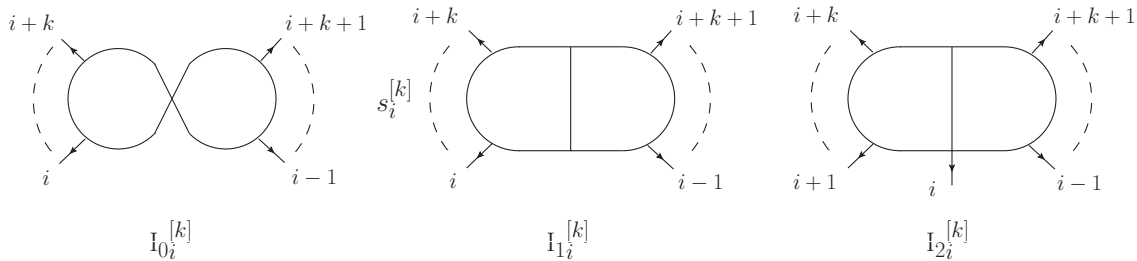}
\caption{General two-loop integrals: $\text{I}_0{}_{i}^{[k]}$, $\text{I}_1{}_{i}^{[k]}$,
$\text{I}_2{}_{i}^{[k]}$.  The integral  $\text{I}_1{}_{i}^{[k]}$  includes the
numerator factor $s_{i}^{[k]}=(\sum_{j=0}^k q_{i+j})^2$. \label{integrals_2loops_npt}}
\end{center}
\end{figure}

The cuts that need to be evaluated are similar to the cuts in fig.~\ref{NLO_cuts} except that each form factor
is replaced with a generalized form factor with arbitrary number of operators subject to the constraint that the
sum of the number of operators is fixed to $n$. It is not difficult to see that the building blocks are identified
by eqs.~\rf{disconnected_4pt}, \rf{connected_ab}, \rf{connected_cd} and they lead to the integrals shown in
fig.~\ref{integrals_2loops_npt}.

Based on the calculation of the 4-point correlation function one can see that integrals of the
type $\text{I}_0{}_{i}^{[k]}$ enter only in the partly-connected (in R-symmetry space) components of the
correlator and integrals of the type $\text{I}_1{}_{i}^{[k]}$ and $\text{I}_2{}_{i}^{[k]}$ appear
in the connected components.
Analyzing the cuts we find that the analog of the $b_i$ coefficients in eq.~\rf{general4CPO}, {\it i.e.} the
coefficient function for the maximally connected $SO(6)$ singlet, defined by the tensor structure
\be
\prod_{i=1}^{n}\delta_{i, i+1} \ ,
\ee
is given by
\be
{\widetilde b_{{\bf 2},\text{max}}}=
\sum_{i=1}^n\sum_{k=0}^{n-2}
\text{I}_1{}_{i}^{[k]}
-\sum_{i=1}^n\sum_{k=1}^{n-2}
\text{I}_2{}_{i}^{[k]}
\nonumber
\label{npfBPS}
\ee
where the integral $\text{I}_1{}_{i}^{[k]}$ has a numerator factor $s_{i}^{[k]}=(q_{i}^{[k]})^2$ with
$q_{i}^{[k]}=\sum_{j=0}^k q_{i+j}$
and the first index on $b$ denotes the fact that the index contraction is analogous to that of $b_2$.

While kinematic restrictions forbid a parity-odd part ({\it i.e.} containing Levi-Civita tensors) in the
four-point correlation function, one may wonder whether such terms should be present in eq.~\rf{npfBPS}.
An analysis of cuts similar to those in figs.~\ref{NLO_cuts}$(a)$, \ref{NLO_cuts}$(b)$ and \ref{NLO_cuts}$(c)$
shows that the integral topologies detected by these cuts do not exhibit a parity-odd component. Cuts involving
two three-point form factors have two components, in which either one of the form factors is MHV while the other
is ${\overline {\rm MHV}}$. While each component has a nontrivial parity-odd part, it cancels in the sum. Thus,
to this order, no parity-odd part should be present in eq.~\rf{npfBPS}.

While it is not difficult (and perhaps useful for specific applications) to Fourier transform this expression
to position space using the details in Appendix~\ref{Fourier} and check its conformal invariance, the result
is rather complicated.
It is more instructive to carry out the Fourier transform under the assumption that we also take the null limit
of the resulting expression and make contact with the $n$-point 1-loop MHV amplitude thus reproducing the results
of \cite{Eden:2010zz}. Indeed, it is not difficult to see that the integrals $\text{I}_0{}_{i}^{[k]}$ and
$\text{I}_2{}_{i}^{[k]}$ are not proportional to $\prod_{i=1}^nx_{i,i+1}^{-2}$ and thus are not sufficiently
singular to contribute to this limit.
%
%
The Fourier  transform of integrals of the type $\text{I}_1{}_{i}^{[k]}$ is (cf. eqs.~\rf{Xpart} and \rf{Ypart})
\be
&&\int (\prod_j d^4 q_j e^{ix_j\cdot q_j}) \text{I}_1{}_{i}^{[k]}
\\
&&\quad
=\frac{x_{i+k, i-1}^2x_{i, i+k+1}^2-x_{i,i+k}^2x_{i-1, i+k+1}^2
+{\cal O}(x_{l, l+1}^2)}{\prod_{j=1}^nx_{j,j+1}^{2}}
\int \frac{dx_0}{x_{i,0}^2x_{i+k,0}^2x_{i+k+1,0}^2x_{i-1,0}^2}
\nonumber
\ee
for $k\ne 1$ and $k\ne n-1$ and subleading for these two cases. Assembling $b_{{\bf 1},\text{max}}$,
the Fourier-transform  of ${\widetilde b_{{\bf 1},\text{max}}}$ and after a trivial change of summation indices
we recover, as expected, eq (4.11) of \cite{Eden:2010zz}.
It is important to note that all contact term integrals in momentum space -- {\it i.e.} integrals with
cancelled propagators -- do not contribute in the null separation limit of the Fourier-transform of the momentum
space correlator. This might have been expected since at this order any contact term
lies along a path between two null-separated operators and such terms have been argued  \cite{Engelund:2011fg}
to be subleading.


\subsection{Correlators of BPS and non-BPS operators \label{BPSNBPS}}

The methods described in previous sections can equally well be applied to correlation
functions involving non-BPS operators, such as a general twist-2 operator. As discussed
in \S~\ref{presentation_and_symmetries},
since these operators are nontrivial sums of monomials with coupling constant-dependent
coefficients, the strategy for the evaluation of their correlation function is to evaluate the correlator
of generic monomials and subsequently sum these components with the appropriate coefficients.
We will illustrate here this construction with the calculation of the three-point function of one non-BPS twist-2 operator and
two BPS scalar operators. We will then generalize this calculation to an arbitrary number of scalar
BPS operators as well as to two twist-2 operators.
Compared to the correlation function of BPS operators, the additional
form factors which are needed are special cases of the general form factor found in \S~\ref{tree_twist2}.

\subsubsection{3-point correlators with one twist-2 operator}

For operators with general R-symmetry indices,
\be
{\cal O}_1
=\mathrm{Tr}(D_+^x\phi^{I_1}D_+^{S-x}\phi^{J_1})
\qquad
{\cal O}_2=\mathrm{Tr}(\phi^{I_2}\phi^{J_2})
\qquad
{\cal O}_3=\mathrm{Tr}(\phi^{I_3}\phi^{J_3})
\qquad
\label{operators}
\ee
the R-index structure of the correlator is
\be
\label{general_3pf}
\langle {\cal O}_1 {\cal O}_2 {\cal O}_3\rangle
= c^x_1\delta^{\{J_2}_{\{J_3|}\delta^{I_2\}}_{I_1}\delta^{J_1}_{|I_3\}}
 +c^x_2\delta^{\{J_2}_{\{J_3|}\delta^{I_2\}}_{J_1}\delta^{I_1}_{|I_3\}}
 \ .
\ee
Depending on the choice of operators, both, one or none of the structures above are allowed. Whenever both
are allowed the coefficient functions $c_1$ and $c_2$ are related by the transformation $x\leftrightarrow S-x$.
%
Here we will choose a particular non-BPS representative in which the two scalar fields are the same,
${I_1}={J_1}$;
in this case the factorization occurs on a term by term basis and its momentum/position dependence is given
by $(c^x_1+c^x_2)$.

\begin{figure}[ht]
\begin{center}
\includegraphics[height=32mm]{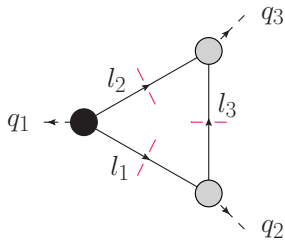}
\caption{Triple cut determining the leading order three-point function in eq.~\rf{general_3pf}. The black dot
denotes the non-BPS operator while the gray dots denote BPS operators.  \label{3pf_cuts_LO}}
\end{center}
\end{figure}

It is convenient to define
\be
\label{Sdef}
S(a,b,x)=(a^-)^{S-x}(b^-)^x + (b^-)^{S-x}(a^-)^x
\  ,
\ee
where the upper index "$-$"denotes a projection onto a null direction ({\it e.g.} $a^-=a_0+a_3$, {\it etc}).
The leading order correlation function is determined by the triple cut in fig.~\ref{3pf_cuts_LO} to be
\be
{\widetilde {c^x_1}}{}^{(0)}+{\widetilde {c^x_2}}{}^{(0)}=\int\prod_{i=1}^3\frac{d^4l_i}{(2\pi)^{4}}
\delta^4(q_3-l_2-l_3)\delta^4(q_2+l_3-l_1)\delta^4(q_1+l_1+l_2)
\frac{
S(l_1, l_2, x)}{l_1^2\,l_2^2\,l_3^2}
\label{tree_nbb}
\ee
The Fourier transform to position space yields the expected result which may then be used to assemble the
correlation function of the full twist-2 spin-$S$ operator with two BPS operators by summing them with the
coefficients in eq.~\rf{nonBPS_1loop}.

\begin{figure}[ht]
\begin{center}
\includegraphics[height=32mm]{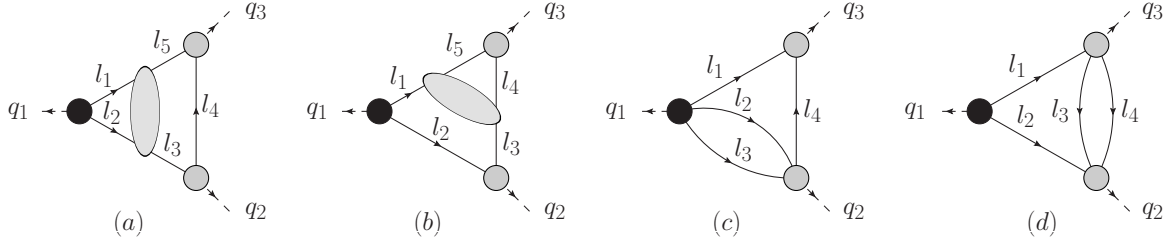}
\caption{Cut topologies which determine the next-to-leading order correction to the three-point function in eq.~\rf{general_3pf}. As in fig.~\ref{3pf_cuts_LO}, the black dot
denotes the non-BPS operator while the gray dots denote BPS operators. \label{3pf_cuts_NLO}}
\end{center}
\end{figure}
\newpage

To construct the next-to-leading order correction to this correlation function we may proceed with the maximal
cut method with the expectation that we will need to consider maximal and next-to-maximal cuts. At this order
however the calculations are sufficiently simple to allow us to proceed directly to next-to-maximal and
next-to-next-to-maximal cuts; up to
the interchange of the two BPS operators (or, alternatively, up to the transformation $q_1\leftrightarrow q_2$)
they are shown in fig.~\ref{3pf_cuts_NLO}.
These cuts are not difficult to evaluate using the form factors described in \S~\ref{formfactors}.
They are:
\be
{\cal C}_{\text{fig.~}\ref{3pf_cuts_NLO}(a)}&=&
S(l_1, l_2, x)\frac{\langle l_1l_2\rangle\langle l_3l_5\rangle}{\langle l_2l_3\rangle\langle l_5l_1\rangle}
=
-S(l_1, l_2, x)\frac{q_1^2}{(l_2+l_3)^2}
\\
{\cal C}_{\text{fig.~}\ref{3pf_cuts_NLO}(b)}&=&
S(l_1, l_2, x)\frac{\langle l_3l_1\rangle\langle l_5l_4\rangle}
{\langle l_1l_5\rangle\langle l_4l_3\rangle}
=
-S(l_1, l_2, x)\frac{q_3^2}{(l_1+l_5)^2} \ .
\ee
Due to the presence of two three-field form factors, the last two cuts receive
contributions from the two configurations in which either one of them is MHV and the other is
$\overline{\rm MHV}$. We find:
\be
{\cal C}_{\text{fig.~}\ref{3pf_cuts_NLO}(c)}
&=&
\Bigg(\frac{l_3^-}{l_2^-}\frac{1}{(l_2+l_3)^2}\Big(
S(l_1, l_2+l_3, x) - S(l_1+ l_2,l_3, x)
\Big)\nonumber\\
&+&\frac{l_4^-}{l_2^-}\frac{1}{(l_2+l_3)^2}\Big(
S(l_1+l_2, l_3, x) - S(l_1, l_2+ l_3, x)
\Big)+S(l_1+l_2, l_3, x)
\nonumber\\
&&\times\bigg(\frac{q^2_1}{(l_2+l_3)^2(l_1+l_2)^2}
-\frac{1}{(l_1+l_2)^2}-\frac{1}{(l_2+l_3)^2}+\frac{q_3^2}{(l_2+l_4)^2(l_1+l_2)^2}\bigg)\nonumber\\
&+&S(l_1, l_2+ l_3, x)
\bigg(\frac{q_2^2}{(l_2+l_3)^2(l_2+l_4)^2}
-\frac{1}{(l_2+l_4)^2}+\frac{1}{(l_2+l_3)^2}\bigg)\Bigg)
\ee

\be
{\cal C}_{\text{fig.~}\ref{3pf_cuts_NLO}(d)}
&=&{}
S(l_1, l_2, x)\bigg(
\frac{q^2_2}{(l_3+l_4)^2(l_2+l_4)^2}
+\frac{q_2^2}{(l_3+l_4)^2(l_2+l_3)^2}+\frac{q_3^2}{(l_3+l_4)^2(l_1+l_3)^2}\nonumber\\
&+&\frac{q_3^2}{(l_3+l_4)^2(l_1-l_4)^2}
+\frac{q_1^2}{(l_2+l_4)^2(l_1-l_4)^2}+\frac{q_1^2}{(l_2+l_3)^2(l_1-l_3)^2}
\nonumber\\
&-&\frac{1}{(l_2+l_4)^2}-\frac{1}{(l_2+l_3)^2}-\frac{1}{(l_1+l_3)^2}-\frac{1}{(l_1-l_4)^2}\bigg) \ .
\ee
Inspecting these expressions and accounting for the symmetries of the cuts it is not difficult to find
the (integrand of the) momentum space correlation  function:
\be
{\widetilde {c^x_1}}{}^{(1)}+{\widetilde {c^x_2}}{}^{(1)}&=&
S(l_1, l_2, x)
\nonumber\\
&&\qquad \times\Bigg(
  q_1^2\parbox{1.5cm}{\includegraphics[width=2.0cm]{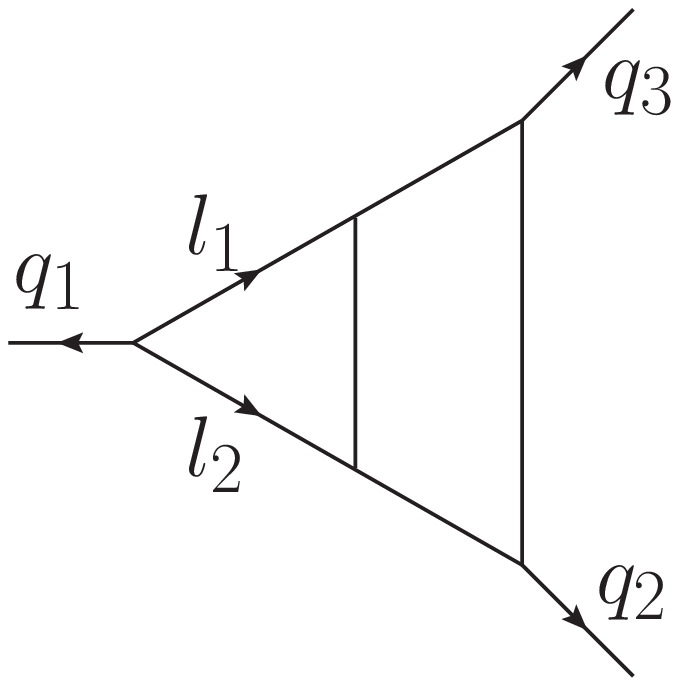}}
+q_3^2\parbox{1.5cm}{\includegraphics[width=2.0cm]{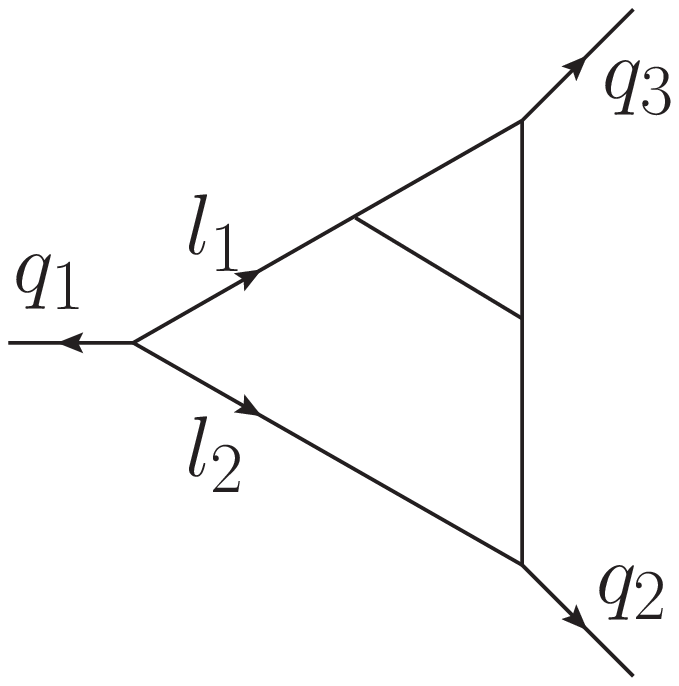}}
+q_2^2\parbox{1.5cm}{\includegraphics[width=2.0cm]{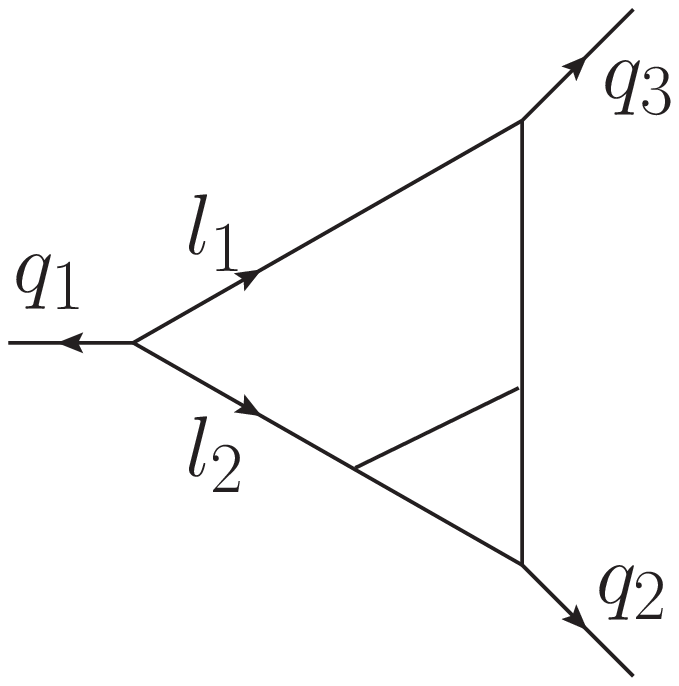}}
       -\parbox{1.5cm}{\includegraphics[width=2.0cm]{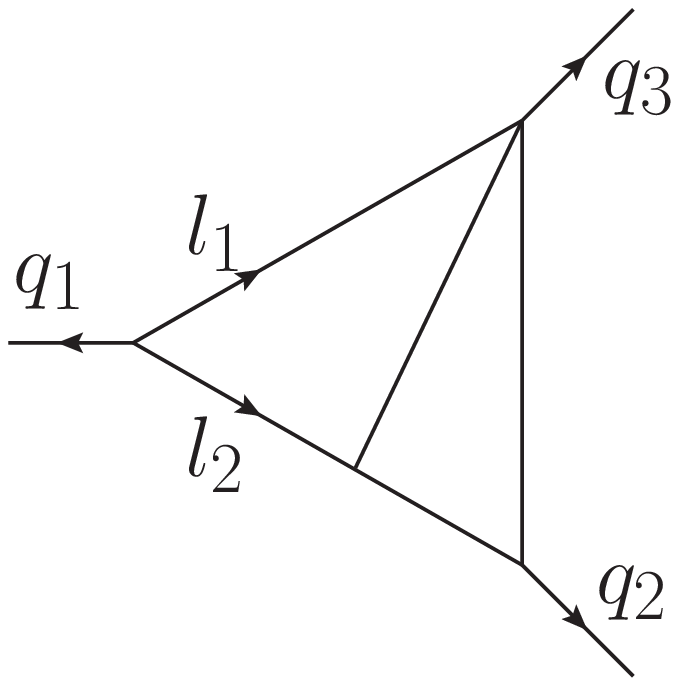}}
       -\parbox{1.5cm}{\includegraphics[width=2.0cm]{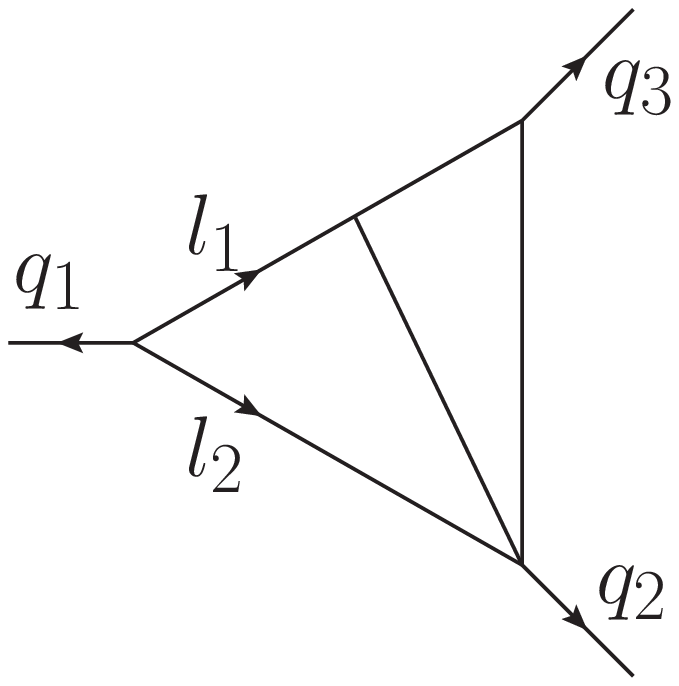}}~~~\Bigg)
\nonumber\\
&&
+\bigg(\frac{l_4^-}{l_2^-}S(l_1+l_2, l_3, x)
+\frac{l_5^-}{l_2^-}
S(l_1, l_2+l_3, x)\bigg)
\parbox{1.5cm}{\includegraphics[width=2.0cm]{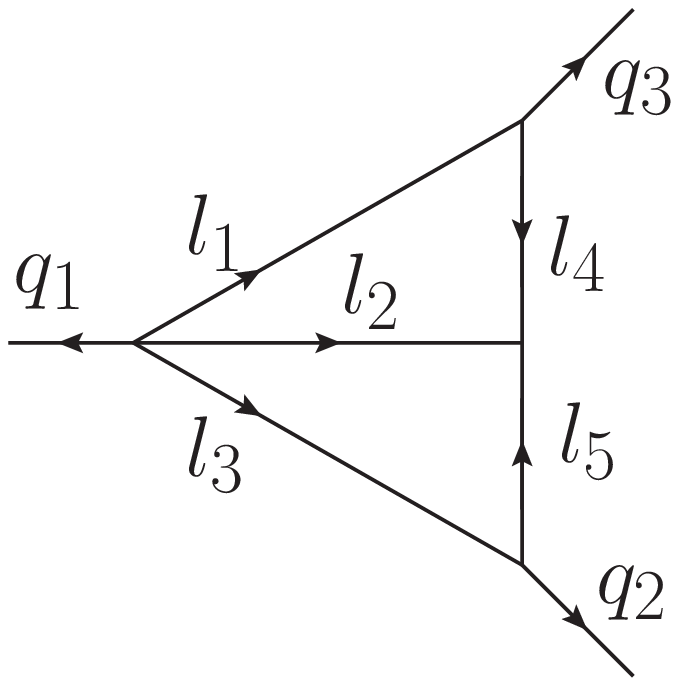}}\nonumber\\
&&
+\frac{l_2^-+l_3^-}{l_2^-}\Big(S(l_1, l_2+l_3, x) - S(l_1+l_2, l_3, x)
\Big)
\parbox{2.0cm}{\includegraphics[width=2.0cm]{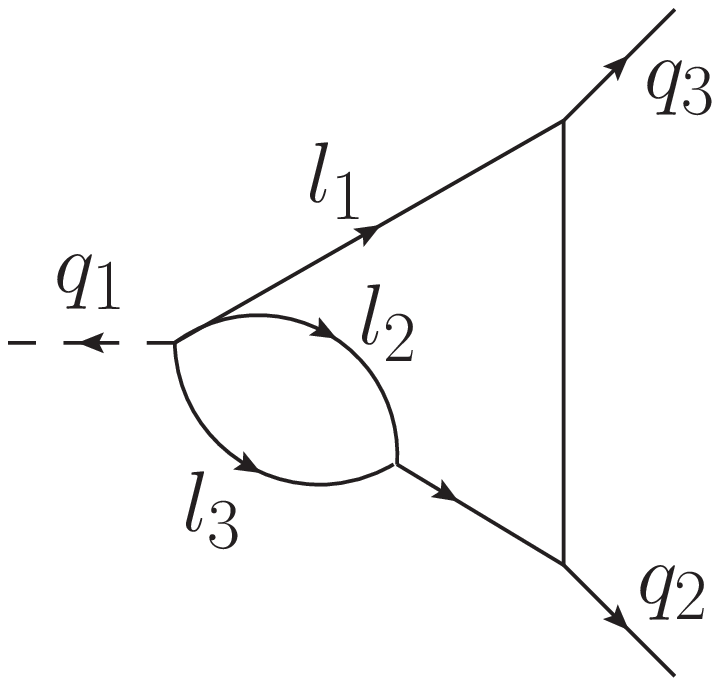}}
\nonumber\\
&&+\frac{l_1^-+l_2^-}{l_2^-}\Big(S(l_1+l_2, l_3, x) - S(l_1, l_2+l_3, x)
\Big)
\parbox{1.5cm}{\includegraphics[width=2.0cm]{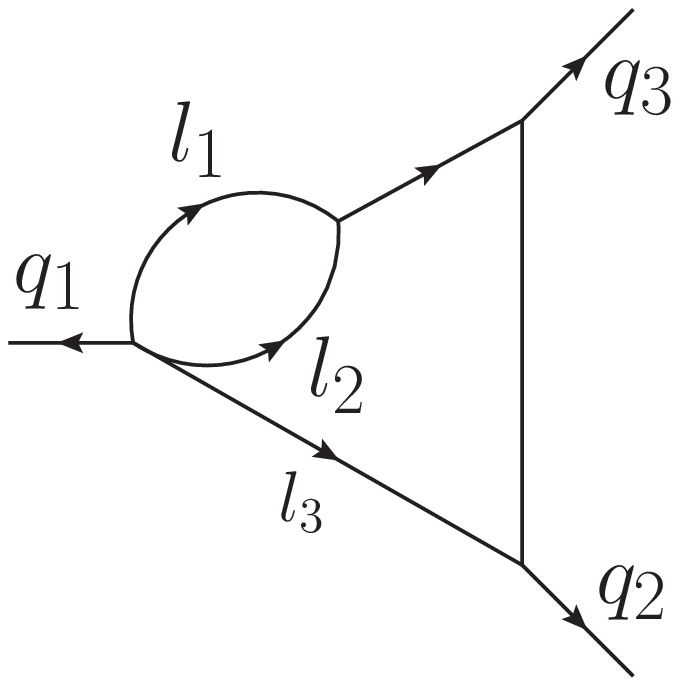}}
\ .
\label{3pf_nBPS2}
\ee
As usual, each diagram stands for the product of scalar propagators corresponding to the graph,
the product of the momentum conserving delta function at each vertex and an integral
over all internal momenta with the numerator factor explicitly shown.

While the spin of the operators can be arbitrarily large and thus the numerator factors can have arbitrarily
high powers of loop momenta (cf. eq.~\rf{Sdef}),
%
%
Lorentz invariance guarantees that
of all the integrals present in eq.~\rf{3pf_nBPS2} only those with an explicit one-loop bubble sub-integral
are divergent in the UV; this divergence is related to the fact that
the non-BPS operator we consider acquires a nontrivial anomalous dimension (and also mixes with
other operators). Thus, these integrals require regularization; we may simply promote to $D$-dimensions
the result of the four-dimensional calculation. The test of the validity of this procedure is the emergence of the correct anomalous dimension contribution to the correlator.

It is not difficult to evaluate the one-loop bubble subintegral in dimensional regularization; from the $l_1^-$ and
$(q_1+l_1)^-$ dependence of the result one can reconstruct the mixing matrix of the monomials
${\cal O}_1={\cal O}_{2,S,x}^{AB, AB}$ in \rf{operators}. Diagonalizing it one finds the correct one-loop
anomalous dimension -- proportional to the harmonic number $h(S)$ -- and the corresponding eigenvectors,
as well as the anomalous dimensions and eigenvectors of the descendants of twist-2 operators of lower spin.
\footnote{Thus, at this order the regulator dependence is completely accounted for by promoting the
integration measure to be $D$-dimensional.}
After taking the appropriate linear combinations \rf{nonBPS_1loop}
of ${\widetilde {c^x_1}}{}^{(1)}+{\widetilde {c^x_2}}{}^{(1)}$
to isolate an operator with definite anomalous dimension, multiplication of the full (LO + NLO) correlator
by the appropriate renormalization factor \rf{Zfactor} will render it finite as the regulator is
removed.~\footnote{Upon Fourier-transform to position space this multiplication also restores the combination
$\gamma_1\ln\frac{\mu |x_{13}||x_{23}|}{|x_{12}|}$ which is required by conformal invariance.}
To complete the calculation of the three-point function it is necessary to also include the tree-level correlation function
\rf{tree_nbb} summed against the two-loop mixing coefficients \cite{Belitsky:2007jp}.

It is relatively easy to check that the expression \rf{3pf_nBPS2} vanishes in the appropriate limits.
In particular, as $S, x\rightarrow 0$ the non-BPS operator becomes BPS and the three-point function should vanish
identically  \cite{Lee:1998bxa}. Indeed, in this limit the coefficients of the last two integrals vanish identically
and thus the correlator is finite; after using momentum conservation the remainder is proportional  to
\be
  q_1^2\parbox{1.5cm}{\includegraphics[width=2.0cm]{twist2_int1.eps}}
+q_3^2\parbox{1.5cm}{\includegraphics[width=2.0cm]{twist2_int2.eps}}
+q_2^2\parbox{1.5cm}{\includegraphics[width=2.0cm]{twist2_int3.eps}}
       -\parbox{1.5cm}{\includegraphics[width=2.0cm]{twist2_int4.eps}}
       -\parbox{1.5cm}{\includegraphics[width=2.0cm]{twist2_int5.eps}}
       -\parbox{1.5cm}{\includegraphics[width=2.0cm]{twist2_int6.eps}} ~~=0\ ;
\label{vanishing_sum}
\ee
We may similarly construct the correlation function of the spin-$S$ descendants of the BPS operator by
noticing that it is given by the sum
\be
\sum_x C_S^x {\cal O}_{2,S,x}^{AB, AB} \ ,
\ee
where $C_S^x$ are the binomial coefficients;
summing ${\widetilde {c^x_1}}{}^{(1)}+{\widetilde {c^x_2}}{}^{(1)}$ over integer values of $x$
between $0$ and $S$ the divergent integrals cancel out and the remainder is again proportional
to \rf{vanishing_sum}.

\subsubsection{Correlators of one non-BPS twist-2 operator and $(n-1)$ BPS}

The calculation of the correlator of one twist-2 and two BPS operators can be generalized to correlators
with any number of BPS operators. As in the case of correlators of only BPS operators, the cut
calculation is almost identical, but in this case we need to evaluate the two additional cuts shown
in fig.~\ref{extra_cut}. These cut are very similar to the cuts in figs.~\ref{3pf_cuts_NLO}(c) and
\ref{3pf_cuts_NLO}(d), respectively, in which the intermediate state is restricted to only gluons and scalars.
Similarly to the discussion in \S~\ref{higherpts_BPS}, the R-index structure is more complicated
than that in eq.~\rf{general_3pf} and it is given by all the $SO(6)$ singlets that appear in the (non-symmetrized)
tensor product of the representations of the BPS and non-BPS operators. We will focus on the completely connected
part of the correlator and, as in the case of the three-point function, we will choose a non-BPS twist-2 operator with
two identical scalar fields.

\begin{figure}[ht]
\begin{center}
\includegraphics[height=33mm]{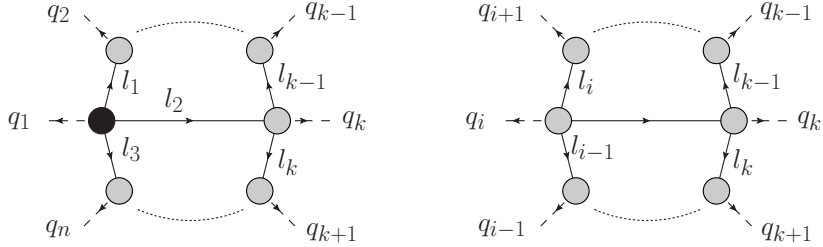}
\caption{Additional cuts necessary to determine the correlator of one twist-2 scalar operator
(with momentum $q_1$) and an arbitrary number of dimension-2 scalar BPS operators. Momentum labels are
slightly changed compared to those in fig.~\ref{3pf_cuts_NLO}.   \label{extra_cut}}
\end{center}
\end{figure}

When non-vanishing, the leading-order connected component of the momentum space
correlator of one spin-S twist-2 scalar operator and $(n-1)$ scalar BPS operators of dimension $2$ is
\be
{\tilde C}^{(0)}=
\int \prod d^4 l_i \delta(q_i+l_i-l_{i+1})
\frac{S(l_1, l_2, x)
}{\prod l_i^2}
\label{nBPS_LO}
\ee
where the external leg $q_1$ corresponds to the non-BPS operator.
%
Denoting by $l_1$ and $l_2$ the momenta of the internal lines immediately adjacent to the external line
$q_1$ -- corresponding to the source-field of the non-BPS operator -- in the integrals in which this line is
not shown explicitly in fig.~\ref{general_nonBPS_ints},
the NLO correction to this correlation function is:
%
\be
{\tilde C}^{(1)}&=&S(l_1, l_2, x)
\left(\sum_{i=1}^n\sum_{k=1}^{n-2}\text{I}_1{}_{i}^{[k]}-\sum_{i=2}^n\sum_{k=1}^{n-2}\text{I}_2{}_{i}^{[k]}\right)
\cr
&&
+\bigg(\frac{p_1^-}{t_2^-}S(t_1+t_2, t_3, x)+\frac{p_2^-}{t_2^-}S(t_1, t_2+t_3, x)\bigg)
\sum_{k=1}^{n-2}\text{I}_2{}_{1}^{[k]}\nonumber\\
&&
+\frac{p_2^-+p_3^-}{p_2^-}\Big(S(p_1, p_2+p_3, x)-S(p_1+p_2, p_3,x)\Big)\;\text{I}_{3;1}
\nonumber\\
&&+\frac{p_1^-+p_2^-}{p_2^-}\Big(S(p_1+p_2, p_3,x)-S(p_1, p_2+p_3, x)\Big)\;\text{I}_{4;1}
\label{nptBPSnonBPS}
\label{nptBPS1nonBPS}
\ .
\ee
The integrals appearing in this expression and their internal momenta are defined in fig.~\ref{general_nonBPS_ints}.
A potentially non-vanishing odd part present in various contributions to the cuts in fig.~\ref{extra_cut} cancels in
the complete expression of the cut. This is consistent with the absence of a parity-odd component of all the other cuts.

\begin{figure}[ht]
\begin{center}
\includegraphics[height=31mm]{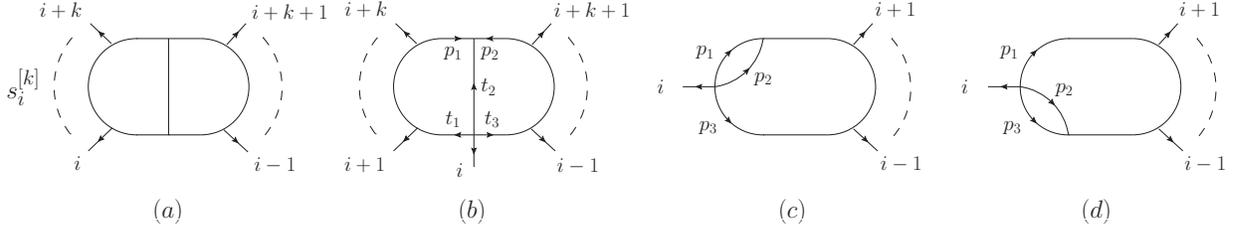}
\caption{Integrals $\text{I}_1{}_{i}^{[k]}$, $\text{I}_2{}_{i}^{[k]}$, $\text{I}_{3;i}$, $\text{I}_{4;i}$
entering the $n$-point correlator of one non-BPS scalar twist-2 operator and any
number of dimension-2 BPS operators. The first two integrals are the same as  $\text{I}_1{}_{i}^{[k]}$
and $\text{I}_2{}_{i}^{[k]}$ in fig.~\ref{integrals_2loops_npt}. \label{general_nonBPS_ints}}
\end{center}
\end{figure}

With this expression in hand we may test whether the non-BPS nature of one operator affects the relation
between the null limit of the correlator and the MHV $n$-point amplitude; general arguments in a specific
(non-standard) regularization scheme \cite{Engelund:2011fg} suggest that the relation is unaffected.
The similarity of the contact term integral $\text{I}_2{}_{1}^{[k]}$ with the integral with the same name
entering the correlation function of BPS operators \rf{npfBPS} and the fact that the Fourier-transform of the latter
is subleading in the null implies that the integral $\text{I}_2{}_{1}^{[k]}$ here will also yield only subleading
terms; as in that case, it is not difficult to check this explicitly.
Inspecting \rf{nptBPSnonBPS} and comparing it with \rf{npfBPS} we notice that, up to a factor
$S(l_1, l_2, x)=(l_1^-)^{S-x}(l_2^-)^{x}+(l_1^-)^{x}(l_2^-)^{S-x}$ the integrals are identical. This factor is
present in the leading order correlator \rf{nBPS_LO} and this cancels out in normalization.
Of the other integrals in \rf{nptBPSnonBPS},  $\text{I}_3$ and $\text{I}_4$ could also give contributions of the
same order. If however the null limit is taken in the presence of a finite dimensional regulator, these integrals are
subleading by $x_{12}^{2\epsilon}$ and $x_{n1}^{2\epsilon}$, respectively, and thus drop out as well
\cite{Eden:2010zz, Engelund:2011fg}.\footnote{If the null limit is taken after the correlator is
expanded in small $\epsilon$ at finite $x_{i, i+1}^2$, the contribution of these integrals should
factorize into a scheme-dependent coefficient function \cite{Eden:2010zz}.}
We therefore explicitly confirm that the null limit of the correlator of one non-BPS and $(n-1)$ BPS twist-2
operators
is the same as that of the correlator of $n$ BPS twist-2 operators and the same as the MHV gluon amplitude.
We also note that, as in the case of correlators of BPS operators, momentum space contact term integrals do
not contribute to this limit.

\subsubsection{The correlation function of two twist-2 and any number of BPS primaries}

Generalizing slightly the calculation it is not difficult to construct the correlator of two non-BPS twist-2 operators
and any number of BPS operators.
For general choices of BPS scalar operators and twist-2 operators one would have several possible choices
of R-index structures; the coefficient of each one of them may be obtained either by choosing suitable
representatives or by using supersymmetric methods.
We will focus here on the structures that have a nonzero position-space tree-level term
if the twist-2 operators are constructed from $\Tr[D_-^{S-x}\phi^{AB}D_-^x\phi^{AB}]$ and
its conjugate.\footnote{I.e. the scalars in the two BPS
operators are conjugates of each other.} There are two classes of R-index structures that can appear: those
in which the two twist-2 operators are adjacent ({\it i.e.} their R-indices are contracted)  and those in which
the two twist-2 operators are not ({\it i.e.} their R-indices are not contracted but rather are contracted
with the R-indices of BPS operators).

\begin{figure}[ht]
\begin{center}
\includegraphics[height=32mm]{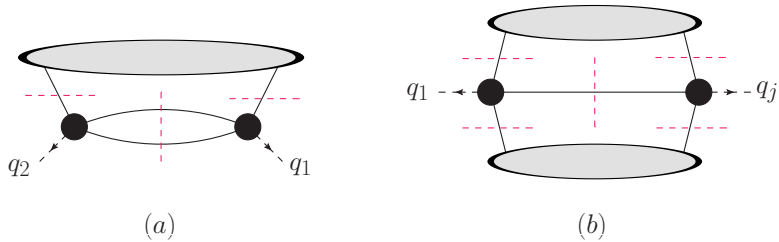}
\caption{ Additional cuts needed to find the correlation function if two non-BPS operators and any
number of BPS operators at NLO.  Cut $(a)$ is absent if the scalar fields in the two twist-2 operators
are not conjugates of each other.\label{extra_2NBPS}}
\end{center}
\end{figure}

Apart from the cuts already used to construct the correlators in the previous section, we need the information provided by the cuts in fig.~\ref{extra_2NBPS}(a) in order to find the correlators in the
first class.
Without loss of generality we will assume that the two non-BPS operators are at
locations 1 and 2; whenever there are only two internal lines attached to the non-BPS operators, we denote
their momenta by $l_1$, $l_2$ and $l_2$, $l_3$, respectively. The result for the NLO contribution to this
correlator is:
\be
{\tilde C}^{(1)}_1&=&S(l_1, l_2, x)S(l_2, l_3, y)
\left(\sum_{i=1}^n\sum_{k=1}^{n-2}\text{I}_1{}_{i}^{[k]}-\sum_{i=3}^n\sum_{k=1}^{n-2}\text{I}_2{}_{i}^{[k]}\right)
\cr
&&+
\sum_{k=2}^{n-2}
\bigg(\frac{p_1^-}{t_2^-}S(t_1+t_2, t_3, x)+\frac{p_2^-}{t_2^-}S(t_1,t_2+ t_3, x)\bigg)
S(l_2,l_3,y)\text{I}_2{}_{1}^{[k]}\nonumber\\
&&+
\sum_{k=1}^{n-2}\bigg(\frac{p_1^-}{t_2^-}S(t_1+t_2, t_3, y)+\frac{p_2^-}{t_2^-}S(t_1,t_2+ t_3, y)\bigg)
S(l_1,l_2,x)\text{I}_2{}_{2}^{[k]}\nonumber\\
&&
+\frac{p_1^-+p_2^-}{p_2^-}\Big(S(p_1, p_2+p_3, y) - S(p_1+ p_2,p_3, y)\Big)
S(p_3, p_3-q_1, x)\;\text{I}_{3;2}
\nonumber\\
&&+\frac{p_2^-+p_3^-}{p_2^-}\Big(S(p_1+p_2, p_3, y)-S(p_1,p_2+ p_3, y)\Big)
S(p_2+p_3, p_2+p_3-q_1, x)\;\text{I}_{4;2}
\nonumber\\
&&
+\frac{p_1^-+p_2^-}{p_2^-}\Big(S(p_1, p_2+p_3, x) - S(p_1+ p_2,p_3, x)\Big)
S(p_1+p_2, p_1+p_2-q_2, y)\;\text{I}_{3;1}
\nonumber\\
&&+\frac{p_2^-+p_3^-}{p_2^-}\Big(S(p_1+p_2, p_3, x)-S(p_1,p_2+ p_3, x)\Big)
S(p_1, p_1-q_2, y)\;\text{I}_{4;1}
\ .
\label{nptBPS2nonBPS}
\ee
Note that the position of the bubble integral in the line connecting the two non-BPS operators is ambiguous;
we have chosen it such that the coefficients of $\text{I}_{4;2}$ and $\text{I}_{3;1}$ are the ones that appear
in the anomalous dimension of the twist-2 operators. Such an organization exists for any operator.

It is not difficult to extend this result to an arbitrary number of BPS and non-BPS twist-2 operators in a split
configuration ({\it i.e.} with all adjacent BPS operators). The only difference compared to eq.~\rf{nptBPS2nonBPS}
is the proliferation of integrals $\text{I}_{3}$ and $\text{I}_{4}$: each twist-2 non-BPS operator contributes two
such integrals in a pattern easily identifiable by comparing the past two terms in eq.~\rf{nptBPS1nonBPS} and
the last four lines of eq.~\rf{nptBPS2nonBPS}.

For the correlators in the second class, in which the non-BPS operators are not adjacent, we need to also
evaluate the cut fig.~\ref{extra_2NBPS}(b), in which both such operators enter via a three-point form factor.
It turns out however that this cut does not reveal any new terms. Let us assume that the two non-BPS
operators are at positions $1$ and $j$ and  that, whenever there are only two internal lines attached to each
of them,  their momenta are $l_1$, $l_2$ and $l_j$, $l_{j+1}$, respectively. The NLO contribution to the correlator
correlator is given by:
%
\be
{\tilde C}^{(1)}_2&=&S(l_1, l_2, x)S(l_j, l_{j+1}, y)
\left(\sum_{i=1}^n\sum_{k=1}^{n-2}\text{I}_1{}_{i}^{[k]}-\sum_{i=2, i\ne j}^n\sum_{k=1}^{n-2}
\text{I}_2{}_{i}^{[k]}\right)
\cr
&&+
\sum_{k=2}^{n-2}
\bigg(\frac{p_1^-}{t_2^-}S(t_1+t_2, t_3, x)+\frac{p_2^-}{t_2^-}S(t_1,t_2+ t_3, x)\bigg)
S(l_j,l_{j+1},y)\text{I}_2{}_{1}^{[k]}\nonumber\\
&&+
\sum_{k=1}^{n-2}
\bigg(\frac{p_1^-}{t_2^-}S(t_1+t_2, t_3, y)+\frac{p_2^-}{t_2^-}S(t_1,t_2+ t_3, y)\bigg)
S(l_1,l_2,x)\text{I}_2{}_{j}^{[k]}\nonumber\\
&&
+\frac{p_2^-+p_3^-}{p_2^-}\Big(S(p_1, p_2+p_3, x)-S(p_1+p_2, p_3, x)\Big)
S(l_j,l_{j+1},y)\;\text{I}_{3;1}
\nonumber\\
&&+\frac{p_1^-+p_2^-}{p_2^-}\Big(S(p_1+p_2, p_3, x)-S(p_1, p_2+p_3, x)\Big)
S(l_j,l_{j+1},y)\;\text{I}_{4;1} \nonumber\\
&&
+\frac{p_2^-+p_3^-}{p_2^-}\Big(S(p_1, p_2+p_3, y)-S(p_1+p_2, p_3, y)\Big)
S(l_1,l_{2},x)\;\text{I}_{3;j}
\nonumber\\
&&+\frac{p_1^-+p_2^-}{p_2^-}\Big(S(p_1+p_2, p_3, y)-S(p_1, p_2+p_3, y)\Big)S(l_1,l_{2},x)\;\text{I}_{4;j}
\ .
\label{nptBPS2nonBPS_nonadj}
\ee
In both cases the potential parity-odd components cancel (up to total derivatives) in the complete expression
of cuts. As in the case of correlators with a single non-BPS operator, it is not difficult to Fourier-transform
eqs.~\rf{nptBPS2nonBPS} and \rf{nptBPS2nonBPS_nonadj} in the limit in which the operator  insertions
are null-separated. As in the case of a single non-BPS operator, the null limit is to be taken at finite
dimensional regulator.

\section{On effective actions and energy flow correlators\label{Tcorrelators}}

As discussed in the introduction, an interesting application of the techniques we described is the construction of
effective actions in background fields --- in particular in non-dynamical gravitational background --- as well as
the construction of correlation functions describing the energy and charge flow in scattering processes of
gauge-singlet states \cite{energy_correlators, Hofman:2008ar}. Since the relevant correlators are somewhat
different in the two cases we will discuss them separately.

\subsection{Effective actions}

The linearized coupling of a field theory with a background gravitational field is universal:
\be
S=S_0+\int d^d x\; h^{\mu\nu}T_{\mu\nu}+\dots
\label{grav_action}
\ee
where $S_0$ is the flat space action, $h^{\mu\nu}$ is the departure of the background metric from
that of the flat Minkowski space, $T_{\mu\nu}$ is the stress tensor and the ellipsis stand for terms
non-linear in $h$. The effective action for the gravitational field is obtained by integrating out all the
fields except for $h^{\mu\nu}$. We may interpret this as the evaluation of the scattering amplitudes
of the background gravitons $h^{\mu\nu}$. As discussed in \S~\ref{general_unitarity}, up to terms
in which two gravitons emerge from the same vertex, this is the same as the evaluation of the stress
tensor correlation functions; that is, up to contact terms that may perhaps be constructed on the basis of
general coordinate invariance and other symmetries, the effective action for the background gravitational
field is nothing but the collection of the stress tensor correlation functions. To construct them in our approach
we need to understand how to extract the stress tensor form factors. We will describe this in the next
section. We will however leave for the future the complete evaluation of effective actions, including the
determination of contact terms either by requiring that all Ward identities are satisfied or by finding
and using the multi-stress tensor form factors.


\subsection{The MHV stress tensor form factor}

Thus, both for the evaluation of effective actions as well as for the evaluation of energy and charge
correlators it is necessary to evaluate correlation functions involving the stress
tensor perhaps together with other operators, perhaps with additional constraints in the structure of
the contributing diagrams.
They may -- in principle -- be constructed  by a careful application of supersymmetry
transformations on correlation functions in which the stress tensors are replaced by dimension-2 BPS
operators and separation of the contributions of the descendants of lower levels in the supersymmetry
multiplet. Extraction of the stress tensor component is, however, not obvious since it is necessary to
separate from the coefficient of $\theta^2{\bar\theta}^2$ all the contribution of the supersymmetry
descendants of the operators appearing on lower levels in the multiplet.

To evaluate such correlators in our approach it is necessary to know the form factors and generalized form
factors of the stress tensor; their MHV components are part of the MHV super-form factor of the complete
stress tensor multiplet constructed in \cite{Brandhuber:2011tv}, which is a natural generalization of the
super-form factor of the chiral stress tensor multiplet:
\be
{\cal G}&=&
\prod_{i=1}^n \frac{1}{\langle i, i+1\rangle}
\;
\delta^{(4)}(q-\sum_i\lambda_i{\tilde\lambda}_i)
\delta^{(4)}(\gamma_+-\sum_i\lambda_i\eta_{+;i})\delta^{(4)}(\gamma_--\sum_i\lambda_i\eta_{-;i})
\\
&=&
\prod_{i=1}^n \frac{1}{\langle i, i+1\rangle}
\;
\delta^{(4)}(q-\sum_i\lambda_i{\tilde\lambda}_i)
\delta^{(8)}(\gamma-\sum_i\lambda_i\eta_{i}) \ .
\label{fullTff}
\ee
This super-form factor and its generalizations may be used to construct super-correlation functions from
which the desired components can then be extracted. Alternatively, one construct directly the component
correlation function by first extracting the relevant form factors.

A naive extraction of the component form factor from \rf{fullTff} as the coefficient of
$\gamma_+\sigma^\mu\gamma_-\gamma_+\sigma^\nu\gamma_-$
leads to expressions which violate the expected conformal Ward identities related to the tracelessness
and conservation of the stress tensor.
This is a reflection of the difficulty mentioned above, that apart from the stress tensor, the coefficient
of $\theta^2{\bar\theta}^2$ also contains descendants of the operators at lower levels in the
supersymmetry multiplet.
It turns out that the stress tensor form factor ({\it i.e.} the coefficient of
$\theta^2{\bar\theta}^2$ from which the descendant contribution is separated) may be extracted as
\be
\langle 0|T^{\mu\nu}|1,\dots,n\rangle = \frac{1}{4!}\int d^8\gamma  \;
\epsilon^{ABCD} \gamma_Aq\llap/ {\bar \sigma}^\mu\gamma_B
                             \gamma_Cq\llap/ {\bar \sigma}^\nu\gamma_D \; {\cal G} \ .
\ee
Conservation and tracelessness follow from the fact that $\epsilon^{ABCD} \gamma_A\gamma_B=0$, which is
a consequence of the Grassmann nature of $\gamma$ and of the antisymmetry of the Lorentz contraction.
We have checked that this expression agrees with a Feynman diagram and BCFW-based evaluation of this
form factor.

As discussed in section~\ref{general_unitarity}, to construct correlation functions though generalized unitarity
we also need generalized form factors with more operator insertions. In the case of the stress tensor they
are particularly important to consider because the derivatives present in the stress tensor lead
to contact terms which are missed if only regular form factors are used; it is possible that such contact terms
-- which are important for the construction of effective actions -- can be determined by requiring that
correlators exhibit conformal invariance.

\subsection{Energy and charge flow correlators \label{efc}}

Energy and charge correlation functions \cite{energy_correlators, energy_correlators_1, KS, BKS} are particular
examples of event shapes, which are important observables in QCD.
For some some initial state $X$ and observable $e(N)$ that depends on the final state, they are defined
as weighted cross sections:
\be
\frac{1}{\sigma_{\text{tot}}}\frac{d\sigma_\text{tot}}{de} =
\sum_{N}d\sigma_{X\rightarrow N} \delta(e-{\hat e}(N))
\label{wcs}
\ee
and describe the distribution of $e(N)$ in the final state of some scattering process while being insensitive
to the  particular composition of this state. Popular choices for ${\hat e}$ is the energy or (some) charge flowing
in some direction, {\it e.g.}
\be
{\cal E}({\vec n})|k_1,\dots,k_N\rangle = \sum_{i=1}^N\delta(\cos\theta-\cos\theta_i)\delta(\phi-\phi_i)\,
k^0_i\,|k_1,\dots,k_N\rangle
\label{def_calE_ini}
\ee
and similarly for some charge ${\cal Q}$. The angles $(\theta_i, \phi_i)$ define the unit vector
${\vec n}_i = {\vec k}_i/|{\vec k}_i|$ with respect to some reference direction $(\theta, \phi)$.
The operator ${\cal E}$ admits presentations \cite{stress_tensor, Hofman:2008ar} in
terms of the stress tensor:
\be
{\hat e} &=& \int d^3 {\vec n} \; \delta({\vec n}^2-1 )\,  w({\vec n}){\cal E}({\vec n})
\nonumber
\\
{\cal E}({\vec n}) &=& \lim_{R\rightarrow\infty} R^2\int_{-\infty}^{+\infty}dt \; n^i \, T^0{}_i(t, R {\vec n}^i)  \ .
\label{def_calE}
\ee
Using the expression of the differential cross section in terms of squared S-matrix elements one
casts~\cite{KS, FlemingSterman}
eq.~\rf{wcs} into an expression whose building blocks are special correlation functions of  the operators
${\cal E}$. In the definition above ${\cal E}$ is understood to contain a projection onto states at $t=+\infty$.

In a conformal field theory the initial state $X$ is created by the action on the in vacuum state of
some local gauge invariant operator:
\be
|X\rangle = {\cal O}|0\rangle_{\text{in}}
\label{state}
\ee
Then, the analog of the total weighted cross section is
\be
\frac{1}{\sigma_{\text{tot}}}\frac{d\sigma_\text{tot}}{de}
\sim \frac{
{}_{\text{in}}\langle 0 |{\cal O}^\dagger \delta(e-\int d^3 {\vec n} \; w({\vec n}){\cal E}({\vec n})){\cal O}|0\rangle_{\text{in}}}{\langle 0 |{\cal O}^\dagger {\cal O}|0\rangle_{\text{in}}} \ .
\label{wcs_cf}
\ee
This expression may be refined by assigning a different $\delta-$function factor to each S-matrix element
Using an integral representation of the $\delta-$function it is easy to see that the expectation value
\rf{wcs_cf} is defined by the more general correlators~\cite{KS}
\footnote{One may similarly define charge correlators, in
which (some of) the energy detectors ${\cal E}$  are replaced by the R-charge detectors
$$
{\cal Q}(\theta) = \lim_{r\rightarrow\infty} r^2\int_{-\infty}^{+\infty}dt \;n^i \, J{}_i(t, r {\vec n}^i) \ .
$$
}.
\be
G({\vec n}_1,\dots {\vec n}_n, {\cal O})=\langle {\cal E}({\vec n}_1)\dots {\cal E}({\vec n}_n)\rangle
\equiv
\frac{{}_{\text{in}}\langle 0|{\cal O}^\dagger {\cal E}({\vec n}_1)\dots {\cal E}({\vec n}_n)
{\cal O}|0\rangle_{\text{in}}}
{{}_{\text{in}}\langle 0|{\cal O}^\dagger {\cal O}|0\rangle_{\text{in}}} \ .
\label{def_energy_correlator}
\ee
The fact that the vacuum states are both in-states makes this correlator different than the standard ones which
were interpreted as scattering amplitudes of source fields. Rather, as mentioned in \cite{BKS}, both the
numerator and denominator have the interpretation of cut diagrams \footnote{This is the same as the
cut diagram interpretation of inclusive cross sections.}; we emphasize that these  are regular unitarity cuts
that have an imaginary part interpretation: in this case it is the imaginary part of the two-point function
$\langle 0|{\cal O}^\dagger\,{\cal O}|0\rangle$.
The fact that ${\cal E}$ contains a projection onto out-states implies that, in these unitarity cuts, the
operators ${\cal E}$ are to be inserted {\it only} on the cut lines; this is a reflection of the definition
\rf{wcs} which assigns the energy weight factors only on the final state $|N\rangle$.
For the same reason, as well as because ${\cal E}$ acts in the final state on widely separated
particles, all operators ${\cal E}$ commute with each other and thus no ordering needs to be specified.
Singularities may nevertheless arise if some ${\vec n}_i={\vec n}_j$ and the corresponding
operators act on the same particle. We will avoid such situations by keeping ${\vec n}_i\ne {\vec n}_j$
for all $i\ne j$.

\subsection{On the evaluation of energy correlators at weak and strong coupling \label{WScoupling}}

In contrast to correlation functions of stress tensors,
energy correlators \cite{energy_correlators, Hofman:2008ar} provide a class of observables which
are insensitive to contact terms. Indeed, as discussed in detail in \S~\ref{efc}, the stress tensors featuring in
eq.~\rf{def_energy_correlator} are connected only to cut lines and therefore, all potential contact terms are set
to zero. Moreover, since the energy measurement is made at infinity it affects only one particle at a time and
thus only the two-point form factor of the stress tensor is necessary.

As described earlier in this section, the energy correlators in the state created by the operator ${\cal O}$
may be interpreted as the unitarity cut of the two-point function of ${\cal O}$ and ${\cal O}^\dagger$
dressed with energy- and orientation-dependent factors. The discussion in
\S~\ref{general_unitarity} implies that this construction may be given an interpretation in terms of the
form factors of ${\cal O}$. Indeed, inserting complete sets of out-states we have
\be
{}_{\text{in}}\langle 0|{\cal O}^\dagger {\cal O}|0\rangle_{\text{in}} &=&
\sum_X {}_{\text{in}}\langle 0|{\cal O}^\dagger |N\rangle \langle N|{\cal O}|0\rangle_{\text{in}} \ ,
\\
{}_{\text{in}}\langle 0|{\cal O}^\dagger {\cal E}({\vec n}_1)\dots {\cal E}({\vec n}_n)
{\cal O}|0\rangle_{\text{in}}
&=&\sum_{N, N'} {}_{\text{in}}\langle 0|{\cal O}^\dagger |N'\rangle
\langle N'| {\cal E}({\vec n}_1)\dots {\cal E}({\vec n}_n)|N \rangle
\langle N|{\cal O}|0\rangle_{\text{in}} \ ,
\label{css_calE}
\ee
where the sums run over all physical Fock space states $|X\rangle$ of some fixed momentum; we recognize
each factor as the form factor of ${\cal O}$ and its conjugate.

To evaluate the right-hand side of \rf{css_calE} one may use the original definition \rf{def_calE_ini} of
${\cal E}$, superficially perhaps at the expense of not using supersymmetric methods to evaluate the sum over
intermediate states.
We may alternatively use the stress tensor super-form factor and carefully take the limit $R\rightarrow\infty$
limit at the end while enforcing the relevant on-shell conditions. As we shall see shortly, the limit
$R\rightarrow\infty$ requires that the momentum of the stress tensor operator vanishes; in this case it turns
out that the form factor is just the traceless part of $p^\mu p^\nu$, where $p$ is the momentum of one of the
external fields and thus factorizes in the sum over intermediate states.
One may show  along the lines of
\cite{FlemingSterman} that, for $x=(t, R {\vec n})$,
\be
&&
\lim_{R\rightarrow\infty}\,R^2\int dt
\int \frac{d^4p_1}{(2\pi)^4} \frac{d^4p_2}{(2\pi)^4}
\delta^{(+)}(p_1^2)\delta^{(+)}(p_2^2)\; (p_1^0 \, n_ip_2^i) \;e^{ip_1\cdot x}e^{-ip_2\cdot x}
F(p_1, p_2, \dots)
\nonumber\\
&&\qquad\qquad\qquad
= \int \frac{d^3 p}{(2\pi)^3 2\omega_p}\; \omega_p\,\delta^2({\vec n}-\frac{\vec p}{|{\vec p}|})
F(p, p, \dots) \ .
\ee
The argument of the integral on the left-hand side of the equation above is the typical term in the
momentum space correlation function corresponding to \rf{def_energy_correlator} after the state sum
was performed and {\it before} the limit $R\rightarrow\infty$ is taken. $F$ represents the contribution
of the two form factors and of the other stress tensors except for one; the contribution of the last stress
tensor is just $(p_1^0 \, n_ip_2^i)$. Evaluating the angular integrals in the saddle-point approximation
(justified by the large value of $R$) leads to the orientation of the space-like parts of $p_1$ and $p_2$
be the same as
the external vector ${\vec n}$. The integral in the time direction sets to zero the energy carried by
stress tensor operator thus setting $p_1^0 = p_2^0$ and thus also sets equal the norms of the space-like
components of the momenta $p_1$ and $p_2$ and thus also $|X\rangle = |X'\rangle$.
This identity therefore allows us to use supersymmetric methods to evaluate \rf{css_calE}.

In \cite{Hofman:2008ar} it was suggested that the energy correlation function may be extracted by
 suitable analytic continuation of the time-ordered correlation function.
It is in principle possible to use the form factor interpretation described above to evaluate them at
strong coupling; the construction the form factors \cite{Maldacena:2010kp} in terms of a $Y-$system would
be an essential ingredient. A difficulty arises however in the treatment and cancellation of infrared divergences: while
the energy correlators are infrared-safe observables, form factors exhibit infrared divergences. In a fixed
order calculation they cancel against real emission contributions -- lower-loop higher-points contributions to
form factors.
It is not obvious how the cancellation proceeds at strong coupling; we leave this important problem for the future.

\subsection{Examples of energy correlators}

We will discuss examples of energy correlators for two-field operators.
The simplest example involves a single operator ${\cal E}$. At leading order the calculation
is trivial; we need the explicit expression of the two-point form factors of  the desired operator
and we need to carry out the  phase space integral. For definiteness let us consider the scalar BPS operator,
also discussed in \cite{Hofman:2008ar}. In this case the form factors are unity; assuming that the
state created by the operator has 4-momentum $(q, {\vec 0})$, using the notation
${\hat p}={\vec p}/|\vec p|$ and using the symmetry under the interchange of the two cut lines
as well as the on-shell conditions we find
\be
{}_{\text{in}}\langle 0|{\cal O}^\dagger {\cal E}({\vec n}){\cal O}|0\rangle_{\text{in}}^{(0)}
&=&
2\int \frac{d^3 p_1}{(2\pi)^3 2\omega_1}\frac{d^3 p_2}{(2\pi)^3 2\omega_2} \,
\omega_1 \delta^{(2)}({\vec n}-{\hat p}_1)
\delta^3({\vec p}_1 + {\vec p}_2)\delta(\omega_1+\omega_2-q)
\cr
&=&\frac{2}{(2\pi)^6}\int \frac{\omega_1^2 d\omega_1}{4\omega_1}\delta(2\omega_1-q)
=\frac{2}{(2\pi)^6}\,\frac{q}{16}
\\
{}_{\text{in}}\langle 0|{\cal O}^\dagger {\cal O}|0\rangle_{\text{in}}^{(0)}
&=&
\int \frac{d^3 p_1}{(2\pi)^3 2\omega_1}\frac{d^3 p_2}{(2\pi)^3 2\omega_2} \,
\delta^3({\vec p}_1 + {\vec p}_2)\delta(\omega_1+\omega_2-q)
\cr
&=&\frac{4\pi}{(2\pi)^6}\int \frac{\omega_1^2 d\omega_1}{4\omega_1^2}\delta(2\omega_1-q)
=\frac{4\pi}{(2\pi)^6}\,\frac{1}{8}
\label{cutLObps}
\ee
Taking the ratio \rf{def_energy_correlator} we find \cite{Hofman:2008ar}
\be
\langle {\cal E}({\vec n}) \rangle_{\rm BPS}^{(0)} = \frac{q}{4\pi} \ .
\ee
The spherical symmetry is, of course, due to the state created by ${\cal O}_{\rm BPS}=\Tr[(\phi^{12})^2]$ being
spherically symmetric \cite{Hofman:2008ar}.

We may introduce an asymmetry by considering the state created by an {\it e.g.}
twist-2 operator. Since in this case the operator is a sum of terms, to reconstruct both the operator and its
conjugate, we need to consider the mode general building block ${}_{\text{in}}\langle 0|{\cal O}_{2,S,y}^\dagger
{\cal E}({\vec n}_1){\cal O}_{2,S,x}|0\rangle_{\text{in}}$ with the unspecified R-symmetry indices chosen such
that the total R-charge vanishes. It evaluates to:
\be
&&{}_{\text{in}}\langle 0|{\cal O}_{2,S,y}^\dagger {\cal E}({\vec n}){\cal O}_{2,S,x}|0\rangle_{\text{in}}^{(0)} \cr
&=&2\int \frac{d^3 p_1}{(2\pi)^3 2\omega_1}\frac{d^3 p_2}{(2\pi)^3 2\omega_2} \,
\omega_1 \delta^{(2)}({\vec n}-{\hat p}_1)
(p_{1+}^{S-x}p_{2+}^x)(p_{1+}^{S-y}p_{2+}^y)
\delta^3({\vec p}_1 + {\vec p}_2)\delta(\omega_1+\omega_2-q)
\cr
&=&
\frac{2}{(2\pi)^6}\,\frac{q}{16}\,\left(\frac{q}{2}\right)^{2S} (1 + n_3)^{2S}
\left(\frac{1 - n_3}{1 + n_3}\right)^{x+y}
\\[1pt]
&&{}_{\text{in}}\langle 0|{\cal O}_{2,S,y}^\dagger {\cal O}_{2,S,x}|0\rangle_{\text{in}}^{(0)}
\cr
&=&
\int \frac{d^3 p_1}{(2\pi)^3 2\omega_1}\frac{d^3 p_2}{(2\pi)^3 2\omega_2} \,
(p_{1+}^{S-x}p_{2+}^x)(p_{1+}^{S-y}p_{2+}^y)
\delta^3({\vec p}_1 + {\vec p}_2)\delta(\omega_1+\omega_2-q)
\cr
&=&
\frac{1}{(2\pi)^6}\frac{1}{8}\,\left(\frac{q}{2}\right)^{2S}\int d^2\Omega \;
(1 + \cos\theta)^{2S}
\left(\frac{1 - \cos\theta}{1 + \cos\theta}\right)^{x+y}
\ ,
\ee
where we used that $p_{i,+} \equiv p_{i,0}+p_{i,3}$. Summing over $x$ and $y$ the ratio of the two
expressions above with coefficients given by \rf{nonBPS_1loop} and using the orthogonality of Gegenbauer
polynomials to evaluate the angular integral
%
%
we find that the leading order of the energy one-point function in the state
created by a non-BPS twist-2 operator  is
\be
\langle {\cal E}({\vec n}) \rangle^{(0)}_{2,S} = \frac{q}{2\pi} \left(S+\textstyle{\frac{1}{2}}\right)
C_S^{1/2}\left(n_3\right) ^2\ .
\ee
For twist-2 operators expressed in terms of an arbitrary null vector $\ell=(\ell_0, {\vec \ell})$ the energy
one-point function is obtained from the above by the replacement $n_3\mapsto {\vec \ell}\cdot {\vec n}$.
By appropriately changing the coefficients we may also construct the same one-point
function in the states created by conformal descendants of twist-2 lower-spin operators.

The correlator of two (or several) energy flow operators is an observable with nontrivial parameter dependence
even in the case of a spherically-symmetric state. Let us briefly discuss the correlator
$\langle  {\cal E}({\vec n}_1){\cal E}({\vec n}_2) \rangle_{\rm BPS}$. We find:
\be
&&{}_{\text{in}}\langle 0| {\cal O}_{\rm BPS}^\dagger {\cal E}({\vec n}_1)
{\cal E}({\vec n}_2) {\cal O}_{\rm BPS}|0\rangle^{(0)}_{\text{in}}
\cr
&=&
\int \frac{d^3 p_1}{(2\pi)^3 2\omega_1}\frac{d^3 p_2}{(2\pi)^3 2\omega_2} \,
\omega_1 \delta^{(2)}({\vec n}_1-{\hat p}_1)
\omega_2\delta^{(2)}({\vec n}_2-{\hat p}_2)
\delta^3({\vec p}_1 + {\vec p}_2)\delta(\omega_1+\omega_2-q)
\cr
&+&\int \frac{d^3 p_1}{(2\pi)^3 2\omega_1}\frac{d^3 p_2}{(2\pi)^3 2\omega_2} \,
\omega_1 \delta^{(2)}({\vec n}_2-{\hat p}_1)
\omega_1\delta^{(2)}({\vec n}_1-{\hat p}_2)
\delta^3({\vec p}_1 + {\vec p}_2)\delta(\omega_1+\omega_2-q)
\cr
&+&\int \frac{d^3 p_1}{(2\pi)^3 2\omega_1}\frac{d^3 p_2}{(2\pi)^3 2\omega_2} \,
\omega_2 \delta^{(2)}({\vec n}_2-{\hat p}_2)
\omega_2 \delta^{(2)}({\vec n}_1-{\hat p}_2)
\delta^3({\vec p}_1 + {\vec p}_2)\delta(\omega_1+\omega_2-q)
\cr
&+&\int \frac{d^3 p_1}{(2\pi)^3 2\omega_1}\frac{d^3 p_2}{(2\pi)^3 2\omega_2} \,
\omega_1 \delta^{(2)}({\vec n}_2-{\hat p}_1)
\omega_1 \delta^{(2)}({\vec n}_1-{\hat p}_1)
\delta^3({\vec p}_1 + {\vec p}_2)\delta(\omega_1+\omega_2-q)
\cr
&=&\frac{2}{(2\pi)^6}\; \frac{q^2}{32} \left(\delta^{(2)}({\vec n}_1 + {\vec n}_2)
                                                                      + \delta^{(2)}({\vec n}_1 - {\vec n}_2)\right)\ ;
\label{79}
\ee
together with \rf{cutLObps} it implies that
\be
\langle {\cal E}({\vec n}_1){\cal E}({\vec n}_2) \rangle^{(0)}_{\rm BPS} =
\frac{q^2}{8\pi}\left(\delta^{(2)}({\vec n}_1 + {\vec n}_2)
                                + \delta^{(2)}({\vec n}_1 - {\vec n}_2)\right) \ .
\ee
Thus, this correlator vanishes identically as long as ${\vec n}_1\ne \pm {\vec n}_2$.  The same
$\delta-$function dependence on the fixed vectors ${\vec n}_i$ seems to be generic for
the leading order two-${\cal E}$ correlator in the state created by any two-field operator.
At higher orders this appears to change; below we will explore the next order -- which is also the
first nontrivial order -- assuming that ${\vec n}_1\ne \pm {\vec n}_2$.

\begin{figure}[ht]
\begin{center}
\includegraphics[height=25mm]{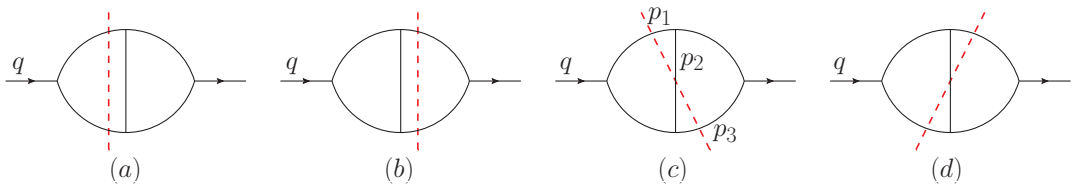}
\caption{The four components of the cut contributing to the energy correlators at the
NLO in the state created by a BPS  two-field operator. The operators ${\cal E}({\vec n})$ are placed
on the cut propagators.\label{NLO_EEcor} }
\end{center}
\end{figure}

The components of the cut contributing to energy correlators at the first nontrivial order (we will refer to it as
the LO) in the state created by a BPS
two-field operator are shown in fig.~\ref{NLO_EEcor}.
They may be evaluated directly using the one-loop
two-point form factor and the tree-level three-point form factor of dimension-2 BPS operators or they may be
obtained by constructing the cut of the momentum-space NLO two-point function. Using the discussion in
secs.~\ref{general_unitarity}, \ref{formfactors} and \ref{examples}
it is easy to find that, up to a factor of the square of the momentum carried by the operator, the two-point
function is given by the two-loop scalar bubble integral\footnote{This is consistent with the non-renormalization
of the two-point function of BPS operators: the two-loop bubble integral is proportional to $q^{-2}$ and
therefore the momentum-space two-point function is a constant. Fourier-transforming to position space yields
only contact terms, proportional to $\delta^4(x_1-x_2)$. It is interesting to note that in this case
the energy correlator is related to such contact terms. This is special to states created by BPS operators.}.
From the integral form of the two-point function is it easy to find the four components of the cut such that
momentum labels are aligned.
%
%

A calculation similar to the one leading to eq.~\rf{79}
shows that the contribution of the first two components is proportional to $\delta^{(2)}({\vec n}_1\pm {\vec n}_2)$,
which vanishes due to our assumption that ${\vec n}_1\ne \pm {\vec n}_2$. The last two components are the
same up to relabeling of momenta; we will analyze only the first one and double it. The contribution of both
energy operators are inserted on the same cut line is proportional to $\delta^{(2)}({\vec n}_1 - {\vec n}_2)$ and
thus may also be ignored. The remaining terms are:
\be
&&{}_{\text{in}}\langle 0| {\cal O}_{\rm BPS}^\dagger {\cal E}({\vec n}_1){\cal E}
({\vec n}_2) {\cal O}_{\rm BPS}|0\rangle^{(1)}_{\text{in}}
\\[1pt]
&=&q^2\int \frac{d^3 p_1}{(2\pi)^3 2\omega_1}\frac{d^3 p_2}{(2\pi)^3 2\omega_2}
\frac{d^3 p_3}{(2\pi)^3 2\omega_3}
\delta^{(3)}({\vec p}_1+{\vec p}_2+{\vec p}_3)\delta(\omega_1+\omega_2+\omega_3-q)
\cr
&&
\qquad\quad\times
\frac{1}{2(\omega_1\omega_2-{\vec p}_1\cdot{\vec p}_2)}
\frac{1}{2(\omega_2\omega_3-{\vec p}_2\cdot{\vec p}_3)}
\cr
&&\quad\times\Big(~\omega_1\omega_2(\delta^{(2)}({\vec n}_1-{\hat p}_1)\delta^{(2)}({\vec n}_2-{\hat p}_2)
                                                   +\delta^{(2)}({\vec n}_2-{\hat p}_1)\delta^{(2)}({\vec n}_1-{\hat p}_2))
                                                   \cr
&&\qquad+\omega_1\omega_3(\delta^{(2)}({\vec n}_1-{\hat p}_1)\delta^{(2)}({\vec n}_2-{\hat p}_3)
                                                   +\delta^{(2)}({\vec n}_2-{\hat p}_1)\delta^{(2)}({\vec n}_1-{\hat p}_3))
                                                   \cr
&&\qquad+\omega_2\omega_3(\delta^{(2)}({\vec n}_1-{\hat p}_2)\delta^{(2)}({\vec n}_2-{\hat p}_3)
                                                   +\delta^{(2)}({\vec n}_2-{\hat p}_2)\delta^{(2)}({\vec n}_1-{\hat p}_3))~\Big)
\nonumber
\ee
The six integrals are essentially the same; we will discuss only the ones on the first line of the parenthesis.
The integral over the angular parts  of ${\vec p}_{1,2}$ sets ${\vec p}_{1,2}=\omega_{1,2}{\vec n}_{a,b}$ with
$a\ne b\in\{1,2\}$. Then the ${\vec p}_3$ integral sets
${\vec p}_3=-(\omega_{1}{\vec n}_{a}+\omega_{2}{\vec n}_{b})$ and $\omega_3 = |{\vec p}_3|$. Last, we can
do the integral over $|{\vec p}_2|=\omega_2$; the $\delta-$function fixes $\omega_2$
and introduces a Jacobian factor $J=\omega_3/(q-\omega_1(1-{\vec n}_1\cdot {\vec n}_2))$.
Most factors cancel and we are left with (for $0< {\vec n}_1\cdot {\vec n}_2 < 1$)
\be
I_{1,2}&=&
2\frac{q^2}{32(2\pi)^9}\frac{1}{1-{\vec n}_1\cdot {\vec n}_2}\int_0^{q/2}
\frac{\omega_1 d\omega_1}{(q-\omega_1(1-{\vec n}_1\cdot {\vec n}_2))^2}
\cr
&=&
\frac{q^2}{16(2\pi)^9}
\frac{(1-{\vec n}_1\cdot {\vec n}_2)
-(1+{\vec n}_1\cdot {\vec n}_2) \ln(2/(1+{\vec n}_1\cdot {\vec n}_2))}
{(1-{\vec n}_1\cdot {\vec n}_2)^3(1+{\vec n}_1\cdot {\vec n}_2)}
%
\ee
Similarly, the integrals on the second and third lines of the parenthesis are
\be
I_{1,3}&=&\frac{q^2}{16(2\pi)^9}
\frac{-2(1-{\vec n}_1\cdot {\vec n}_2)+(3+{\vec n}_1\cdot {\vec n}_2)\ln(2/(1+{\vec n}_1\cdot {\vec n}_2))}
{(1-{\vec n}_1\cdot {\vec n}_2)^3(1+{\vec n}_1\cdot {\vec n}_2)}
\\
I_{2,3}&=& I_{1,2}
\ee
Combining them, accounting for the component in fig.~\ref{NLO_EEcor}(d)
and normalizing\footnote{Since the LO correlator
${}_{\text{in}}\langle 0| {\cal O}_{\rm BPS}^\dagger {\cal E}({\vec n}_1){\cal E}({\vec n}_2)
{\cal O}_{\rm BPS}|0\rangle^{(0)}_{\text{in}}$ the cut of the NLO 2-point function contributes
only to NLO energy correlator.} by dividing by \rf{cutLObps} leads to\footnote{This
logarithmic dependence is the same as in QCD while the coefficient of the logarithm is different.
We thank Lance Dixon for emphasizing this point to us.}
\be
\label{EEBPS_2loop}
\langle {\cal E}({\vec n}_1){\cal E}({\vec n}_2) \rangle^{(1)}_{\text{BPS}}
=\frac{q^2}{2\, (2\pi)^4}
\frac{\ln(2/(1+{\vec n}_1\cdot {\vec n}_2))}{(1-{\vec n}_1\cdot {\vec n}_2)^2(1+{\vec n}_1\cdot {\vec n}_2)} \ .
\ee
It should not be difficult to proceed to higher orders or to evaluate the energy correlators in states created by
other operators. For example, at the next order one may start from the presentation of the two-point function in
terms of the three-loop ladder and crossed-ladder propagator integrals, construct their three- and four-particle cuts
and dress them with energy factors. In evaluating the remaining integrals one sound observe the cancellation of
IR divergences between the contribution of these two cuts.

We may also easily evaluate the two-energy correlator to first nontrivial order in the state created by a
non-BPS twist-2 spin-$S$ operator. The relevant dressed cut is:
\be
&&{}_{\text{in}}\langle 0| {\cal O}_{2,S,y}^\dagger {\cal E}({\vec n}_1)
{\cal E}({\vec n}_2) {\cal O}_{2,S,x}|0\rangle^{(0)}_{\text{in}}\cr
&=&\int\frac{d^3p_1}{(2\pi)^32\omega_1}\frac{d^3p_2}{(2\pi)^32\omega_2}\frac{d^3p_3}
{(2\pi)^32\omega_3}\delta^3(\vec{p}_1+\vec{p}_2+\vec{p}_3)\delta(\omega_1+\omega_2+\omega_3-q)\cr
&&\Bigg[\frac{q^2}{16(\omega_1\omega_2-\vec{p}_1\cdot\vec{p}_2)(\omega_2\omega_3-\vec{p}_2\cdot\vec{p}_3)}
S(p_1+p_2, p_3, x)S(p_1,p_2+ p_3, y)\cr
&& -\frac{p_{1+}+p_{2+}}{8p_{2+}(\omega_1\omega_2-\vec{p}_1\cdot\vec{p}_2)}
\bigg(S(p_1+p_2, p_3, x)\big(S(p_1, p_2+p_3, y)-S(p_1+p_2, p_3, y)\big)
\cr
&&\qquad\qquad\qquad
+S(p_1+p_2, p_3, y)\big(S(p_1,p_2+ p_3, x)-S(p_1+p_2, p_3, x)\big)\bigg)\Bigg]
\cr
&&\qquad
\times\Big(\;\omega_1\omega_2(\delta^2(\vec{n}_1-\hat{p}_1)\delta^2(\vec{n}_2-\hat{p}_2)
+\delta^2(\vec{n}_1-\hat{p}_2)\delta^2(\vec{n}_2-\hat{p}_1))\cr
&&\qquad~~
+\omega_1\omega_3(\delta^2(\vec{n}_1-\hat{p}_1)\delta^2(\vec{n}_2-\hat{p}_3)
+\delta^2(\vec{n}_1-\hat{p}_3)\delta^2(\vec{n}_2-\hat{p}_1))\cr
&&\qquad~~
+\omega_2\omega_3(\delta^2(\vec{n}_1-\hat{p}_2)\delta^2(\vec{n}_2-\hat{p}_3)
+\delta^2(\vec{n}_1-\hat{p}_3)\delta^2(\vec{n}_2-\hat{p}_2))\Big)
\ .
\ee
The evaluation of integrals proceeds following the same steps as for
$\langle {\cal E}({\vec n}_1){\cal E}({\vec n}_2)\rangle_{\text{BPS}}$. The resulting expressions, which
are now functions of four variables -- $S$, ${\vec n}_1\cdot {\vec n}_2$, ${\vec \ell} \cdot {\vec n}_1$ and
${\vec \ell} \cdot {\vec n}_2$ where ${\vec \ell}$ defines the null direction in the twist-2 spin-$S$ operators --
are quite lengthy. Multiplication by $8/(2\pi)^2 (2S+1)/q^{2S}$ yields the desired two-energy correlator,
$\langle {\cal E}({\vec n}_1){\cal E}({\vec n}_2)\rangle_{2, S}$.

\begin{figure}
  \centering
  \subfigure[]{%
    \includegraphics[height=45mm]{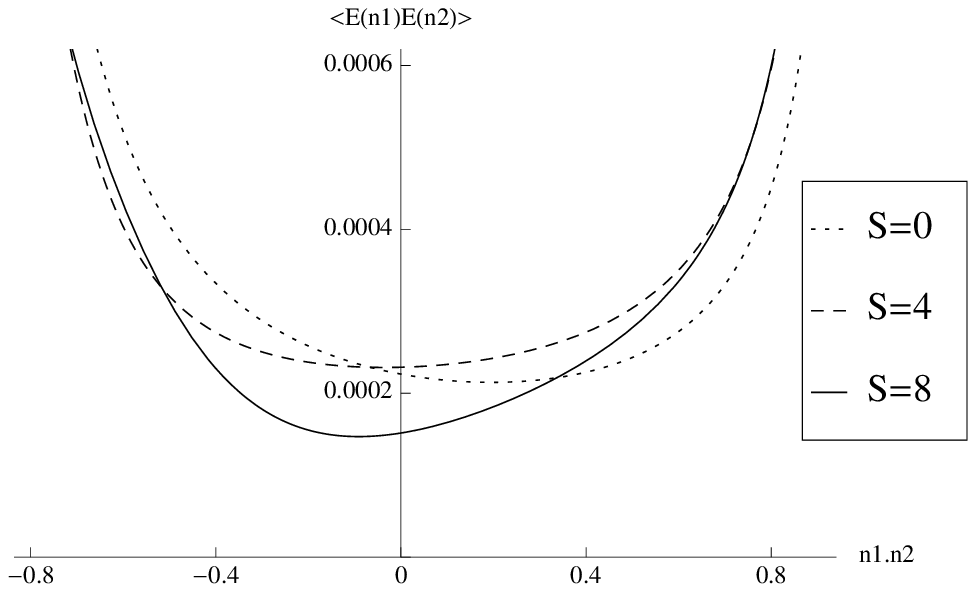}
    \label{EEplotLow}
  }
  ~~
  \subfigure[]{%
    \includegraphics[height=45mm]{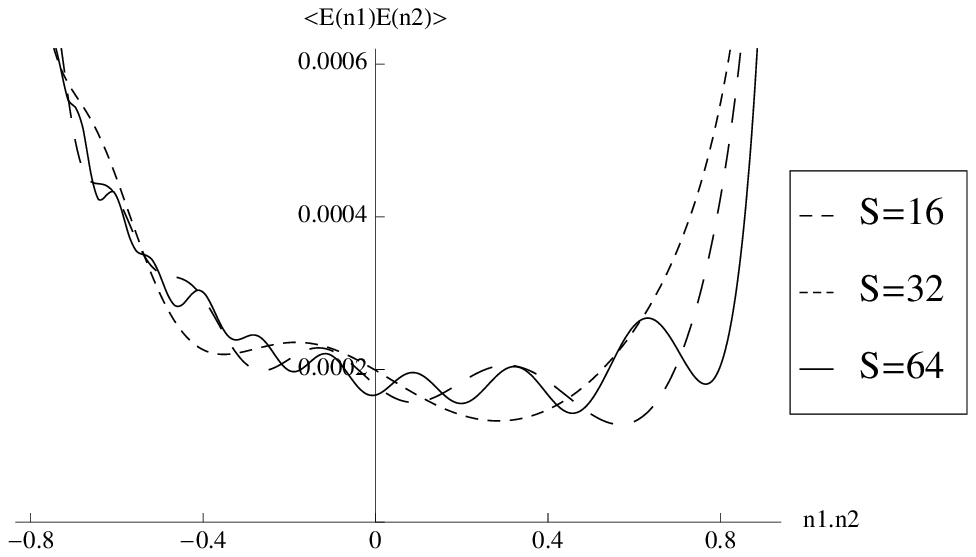}
    \label{EEplotHigh}
  }
  \caption{Plot of the two-energy correlator in the state created by a twist-2 spin-$S$ operator for $S=0$ (BPS),
$S=4$ and $S=8$ and for $S=16$, $S=32$ and $S=64$ as a function of ${\vec n}_1\cdot {\vec n}_2$,
for ${\vec\ell}\cdot {\vec n}_1=0={\vec\ell}\cdot {\vec n}_2$. The energy scale is set to $q=1$ and may be 
restored to arbitrary values by conformal invariance.}
  \label{EEplot}
\end{figure}

Its expression simplifies somewhat by choosing the vector $\vec \ell$ defining the  light-like direction to be
orthogonal to the two unit vectors ${\vec n}_{1,2}$, {\it i.e} for
${\vec\ell}\cdot {\vec n}_1=0={\vec\ell}\cdot {\vec n}_2$.
In fig.~\ref{EEplot} we show the graph of $\langle {\cal E}({\vec n}_1){\cal E}({\vec n}_2)\rangle_{2, S}$
for $S=0, 4, 8$ and for $S=16, 32, 64$ for this choice of $\vec\ell$, as a function of ${\vec n}_1\cdot {\vec n}_2$.
Similarly to the two-energy correlation in the state created by a BPS operator (recovered here as the $S=0$ case),
it  has a simple
pole at ${\vec n}_1\cdot {\vec n}_2=-1$ and a pole of order $(2S+4)$ $(S\ne 0 \text{ and even})$ at
${\vec n}_1\cdot {\vec n}_2=+1$. The numerator exhibits the same logarithmic dependence as in the
case of a state created by BPS operators, {\it i.e.} it depends on $\ln(2/(1+{\vec n}_1\cdot {\vec n}_2))$.
Unlike that case however, the coefficient of the logarithm is a polynomial of degree $(2S+2)$ in
${\vec n}_1\cdot {\vec n}_2$; the numerator also contains a polynomial of degree $(2S+3)$ in
${\vec n}_1\cdot {\vec n}_2$ which factorizes as $(1-{\vec n}_1\cdot {\vec n}_2)P_{2S+2}({\vec n}_1\cdot {\vec n}_2)$ and thus lowers the strength of the pole at ${\vec n}_1\cdot {\vec n}_2=1)$.
These polynomial factors are a consequence of  the momentum factors present in the operators. Indeed,
setting $S=0$ we recover the expression in eq.~\rf{EEBPS_2loop}.
By comparing figs.~\ref{EEplotLow} and \ref{EEplotHigh} we also note that the details of the
operator with definite anomalous dimension are important for the two-energy correlator in a large spin state
and are not important at low spin. Indeed, it is easy to see that the oscillatory behavior of Gegenbauer 
polynomials is not reflected by $\langle {\cal E}({\vec n}_1){\cal E}({\vec n}_2)\rangle_{2, S}$ at low spin, 
fig.~\ref{EEplotLow},  while it is obvious at larger spins, fig.~\ref{EEplotHigh}.

The same general features persist also if ${\vec\ell}\cdot {\vec n}_1\ne 0 \ne{\vec\ell}\cdot {\vec n}_2$. These
scalar products do not affect the pole structure of the logarithmic terms of the two-energy correlator; they appear
only in the various polynomial factors due to the momentum dependence of the operator creating the state.

\section{Outlook \label{outlook}}

Correlation functions of local gauge-invariant operators are an important class of observables in both
conformal and non-conformal field theories. In this paper we extended the generalized unitarity method
to the calculation of momentum space correlation functions; apart from the tree-level amplitudes of fundamental
fields, correlation functions of gauge-invariant operators ${\cal O}_i$ are determined by their on-shell form
factors and generalized (multi-operator)  form factors. This approach makes use of the methods developed for
efficient construction of on-shell scattering amplitudes and exposes properties of some of the contributions to
the integrand of momentum space correlation functions.
In particular, we have argued for the presence of color/kinematics duality for all internal
propagators which do not begin or end at an operator insertion up to contact terms that collapse
at least one line connected to an operator.
Using this approach we have recovered the known expression \cite{Eden:1998hh, Eden:1999kh} for the
four-point function of scalar BPS operators of dimension 2, we have found the $n-$point  functions
of such operators at the next-to-leading order and discussed the limit in which the operator insertion
points are light-like separated. As expected \cite{Eden:2010zz}, we found that in this limit the correlator
reduces to the one-loop $n-$point MHV amplitudes of $\NeqFour$ theory.
The methods described in this paper are not restricted to correlators of BPS operators. To illustrate this
we have computed the next-to-leading order correction to correlation functions of one and two
non-BPS twist-$2$ spin-$S$ operators and any number of scalar BPS operators of dimension 2 and
described the features of the correlator of $m$ twist-2 and $n$ BPS operators in a split configuration.
We have moreover discussed the ${\cal N}=4$ realization of the energy correlators and evaluated
the one-point and two-point correlators to leading nontrivial order in states created by the action of
the chiral primary and twist-2 non-BPS operators. It is not difficult to extend these calculations to higher
orders. It would be very interesting to understand the relation between the standard approach we followed
and the analytic continuation prescription suggested in \cite{Hofman:2008ar}. It would also be very
interesting to understand how to carry out the same calculation at strong coupling,
perhaps by making use of the available techniques \cite{Maldacena:2010kp} for the evaluation of
form factors.

The use of generalized unitarity for the construction of correlation functions follows closely the construction of
scattering amplitudes and thus it produces momentum space correlation functions in an algorithmic way.
It would be interesting to formulate generalized unitarity directly in position space; unlike its momentum
space version, such a formulation would likely manifestly preserve conformal symmetry. It is possible
that it will bring further simplifications to the Lagrangian insertion method.

As suggested and illustrated in \cite{Eden:2010zz, Eden:2011yp, Eden:2011ku}
position space $n$-point super-correlation functions of the chiral
stress tensor multiplet are, in the null-separation limit,
proportional to $n$-point superamplitudes. We have reproduced this result to leading nontrivial
order using the momentum space correlation functions constructed here; we have also
explicitly shown that the presence of non-BPS operators does not spoil this conclusion.

The calculation of general correlation functions at large 't~Hooft coupling remains an interesting open problem.
The use of form factors, for which Y-system-based methods have been developed \cite{Maldacena:2010kp},
together with some variant of  unitarity is a possible approach, as discussed in \S~\ref{WScoupling}.
Worldsheet form factors, recently discussed in \cite{Klose:2012ju}, should provide an alternative approach to
the same quantities. To this end it is necessary to find the worldsheet form factors of the vertex
operators  dual to gauge theory local operators and use them to construct correlation functions of vertex
operators using worldsheet generalized unitarity.

Our approach suggests that, at least in certain situations, recursive
methods could be used to construct the higher-loop integrand of momentum space correlation functions.
To this end a necessary ingredient is the existence of a BCFW shift of operator momenta for which the
tree-level form factors and generalized form factors are well-defined at infinity. Most operators explicitly
discussed in this paper have this property, as can be seen from secs.~\ref{NMHV_2T}, \ref{tree_su2},
\ref{tree_twist2} and Appendix~\ref{twist2Appendix}.
Clearly, if such a recursion relation exists it will determine simultaneously scattering amplitudes,
form factors of operators, as well as their correlation functions. While we believe that a recursion relation
exists -- if only because, for correlation functions of BPS operators at generic positions, recursion relations
have been argued to exist at strong coupling~\cite{Raju:2010by, Raju:2011mp} --
the nonplanar nature of correlation functions appears to be a (hopefully temporary) obstruction to formulating
such a recursion relation in weak coupling perturbation theory due to the non-uniqueness of the parametrization
of non-planar integrands.

As we illustrated briefly in \S~\ref{susy_methods}, supersymmetric methods initially developed for
the construction of form factors \cite{Brandhuber:2011tv, Bork:2011cj} apply
naturally to the calculation of correlation functions and are quite efficient if all operators are annihilated
by the same supersymmetry generators. In this case the presence of Grassmann $\delta-$functions in form factors
and scattering amplitudes simplifies the evaluation of sums over internal states and reduces\footnote{This holds both
in $\NeqFour$ sYM theory as well as in theories with fewer supercharges.}
them to simply an exercise in solving linear algebraic equations \cite{Bern:2009xq, Drummond:2008bq}.
While for non-BPS operators such convenient presentations of form factors do not exist, it is possible to construct
a coherent state representation for them once a component is found by some means. Adopting such a
representation \cite{ArkaniHamed:2008gz} for both the form factors of BPS operators amplitudes seems
likely to simplify the evaluation of state sums for correlation functions involving non-BPS operators.

\section*{Acknowledgments}
We are grateful to Z.~Bern, J.C.~Collins, J.~Maldacena, G.~Travaglini and A.~Tseytlin  for useful discussions
and to A.~Belitsky, J.J.~Carrasco, L.~Dixon, H.~Johansson, G.~Korchemsky, E.~Sokatchev and A.~Tseytlin
for comments on a preliminary version of this paper.
This work  is supported by the US Department of Energy under contract DE-SC0008745.


\appendix

\refstepcounter{section}
\section*{Appendix A: Harmonic superspace conventions \label{harmonic_superspace} }

\def\theequation{A.\arabic{equation}}
\setcounter{equation}{0}

The harmonic superspace \cite{Harmonic_superspace}
provides a bridge
between a non-chiral $\NeqFour$ off-shell superspace and the chiral on-shell momentum superspace
used to specify external states of scattering amplitudes and form factors. By introducing harmonic variables
$(u_{Aa}^+,u^-_{Aa'})$ valued in the coset
\be
\frac{SU(4)}{SU(2)\times SU(2)'\times U(1)}
\ee
an $SU(4)$ index is split into indices transforming under a {\em local} $SU(2)\times SU(2)'\times U(1)$ subgroup:
\be
\theta^A\;\mapsto\;\theta^{+a} = \theta^Au_A^{+a}\qquad\theta^{-a} = \theta^Au_A^{-a} \ .
\label{projection}
\ee
By gauge-fixing the local symmetries one obtains a parametrization of the coset in which the $SU(4)$
symmetry is broken. Keeping the harmonic variables generic allows one to carry out the projections  \rf{projection}
while maintaining an unbroken $SU(4)$ symmetry.

As an element of $SU(4)$ the $4\times 4$ matrix $(u_{Aa}^+,u^-_{Aa'})$ is unitary and thus its $4\times 2$
blocks  have the following properties:
\begin{align}
\bar{u}_+^{Aa}u_{Ab}^+&=\delta^a_b,&\bar{u}_-^{Aa'}u_{Ab'}^-&=\delta^{a'}_{b'},
&\bar{u}_-^{Aa'}u_{Ab}^+=\bar{u}_+^{Aa}u^-_{Ab'}&=0 \ ,
\end{align}
where the bar stands for complex conjugation. Similarly,
\begin{align}
u^+_{Aa}\bar{u}_+^{Ba}+u^-_{Aa'}\bar{u}_-^{Ba'}&=\delta^B_A \ .
\label{id}
\end{align}
Unitarity also requires that the matrix have unit determinant:
\begin{align}
\tfrac{1}{4}\epsilon^{ABCD}u^+_{Aa}\epsilon^{ab}u^+_{Bb}u^-_{Cc'}\epsilon^{c'd'}u^-_{Dd'}&=1 \ ,
\label{determinant harmonic variables}
\end{align}
where we use the convention:
\begin{align}
\epsilon^{12}=\epsilon^{1'2'}&=1=-\epsilon_{12}=-\epsilon_{1'2'}.
\end{align}
Eq.~\eqref{determinant harmonic variables} leads to the relation:
\begin{align}
\tfrac{1}{2}\epsilon^{ABCD}u_{Aa}^+\epsilon^{ab}u^+_{Bb}&=-\bar{u}_-^{Cc'}\epsilon_{c'd'}\bar{u}_-^{Dd'} \ .
\end{align}

There exists one set of harmonic variables for each operator; rather than attaching another decoration to $u$,
it is convenient to use the operator label (or the insertion point label) instead:
$u^+_{Aa}\mapsto k^+_{Aa}$. A quantity which commonly appears in calculations manifestly preserving the
$SU(4)$ symmetry is
 \begin{align}
\mathbf{(12)}&=\tfrac{1}{4}\epsilon^{ABCD}k^+_{Aa}\epsilon^{ab}k^+_{Bb}l^+_{Cc}\epsilon^{cd}l_{Dd}^+
= -\tfrac{1}{2} {\bar k}_-^{Cc'}\epsilon_{c'd'}{\bar k}_-^{Dd'}\, l^+_{Cc}\epsilon^{cd}l_{Dd}^+
\ .
\end{align}

The on-shell superspace coordinates $\eta$ can be projected similarly to \rf{projection}:
\be
\eta_A\;\mapsto\;\eta_+^{a} = {\bar u}_+^{Aa} \eta_A  \qquad\eta_-^{a'} = {\bar u}_-^{Aa'} \eta_{A} \ .
\ee
and similarly project the $\NeqFour$  fields such that the on-shell superfield is invariant; this amounts to
inserting a suitable number of identity matrices in the form \rf{id} in the superfield in footnote~
\ref{superfieldfootnote}. Fourier-transforming the coordinates $\eta_-^{a}$ leads to the projective (non-chiral)
superspace of ref.~\cite{Huang:2011um} which may be obtained by dimensional reduction from a six-dimensional
$(1,1)$ superspace. The superfield is then given by \cite{Huang:2011um}:
\be
\Phi &=& \phi^{21}+\eta^{a'}{\bar\psi}_{a'}+ {\tilde \eta}_a\psi^{a+2}
                           +{\tilde \eta}_a\eta^{a'}\phi_{a'}{}^{a+2}+\eta^2 g^- +{\tilde \eta}^2 g^+
\nonumber\\[1pt]
&&            \qquad
               + \eta^2{\tilde \eta}_a{\bar\psi}^{a+2} +{\tilde\eta}^2\eta^{a'}\psi_{a'}
                           +\tfrac{1}{4}\eta^2{\tilde\eta}^2\phi^{43}
\ee
where $\eta\equiv \eta_+$, ${\tilde \eta}$ is the Fourier conjugate of $\eta_-$,
$\eta^2=\tfrac{1}{2}\eta^a\eta_a$ and ${\tilde\eta}^2 = \tfrac{1}{2}{\tilde\eta}_{a'}{\tilde\eta}^{a'}$.

\refstepcounter{section}
\section*{Appendix B:   \label{evaluation_of_formfactors} }

\def\theequation{B.\arabic{equation}}
\setcounter{equation}{0}


In this Appendix we summarize the calculation of the various form factors and generalized form factors
used in our examples of calculation of correlation functions of local operators through the generalized
unitarity method. The following shorthand will be convenient:
\be
P_{ij}=-(p_i+p_{i+1}\cdots+p_j)
~,\qquad
Q^{\tilde{k}}_{ij}=q_{\tilde{k}}+P_{ij} \ .
\ee
Furthermore, we will use $\varepsilon$ for spinor indices and $\epsilon$ for the $SU(2)$ indices of the
harmonic variables and in addition to the definitions from Appendix~\ref{harmonic_superspace} we introduce the following conventions:
\begin{align}
\lambda_i^\alpha\varepsilon_{\alpha\beta}\lambda_j^\beta=\langle ij\rangle
\qquad
\epsilon^1_{\phantom{1}2}=+1
\ .
\end{align}

\subsection{Tree-level (NMHV) generalized form factor of the stress two tensor multiplets}


As may be easily seen using Feynman diagrams, the simplest component of the generalized form
factor with two insertions of the chiral stress tensor multiplet involves two scalar operators connected by a
propagator. We may also interpret this as the result of an MHV vertex expansion; this particular
generalized form factor arises from two MHV vertices, as shown in fig.~\ref{2T_nmhv}(a); we will
therefore refer to it as NMHV.

In computing the NMHV form factor, one $SU(4)$ index will be treated differently than the others.
We will denote it by $T$, while we will sum over the indices $A$, $B$, and $C$. By breaking manifest
$SU(4)$ symmetry we can get a relatively compact expression while still having the possibility of making
a convenient choice of $T$.
For the derivation of the non-contact part of the color-dressed components of this form factor we
will use the MHV vertex expansion, in which one sums up the contribution of MHV diagrams with
the topologies shown in fig.~\ref{2T_nmhv}(a). It is not difficult to see
that
\begin{align}
&\mathcal{F}^{NMHV}_{(a)}\nonumber\\
=&{}\frac{1}{\prod_{m=1}^n\langle m, m+1\rangle}\delta^{(8)}(1^+_a\gamma^a_{\tilde{1}+}+2^+_a
\gamma_{\tilde{2}+}^a-\sum_{m=1}^n\eta_m\lambda_m)
\delta^{(4)}(q_{\tilde{1}}+q_{\tilde{2}}-p_1\cdots-p_n)\nonumber\\
&\sum_{i=1}^n\sum_{j=i}^{i-2}\delta^4(1^+_a\langle\gamma_{\tilde{1}+}^aQ^{\tilde{1}}_{ij}\rangle
-\sum_{r=i}^j\eta_r\langle rQ^{\tilde{1}}_{ij}\rangle)\frac{1}{(Q^{\tilde{1}}_{ij})^2}
\frac{\langle jj+1\rangle\langle i-1i\rangle}{\langle Q^{\tilde{1}}_{ij}i\rangle
\langle jQ^{\tilde{1}}_{ij}\rangle\langle Q^{\tilde{1}}_{ij}j+1\rangle\langle i-1Q^{\tilde{1}}_{ij}\rangle},\\
\end{align}
where we have already combined the Grassmann $\delta-$functions of the two MHV form
factors to expose the overall super-momentum conservation constraint.
The factor $\delta^{(4)}$ may also be written as the four-fold antisymmetric product of its argument. Using two
such factors we partial fraction the ratio of spinor products as
\begin{align}
\mathcal{F}^{NMHV}_{(a)}
 =&{}\frac{1}{3!\prod_{m=1}^n\langle m, m+1\rangle}
 \delta^8(1^+_a\gamma_{\tilde{1}+}^a+2^+_a\gamma_{\tilde{2}+}^a-\sum_{m=1}^n\eta_m\lambda_m)
 \delta^4(q_{\tilde{1}}+q_{\tilde{2}}-p_1\cdots-p_n)\nonumber\\
 &\times\epsilon^{TABC}\sum_{i=1}^n\sum_{j=i}^{i-2}
 (1^+_{Ta}\langle\gamma^a_{\tilde{1}+}Q^{\tilde{1}}_{ij}\rangle-\sum_{r=i}^j\eta_{Tr}
 \langle rQ_{ij}^{\tilde{1}}\rangle)
 (1^+_{Ab}\langle\gamma^b_{\tilde{1}+}Q^{\tilde{1}}_{ij}\rangle-\sum_{r=i}^j\eta_{Ar}
 \langle rQ_{ij}^{\tilde{1}} \rangle)\nonumber\\
 &\times(1^+_{Bc}\gamma^{c\alpha}_{\tilde{1}+}-\sum_{r=i}^j\eta_{Br}\lambda_r^\alpha)
 (1^+_{Cd}\gamma^{d\gamma}_{\tilde{1}+}-\sum_{r=i}^j\eta_{Cr}\lambda_r^\gamma)
 \frac{\varepsilon_{\alpha\beta}\varepsilon_{\gamma\delta}}{(Q_{ij}^{\tilde{1}})^2}
 \nonumber\\
 &\times
 \Bigg[\frac{\lambda_j^\beta\lambda_{i-1}^\delta}{\langle jQ_{ij}^{\tilde{1}}\rangle\langle i-1Q^{\tilde{1}}_{ij}
 \rangle}+\frac{\lambda_j^\beta\lambda_i^\delta}{\langle Q^{\tilde{1}}_{ij}i\rangle\langle jQ_{ij}^{\tilde{1}}\rangle}
 +\frac{\lambda_{j+1}^\beta\lambda_{i-1}^\delta}{\langle Q^{\tilde{1}}_{ij}j+1\rangle\langle i-1Q^{\tilde{1}}_{ij}
 \rangle}
 +\frac{\lambda_{j+1}^\beta\lambda_i^\delta}{\langle Q^{\tilde{1}}_{ij}i\rangle
 \langle Q^{\tilde{1}}_{ij}j+1\rangle} \Bigg] \ .
\end{align}

To simplify somewhat this expression we introduce a factor of $\langle ij\rangle/\langle ij\rangle$ or similar
except in the cases where this is factor takes the value $0/0$ (perhaps due to restricted kinematic configuration).
Using repeatedly Schouten's identity as well as the relation
\begin{align}
\gamma^{1\alpha}_{\tilde{w}+}\gamma^{1\gamma}_{\tilde{w}+}\varepsilon_{\alpha\beta}\varepsilon_{\gamma\delta}&=\tfrac{1}{2}\langle \gamma^1_{\tilde{w}+}\gamma^1_{\tilde{w}+}\rangle\varepsilon_{\beta\delta}.
\end{align}
we find, after a lengthy calculation, that
\begin{align}
&\mathcal{F}^{NMHV}_{(a)}\nonumber\\
=&{}\frac{1}{3!\prod_{m=1}^n\langle m, m+1\rangle}\delta^8(1^+_a\gamma_{\tilde{1}+}^a+2^+_a
\gamma_{\tilde{2}+}^a-\sum_{m=1}^n\eta_m\lambda_m)
\delta^4(q_{\tilde{1}}+q_{\tilde{2}}-p_1\cdots-p_n)\nonumber\\
&\epsilon^{TABC}\Bigg[\sum_{i=1}^n\sum_{j=i+1}^{i-2}
(1^+_{Ta}\langle\gamma^a_{\tilde{1}+}
Q^{\tilde{1}}_{ij}\rangle-\sum_{r=i}^j\eta_{Tr}\langle rQ_{ij}^{\tilde{1}}\rangle)
(1^+_{Ab}\gamma^{b\kappa}_{\tilde{1}+}-\sum_{r=i}^j\eta_{Ar} \lambda_r^\kappa)\nonumber\\
&(1^+_{Bc}\gamma^{c\alpha}_{\tilde{1}+}-\sum_{r=i}^j\eta_{Br}\lambda_r^\alpha)
(1^+_{Cd}\gamma^{d\gamma}_{\tilde{1}+}-\sum_{r=i}^j\eta_{Cr}\lambda_r^\gamma)
\frac{\varepsilon_{\alpha\beta}\varepsilon_{\gamma\delta}\varepsilon_{\kappa\lambda}}
{(Q_{ij}^{\tilde{1}})^2}\bigg(\frac{\lambda_i^\beta\lambda_i^\delta\lambda_i^\lambda}
{\langle Q^{\tilde{1}}_{ij}i\rangle}\frac{\langle jj+1\rangle}{\langle ji\rangle\langle ij+1\rangle}\nonumber\\
&+\frac{\lambda_j^\beta\lambda_j^\delta\lambda_j^\lambda}{\langle jQ^{\tilde{1}}_{ij}\rangle}
\frac{\langle i-1i\rangle}{\langle i-1j\rangle\langle ji\rangle}\bigg)
+\sum_{i=1}^n\sum_{j=i}^{i-3}(1^+_{Ta}\langle\gamma^a_{\tilde{1}+}Q^{\tilde{1}}_{ij}\rangle
-\sum_{r=i}^j\eta_{Tr}\langle rQ_{ij}^{\tilde{1}}\rangle)
\\
&(1^+_{Ab}\gamma^{b\kappa}_{\tilde{1}+}-\sum_{r=i}^j\eta_{Ar} \lambda_r^\kappa)
(1^+_{Bc}\gamma^{c\alpha}_{\tilde{1}+}-\sum_{r=i}^j\eta_{Br}\lambda_r^\alpha)
(1^+_{Cd}\gamma^{d\gamma}_{\tilde{1}+}-\sum_{r=i}^j\eta_{Cr}\lambda_r^\gamma)
\frac{\varepsilon_{\alpha\beta}\varepsilon_{\gamma\delta}\varepsilon_{\kappa\lambda}}
{(Q_{ij}^{\tilde{1}})^2}\nonumber\\
&\bigg(\frac{\lambda_{i-1}^\beta\lambda_{i-1}^\delta\lambda_{i-1}^\lambda}
{\langle i-1Q^{\tilde{1}}_{ij}\rangle}\frac{\langle jj+1\rangle}{\langle ji-1\rangle\langle i-1j+1\rangle}
+\frac{\lambda_{j+1}^\beta\lambda_{j+1}^\delta\lambda_{j+1}^\lambda}
{\langle Q^{\tilde{1}}_{ij}j+1\rangle}\frac{\langle i-1i\rangle}
{\langle i-1j+1\rangle\langle j+1i\rangle}\bigg)\nonumber\\
&-\sum_{i=1}^n\frac{3\epsilon^{b'c'}1^+_{Ab'}1^+_{Bc'}}{2(Q_{ii}^{\tilde{1}})^2}
\bigg(-\eta_{Ti}\eta_{Ci}\epsilon_{bc}\langle
\gamma^b_{\tilde{1}+}i\rangle\langle\gamma^c_{\tilde{1}+}i\rangle-\tfrac{1}{4}
\epsilon^{a'd'}1^+_{Ta'}1^+_{Cd'}\langle
\gamma^1_{\tilde{1}+}\gamma^1_{\tilde{1}+}\rangle\langle\gamma^2_{\tilde{1}+}
\gamma^2_{\tilde{1}+}\rangle\nonumber\\
&+\eta_{Ti}1^+_{Db}
\epsilon^b_{\phantom{b}c}\langle\gamma^c_{\tilde{1}+}i\rangle\delta_{ad}\langle
\gamma^a_{\tilde{1}+}\gamma^d_{\tilde{1}+}
\rangle-\eta_{Di}1^+_{Tb}\epsilon^b_{\phantom{b}c}\langle\gamma^c_{\tilde{1}+}i\rangle\delta_{ad}
\langle\gamma^a_{\tilde{1}+}\gamma^d_{\tilde{1}+}
\rangle\bigg)\nonumber\\
&-\sum_{i=1}^n\frac{3\epsilon^{b'c'}2^+_{Ab'}2^+_{Bc'}}{2(Q_{ii}^{\tilde{2}})^2}
\bigg(-\eta_{Ti}\eta_{Ci}\epsilon_{bc}\langle
\gamma^b_{\tilde{2}+}i\rangle\langle\gamma^c_{\tilde{1}+}i\rangle
-\tfrac{1}{4}\epsilon^{a'd'}2^+_{Ta'}2^+_{Cd'}\langle
\gamma^1_{\tilde{2}+}\gamma^1_{\tilde{2}+}\rangle\langle
\gamma^2_{\tilde{2}+}\gamma^2_{\tilde{2}+}\rangle\nonumber\\
&+\eta_{Ti}2^+_{Db}
\epsilon^b_{\phantom{b}c}\langle\gamma^c_{\tilde{2}+}i\rangle
\delta_{ad}\langle\gamma^a_{\tilde{2}+}\gamma^d_{\tilde{2}+}
\rangle-\eta_{Di}2^+_{Tb}\epsilon^b_{\phantom{b}c}\langle
\gamma^c_{\tilde{2}+}i\rangle\delta_{ad}
\langle\gamma^a_{\tilde{2}+}\gamma^d_{\tilde{2}+}
\rangle\bigg)\nonumber\Bigg]
\end{align}
The relation $\langle iQ_{ij}^{\tilde{1}}\rangle=\langle iQ_{i+1j}^{\tilde{1}}\rangle$ implies that we can shift
the summation index and further simplify the expression of the form factor. In the end we find that the color-ordered
${\overline {\rm MHV}}$ form factor is given by
\begin{align}
&\mathcal{F}^{NMHV}_{(a)}\\
=&{}\frac{1}{3!\prod_{m=1}^n\langle m, m+1\rangle}\delta^8(1^+_a\gamma_{\tilde{1}+}^a+2^+_a\gamma_{\tilde{2}+}^a-\sum_{m=1}^n\eta_m\lambda_m)
\delta^4(q_{\tilde{1}}+q_{\tilde{2}}-p_1\cdots-p_n)\nonumber\\
&\Bigg[\sum_{w=1}^2\sum_{i=1}^n\sum_{j=i+1}^{i-2}A^{\tilde{w}}_{ij}+\sum_{w=1}^2\sum_{i=1}^nB^{\tilde{w}}_i\Bigg]\nonumber,
\end{align}
where we have used the definitions:
\begin{align}
A^{\tilde{w}}_{ij}
=&{}\epsilon^{TABC}(\eta_{Ti}(Q^{\tilde{w}}_{ij})^2-w^+_{Ta}\langle\gamma^a_{\tilde{w}+}|Q^{\tilde{w}}_{ij}|i]+\sum_{r=i}^j\eta_{Tr}\langle r|Q^{\tilde{w}}_{ij}|i])
(w^+_{Ab}\langle\gamma^{b}_{\tilde{w}+}i\rangle-\sum_{r=i}^j\eta_{Ar} \langle ri\rangle)\nonumber\\
&\times(w^+_{Bc}\langle\gamma^{c}_{\tilde{w}+}i\rangle-\sum_{r=i}^j\eta_{Br}\langle ri\rangle)
(w^+_{Cd}\langle\gamma^{d}_{\tilde{w}+}i\rangle-\sum_{r=i}^j\eta_{Cr}\langle ri\rangle)\frac{1}{(Q^{\tilde{w}}_{ij})^2(Q^{\tilde{w}}_{i+1j})^2}\frac{\langle jj+1\rangle}{\langle ji\rangle\langle ij+1\rangle} \nonumber
\end{align}
\begin{align}
B_i^{\tilde{w}}&=-\frac{3\epsilon^{b'c'}w^+_{Ab'}w^+_{Bc'}\epsilon^{TABC}}{2(Q_{ii}^{\tilde{w}})^2}\bigg(-\eta_{Ti}\eta_{Ci}\epsilon_{bc}\langle
\gamma^b_{\tilde{w}+}i\rangle\langle\gamma^c_{\tilde{w}+}i\rangle-\tfrac{1}{4}\epsilon^{a'd'}w^+_{Ta'}w^+_{Cd'}\langle
\gamma^1_{\tilde{w}+}\gamma^1_{\tilde{w}+}\rangle\langle\gamma^2_{\tilde{w}+}\gamma^2_{\tilde{w}+}\rangle\nonumber\\
&+\eta_{Ti}w^+_{Cb}
\epsilon^b_{\phantom{b}c}\langle\gamma^c_{\tilde{w}+}i\rangle\delta_{ad}\langle\gamma^a_{\tilde{w}+}\gamma^d_{\tilde{w}+}
\rangle-\eta_{Ci}w^+_{Tb}\epsilon^b_{\phantom{b}c}\langle\gamma^c_{\tilde{w}+}i\rangle\delta_{ad}
\langle\gamma^a_{\tilde{w}+}\gamma^d_{\tilde{w}+}
\rangle\bigg)\nonumber \ .
\end{align}
Here the harmonic variable $w$ takes two values corresponding the the harmonic variables of the two
chiral stress-tensor multiplets.

This analysis is sufficient for the construction of the generalized form factor of two CPO-s. For operators
at higher levels in the chiral stress tensor multiplet it is possible that MHV generalized form factors exist.
In this case it is necessary to supplement the terms above by the contribution of MHV diagrams in
fig.~\ref{2T_nmhv}$(b)$. We leave the construction of the relevant MHV form factor and of
${\cal F}^{NMHV}_{(b)}$ to the interested reader.

\subsection{MHV form factor of a scalar non-BPS operator}

The $SU(2)$ sector of $\NeqFour$ sYM theory contains operators constructed out of two complex scalar fields
whose two-point function vanishes\footnote{While in its "standard" presentation all operators are constructed
out of two holomorphic scalar fields, this sector may be embedded in the $SO(6)$ R-symmetry in many different
ways.}; it is perhaps the simplest sector which is closed under renormalization group flow. Operators in this sector are
labeled by the number of fields of each type; operators with definite anomalous dimensions are linear combinations
of single-trace monomials with fields ordered in different ways.
In this Appendix we consider a generic single-trace monomial,
${\cO}_{{A_1B_1}\dots{A_kB_k}}=\mathrm{Tr}(\phi_{A_1B_1}\cdots\phi_{A_kB_k})$; and construct
its MHV\footnote{Here MHV stands for $2k$ Grassmann parameters -- two for each scalar field. } super-form
factor; we will find the result quoted in eq.~\rf{scalarNBPS}.
By restricting the pairs $(A_i,B_i)$ to take only two values one finds the building blocks of operators in the
$SU(2)$ sector.

The MHV form factor can be computed easily when there are no gluons by just computing a single
Feynman diagram; the minimum number of external legs is $k$. To add gluons, we will rely on a
BCFW recursion relation. To preserve the MHV character of the form factor the added gluons must all have
positive helicity. We will choose to shift the momenta of two scalar fields; because of this only three-point
scalar-gluon amplitudes can appear in the recursion relation.

Similarly to amplitudes, form factors carry color structure. It may be possible to generalize the construction
of \cite{Duhr:2006iq} to color-dressed form factors (the generalization at least partly bypass the use of
multi-peripheral color decomposition \cite{DelDuca:1999rs} since the presence of the operator makes the form
factor potentially depend on other color tensors apart form the antisymmetric structure constants). Here we
will find the color-ordered form factors from which the complete color-dressed one may be constructed.

Let us focus on the form factor with external gluons and scalars and start by adding a gluon between two
adjacent scalar legs, $i$ and $j$, (in the sense that there is no other scalar between them); the rest of the
external legs will be other scalars. We choose to shift the momenta of the scalars,:
\begin{align}
|\hat{i}\rangle&=|i\rangle+z|j\rangle,\\
|\hat{j}]&=|j]-z|i] \ .
\end{align}
This
ensures that there is no boundary term at infinity. Indeed, since the gluon has to be attached to either the
leg $i$ or the leg $j$, there will  be one propagator dependent on $z$ and the form factor goes to 0 for
$z\rightarrow\infty$. The two BCFW contributions, $T_i$ and $T_j$, are:
\begin{align}
T_i=\frac{[\hat{i}i+1][i+1\hat{P}_{ii+1}]}{[\hat{i}\hat{P}_{ii+1}]}\frac{1}{s_{ii+1}}&
=\frac{-[i+1i]\langle ij\rangle}{[ii+1]\langle i+1j\rangle\langle ii+1\rangle}
=\frac{\langle ij\rangle}{\langle ii+1\rangle\langle i+1j\rangle} \ ,\\
T_j=\frac{[\hat{P}_{i+1j}i+1][i+1\hat{j}]}{[\hat{P}_{i+1j}\hat{j}]}\frac{1}{s_{i+1j}}&=0 \ .
\end{align}
The vanishing of the second contribution is a consequence of the fact that $|\hat j\rangle = |j\rangle$
and the existence of the necessary three-point amplitude requires that $\langle i+1,j\rangle=0$ {\it i.e.} the
external legs cannot be in a generic kinematic configuration.

\begin{figure}
\begin{center}
\includegraphics[height=50mm]{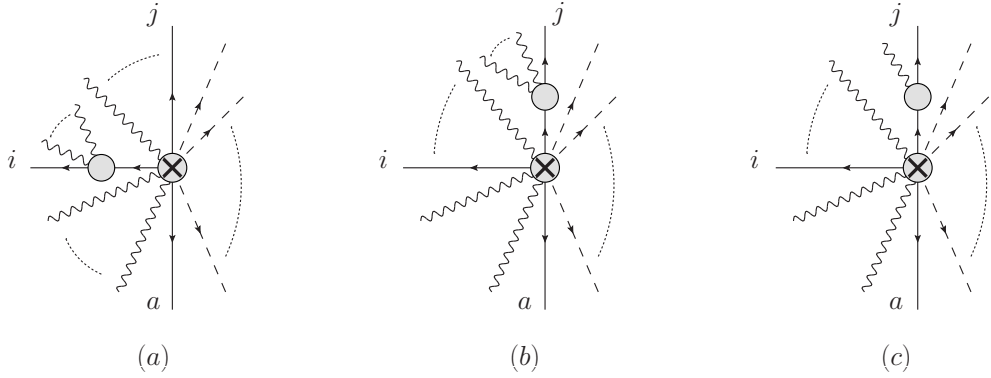}
\caption{Vanishing BCFW diagrams\label{BCFW4}}
\end{center}
\end{figure}

Proceeding to add one more gluon suggests that the expected expression for the color-ordered form factor
with external scalars and gluons is
\be
F = \frac{\langle ai\rangle\langle ij\rangle}{\prod_{k=a}^{j-1}\langle kk+1\rangle} \;C
\label{expect1}
\ee
where $i$ and $j$ are the adjacent shifted legs, $a$ labels the first scalar line before $i$ (see
figs.~\ref{BCFW4}, \ref{BCFW3}) and $C$ denotes the contribution to the form factor that depends on the
external lines  $j, j+1,\dots a$. To prove this conjecture we will show that the BCFW recursion relation
preserves its form.

Let us consider an arbitrary configuration of external lines such that there are at
least two gluons between $i$ and $j$ and at least two gluons between $\kk$ and $i$ (other configurations
must be considered separately).
The possible contributions shown in figs.~\ref{BCFW4}(a) and \ref{BCFW4}(b), in which there exists a
tree-level amplitude with at least two gluons, vanish identically because the (desired) MHV nature of the
form factor required all gluons to have positive helicities and thus requires that
the amplitude factor vanishes identically (being related to am amplitude with a single negative helicity gluon).
The BCFW contribution shown in fig.~\ref{BCFW4}(c) vanished because it requires that the momentum of the
new gluon be related to the momentum of the leg $j$.
The only non-zero diagrams are the ones shown in fig.~\ref{BCFW3}. Notice also that the computation does
not depend on what exactly is on the right side of the graph between $j$ and $\kk$; we will denote this part by
$C$.
The two contributions are:
\begin{align}
T_1^{\text{fig.}~\ref{BCFW3}(a)}=&{}\frac{[\hat{i}i+1][i+1\hat{P}_{ii+1}]}{[\hat{i}\hat{P}_{ii+1}]}
\frac{1}{s_{ii+1}}
       \frac{\langle a{-}\hat{P}_{ii+1}\rangle\langle{-}\hat{P}_{ii+1}\hat{j}\rangle}
               {\langle i-1{-}\hat{P}_{ii+1}\rangle\langle{-}\hat{P}_{ii+1}i+2\rangle}
\frac{C}{\prod_{k=a}^{i-2}\langle kk+1\rangle\prod_{l=i+2}^{j-1}\langle ll+1\rangle}\nonumber\\
=&{}\frac{C\langle ij\rangle}{\prod_{k=a}^{j-1}\langle kk+1\rangle}
\frac{\langle i-1i\rangle\langle ai+1\rangle}{\langle i-1i+1\rangle}\\[1pt]
T_2^{\text{fig.}~\ref{BCFW3}(b)}=&{}\frac{[\hat{P}_{i-1i}i-1][i-1\hat{i}]}{[\hat{P}_{i-1i}\hat{i}]}
\frac{1}{s_{i-1i}}\frac{\langle a{-}\hat{P}_{i-1i}\rangle\langle{-}\hat{P}_{i-1i}\hat{j}\rangle}
 {\langle i-2{-}\hat{P}_{i-1i}\rangle\langle{-}\hat{P}_{i-1i}i+1\rangle}
 \frac{C}{\prod_{k=a}^{i-3}\langle kk+1\rangle\prod_{l=i+1}^{j-1}\langle ll+1\rangle}\nonumber\\
=&{}\frac{C\langle ij\rangle}{\prod_{k=a}^{j-1}\langle kk+1\rangle}
\frac{\langle ai-1\rangle\langle ii+1\rangle}{\langle i-1i+1\rangle}
\end{align}
In these expressions we used \rf{expect1} for the necessary form factors.

\begin{figure}
\begin{center}
\includegraphics[height=50mm]{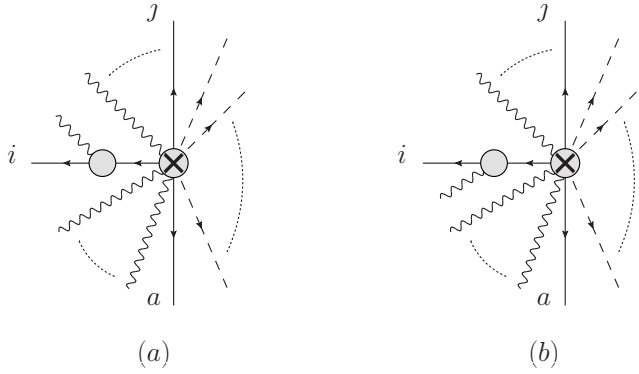}
\caption{The only non-vanishing BCFW diagrams\label{BCFW3}}
\end{center}
\end{figure}

Adding the two contributions $T_1^{\text{fig.}~\ref{BCFW3}(a)}$ and $T_1^{\text{fig.}~\ref{BCFW3}(b)}$
leads to
\be
F=\frac{\langle ai\rangle\langle ij\rangle}{\prod_{k=a}^{j-1}\langle kk+1\rangle} \;C
\label{final1}
\ee
which is consistent with the expectation \rf{expect1}.
It is not difficult to repeat the analysis above for the special cases when the number of gluons between
legs $a$ and $i$ and/or between $i$ and $j$ is $0$ or $1$ and confirm the expression \rf{final1} for these
cases as well.

The computations with fermion legs are very similar. Making use of the fact that \rf{final1} is not sensitive
to the R-symmetry index structure of the operator ${\cal O}_{{A_1B_1},\dots,{A_kB_k}}=
\mathrm{Tr}(\phi_{A_1B_1}\cdots\phi_{A_kB_k})$, the
form factors with $n$ external scalars, fermions and gluons may be combined with \rf{final1} into the
MHV super-form factor
\begin{align}
{\cal F}_{{\cal O}_{A_1B_1},\dots,{A_kB_k}}
=\frac{\delta^4(q-\sum_{l=1}^np_l)}{\prod_{m=1}\langle m, m+1\rangle}\sum_{\{a_1,b_1\cdots,a_k,b_k\}}\left(\prod_{i=1}^k\mathrm{H}_{a_iA_ib_iB_i}\right)\mathrm{Sp}\left(\prod_{j=1}^k\Sigma_{a_jb_j}\right),
\end{align}
where
\begin{align}
\mathrm{H}_{aAbB}&\equiv\eta_{Aa}\eta_{Bb}-\eta_{Ba}\eta_{Ab}+\delta_{ab}\eta_{Ba}\eta_{Ab},\\
\left(\Sigma_{a_1b_1}\right)^{\alpha}_{\phantom{\alpha}\gamma}&\equiv\lambda_{a_1}^{\alpha}
\lambda^{\beta}_{b_1}\varepsilon_{\beta\gamma}
\end{align}
and the first sum runs over all over the sets $\{a_1,b_1\cdots,a_k,b_k\}$ with particle labels valued
between $1$ and $k$ and ordered as  $a_1\leq b_1<a_2\leq b_2\cdots b_{k-1}<a_k\leq b_k$ or
$b_k<a_1\leq b_1<a_2\leq b_2\cdots b_{k-1}<a_k$ etc, as stated in \S~\ref{tree_su2}.

\subsection{Tree-level MHV form factor of twist-2 operators \label{twist2Appendix}}

Another set of operators that is closed under renormalization group flow forms the so-called $SL(2)$ sector;
these operators are linear combinations of monomials of the type
\be
{\cal O}^{AB, CD}_{2, S, x}=\mathrm{Tr}(D_+^x\phi^{AB}D_+^{S-x}\phi^{CD}) \ ,
\label{twist2_fixed_dim}
\ee
as stated in eq.~\rf{twist2_fixed_dim}. In this appendix we show that the tree-level form factor of these
monomials are given by eq.~\rf{twist_2_form_factor}
\begin{align}
\label{form factor}
{\cal F}_{{\cal O}^{AB, CD}_{2, S, x}}=&
\frac{\delta^4(q-\sum_{l=1}^np_l)}{\prod_{m=1}^n\langle m, m+1\rangle}\sum_{\{a,b,c,d\}}
\mathrm{H}_{aAbB}\mathrm{H}_{cCdD}\sum_{k=b}^{c-1}\sum_{l=d}^{a-1}
\left(\sum_{r=l+1}^k p_r^-\right)^x\left(\sum_{s=k+1}^lp_s^-\right)^{S-x}              \\
&\left(\frac{\langle b|\sigma^-\cancel{p}_k|c\rangle}{2p_k^-}+
\frac{\langle b|\cancel{p}_{k+1}\sigma^-|c\rangle}{2p_{k+1}^-}-\langle bc\rangle\right)
\Bigg(\frac{\langle d|\sigma^-\cancel{p}_l|a\rangle}{2p_l^-}
+\frac{\langle d|\cancel{p}_{l+1}\sigma^-|a\rangle}{2p_{l+1}^-}-\langle da\rangle\Bigg) \ ;
\nonumber
\end{align}
the first sum runs over all sets $\{a,b,c,d\}$ where $a\leq b<c\leq d$ or $d<a\leq b<c$ etc.
Fig.~\ref{eksempel} summarizes the notation we use.

\begin{figure}
\begin{center}
\includegraphics[height=40mm]{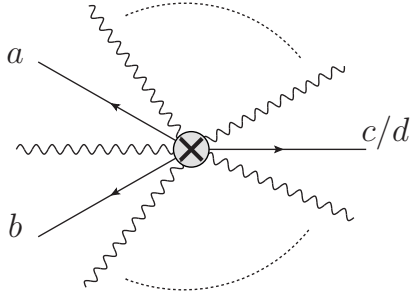}
\caption{An example of a form factor with two external fermions and one external scalar\label{eksempel}}
\end{center}
\end{figure}

We will focus on the case when all external lines are either scalars of gluons. The form factor with
two external scalars and no gluons can be easily constructed by inspection: it is just the product of the
"$-$" components of the momenta of the two scalars to the appropriate powers.

\begin{figure}
\begin{center}
\includegraphics[height=15mm]{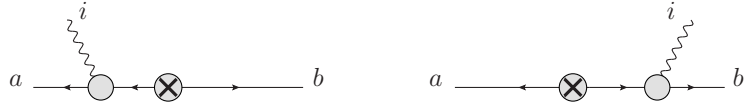}
\caption{The two classes of BCFW terms contributing to the form factor of twist-2 operators. \label{BCFW en gluon}}
\end{center}
\end{figure}

To prove eqs.~\rf{twist_2_form_factor}, \rf{form factor} we will use a variant of a BCFW shift.
In choosing the shift we would like to avoid a boundary term arising from
$z\to\infty$ region; this can be accomplished by the following shift that involves the momentum
of a gluon and of the operator itself:
\begin{align}
|\hat{i}\rangle&=|i\rangle+z|\phi\rangle,\\
\hat{q}^\mu&=q^\mu+\tfrac{1}{2}z[i|\sigma^\mu|\phi\rangle,
\end{align}
where $|\phi\rangle$ is chosen to satisfy $[i|\sigma^-|\phi\rangle=0$; such a choice is always possible.
With this shift the BCFW recursion relation for the MHV form factor of ${\cal O}^{AB, CD}_{2, S, x}$ has
only the two contributions shown in fig.~\ref{BCFW en gluon}:
\begin{align}
T_1^{\text{fig.~}\ref{BCFW en gluon}(a)}
=&{}\frac{[a\hat{i}][\hat{i}{-}\hat{P}_{ia}]}{[a{-}\hat{P}_{ia}]}\frac{1}{P_{ia}^2}
\left(\hat{P}_{ia}^-\right)^x(p_b^-)^{S-x}
={}\frac{\langle a\phi\rangle}{\langle ai\rangle\langle i\phi\rangle}(p_a^-+p_i^-)^x(p_b^-)^{S-x}\nonumber\\
=&{}\frac{\langle a|\sigma^-\cancel{p}_i|b\rangle}{2p_i^-\langle ai\rangle
\langle ib\rangle}(p_a^-+p_i^-)^x(p_b^-)^{S-x}\\
T_2^{\text{fig.~}\ref{BCFW en gluon}(b)}=&
{}\frac{[{-}\hat{P}_{ib}\hat{i}][\hat{i}b]}{[{-}\hat{P}_{ib}b]}\frac{1}{P_{ib}^2}(p_a^-)^x
\left(\hat{P}_{ib}^-\right)^{S-x}
={}\frac{\langle \phi b\rangle}{\langle ib\rangle\langle\phi i\rangle}(p_a^-)^x(p_i^-+p_b^-)^{S-x}\nonumber\\
=&{}\frac{\langle a|\cancel{p}_i\sigma^-|b\rangle}{2p_i^-\langle ai\rangle\langle ib\rangle}(p_a^-)^x(p_i^-+p_b^-)^{S-x} \ .
\end{align}
in deriving these expressions we used that
\begin{align}
\frac{\langle a\phi\rangle}{\langle i\phi\rangle}
=\frac{\langle i|\sigma^-|i]\langle a\phi\rangle}{2p_i^-\langle i\phi\rangle}
=\frac{\langle a|\sigma^-|i]}{2p_i^-}
\end{align}
which follows from the Schouten identity and the choice of $|\phi\rangle$ such that $[i|\sigma^-|\phi\rangle=0$.
Adding the $T_1^{\text{fig.~}\ref{BCFW en gluon}(a)}$ and $T_2^{\text{fig.~}\ref{BCFW en gluon}(b)}$
we find a result that is indeed consistent with the general expression shown in
eqs.~\rf{twist_2_form_factor} and \rf{form factor}:
\be
\label{3fNBPS}
&&\langle  {\widetilde{\cal O}}^{AB, CD}_{2, S, x}| \phi_{AB} A^+ \phi_{CD}\rangle \\
&&=
\frac{\langle ba\rangle}{\langle ai\rangle\langle ib\rangle\langle ba\rangle}
\Big((p_a^-)^x(p_i^-+p_b^-)^{S-x}\frac{\langle a|\cancel{p}_i\sigma^-|b\rangle}{2p_i^-}
      +(p_a^-+p_i^-)^x(p_b^-)^{S-x}\frac{\langle a|\sigma^-\cancel{p}_i|b\rangle}{2p_i^-}\Big) \ .
\nonumber
\ee
This expression was used in the evaluation of some of the cuts in \S~\ref{BPSNBPS}.
It is also possible to check that this agrees with the
Feynman diagram evaluation of
$\langle  {\widetilde{\cal O}}^{AB, CD}_{2, S, x}| \phi_{AB} A^+ \phi_{CD}\rangle$.
It is also easy to see that in the limit $S, x\rightarrow 0$ eq.~\rf{3fNBPS} reduces, as it should, to the
three-field form factor of the BPS scalar operator.

\begin{figure}
\begin{center}
\includegraphics[height=28mm]{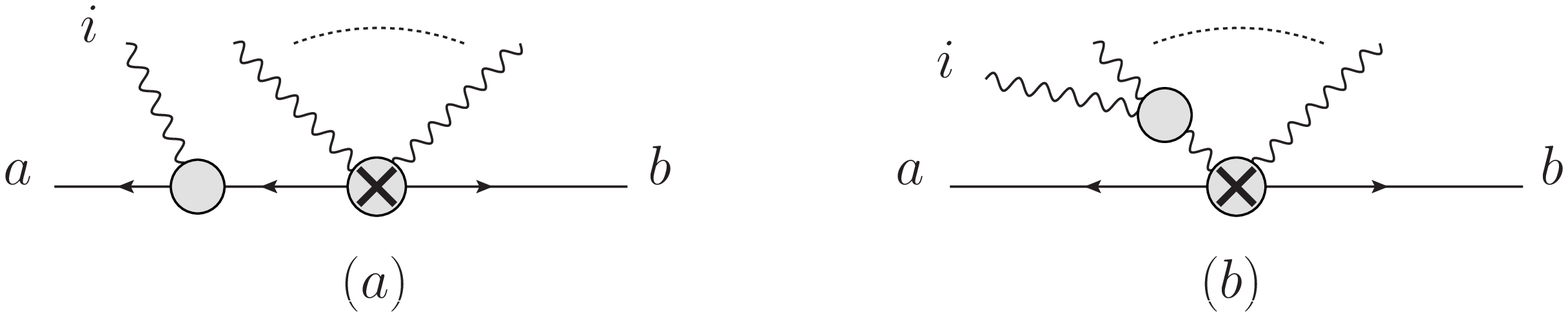}
\caption{\label{BCFW gluoner til en side}}
\end{center}
\end{figure}

Let us proceed to the general case, in which we add one gluon to a form factor which already has some
arbitrary number of gluons present;  the two non-vanishing contributions to the color-ordered form factor
in the ordering in which all gluons are adjacent are shown in fig.~\ref{BCFW gluoner til en side}:
\begin{align}
T_1^{\text{fig.~}\ref{BCFW gluoner til en side}(a)}
=&{}\frac{[a\hat{i}][\hat{i}{-}\hat{P}_{ia}]}{[a{-}\hat{P}_{ia}]}\frac{1}{P_{ia}^2}\frac{1}{\prod_{m=i+1}^b\langle m, m+1\rangle}\frac{1}{\langle\hat{P}_{ia}i+1\rangle}\Bigg[\sum_{k=i+1}^{b-1}\left(\hat{P}_{ia}^-+\sum_{r=i+1}^kp_l^-\right)^x\left(\sum_{s=k+1}^bp_s^-\right)^{S-x}\nonumber\\
&\left(\frac{\langle\hat{P}_{ia}|\sigma^-\cancel{p}_k|b\rangle}{2p_k^-}+\frac{\langle\hat{P}_{ia}\cancel{p}_{k+1}\sigma^-|b\rangle}{2p_{k+1}^-}-\langle\hat{P}_{ia}b\rangle\right)+\left(\hat{P}_{ia}^-\right)^x
\left(\sum_{s=i+1}^bp_s^-\right)^{S-x}\frac{\langle\hat{P}_{ia}|\cancel{p}_{i+1}\sigma^-|b\rangle}{2p_{i+1}^-}\Bigg]\nonumber\\
=&{}\frac{1}{\prod_{m=a}^{b-1}\langle m, m+1\rangle}(p_a^-+p_i^-)^x
\left(\sum_{s=i+1}^bp_s^-\right)^{S-x}\bigg(\frac{\langle a|\sigma^-\cancel{p}_i|b\rangle}{2p_i^-}
+\frac{\langle a|\cancel{p}_{i+1}\sigma^-|b\rangle}{2p_{i+1}^-}-\langle ab\rangle\bigg)\nonumber\\
&+\frac{1}{\prod_{m=a}^{b-1}\langle m, m+1\rangle}\frac{\langle a|\sigma^-|i]\langle ii+1\rangle}
{2p_i^-\langle ai+1\rangle}\sum_{k=i+1}^{b-1}\left(\sum_{r=a}^kp_r^-\right)^x
\left(\sum_{s=k+1}^bp_s^-\right)^{S-x}
\nonumber\\
&\qquad\times\bigg(\frac{\langle a|\sigma^-\cancel{p}_k|b\rangle}{2p_k^-}+\frac{\langle a|\cancel{p}_{k+1}\sigma^-|b\rangle}{2p_{k+1}^-}-\langle ab\rangle\bigg)
\end{align}
\begin{align}
&T_2^{\text{fig.~}\ref{BCFW gluoner til en side}(b)}
={}\frac{[\hat{i}i+1]^4}{[{-}\hat{P}_{ii+1}\hat{i}][\hat{i}i+1][i+1{-}\hat{P}_{ii+1}]}\frac{1}{P_{ii+1}^2}\frac{1}{\prod_{m=i+2}^{b-1}\langle m, m+1\rangle}\frac{1}{\langle a\hat{P}_{ii+1}\rangle\langle\hat{P}_{ii+1}i+2\rangle}\nonumber\\
&\Bigg[(p_a^-)^x\left(\hat{P}_{ii+1}^-+\sum_{s=i+2}^bp_m^-\right)^{S-x}\frac{\langle a|\hat{\cancel{P}}_{ii+1}\sigma^-|b\rangle}{2\hat{P}_{ii+1}^-}\nonumber\\
&+(p_a^-+\hat{P}_{ii+1}^-)^x\left(\sum_{s=i+2}^bp_m^-\right)^{S-x}
\left(\frac{\langle a|\sigma^-\hat{\cancel{P}}_{ii+1}|b\rangle}{2\hat{P}_{ii+1}^-}+\frac{\langle a|\cancel{p}_{i+2}\sigma^-|b\rangle}{2p_{i+2}^-}-\langle ab\rangle\right)\nonumber\\
&+\sum_{k=i+2}^{b-1}\left(p_a^-+\hat{P}^-_{ii+1}+\sum_{r=i+2}^kp_r^-\right)^x\left(\sum_{s=k+1}^bp_s^-\right)^{S-x}\left(\frac{\langle a|\sigma^-\cancel{p}_k|b\rangle}{2p_k^-}+\frac{\langle a|\cancel{p}_{k+1}\sigma^-|b\rangle}{2p_{k+1}^-}-\langle ab\rangle\right)\Bigg]\nonumber\\
=&{}\frac{1}{\prod_{m=a}^{b-1}\langle m, m+1\rangle}(p_a^-)^x\left(\sum_{s=i}^bp_s^-\right)^{S-x}\frac{\langle a|\cancel{p}_i\sigma^-|b\rangle}{2p_i^-}+\frac{1}{\prod_{m=i}^{b-1}\langle m, m+1\rangle}\frac{\langle i+1|\sigma^-|i]}{2p_i^-\langle ai+1\rangle}
\nonumber\\
&\qquad\times\sum_{k=i+1}^{b-1}\left(\sum_{r=a}^kp_r^-\right)^x\left(\sum_{s=k+1}^bp_s^-\right)^{S-x}
\bigg(\frac{\langle a|\sigma^-\cancel{p}_k|b\rangle}{2p_k^-}
+\frac{\langle a|\cancel{p}_{k+1}\sigma^-|b\rangle}{2p_{k+1}^-}-\langle ab\rangle\bigg)
\end{align}
Adding together the two contributions we find exactly the scalar-gluon component  of eqs.~\rf{twist_2_form_factor}
and \rf{form factor}. We described explicitly here the construction of the form factor with adjacent gluons; a very
similar calculation yields the form factor with some gluons above and some gluons below the scalar line.
Among the MHV form factors there also some with two external fermions; For the appropriate color ordering it
combines with the scalar-gluon form factor and yield eqs.~\rf{twist_2_form_factor} and \rf{form factor}.

\refstepcounter{section}
\section*{Appendix C:  On the Fourier-transform of the NLO four-point function of CPO-s\label{Fourier} }

\def\theequation{C.\arabic{equation}}
\setcounter{equation}{0}

Let us begin by Fourier-transforming the double-box integral to position space.
\be
\text{DB}(1,2|3,4) &=& \int \prod_{i=1}^7\frac{d^4 l_i}{(2\pi)^4}\frac{1}{l_i^2}
(2\pi)^{24}\delta^4(q_1+l_1+l_2)\delta^4(q_2+l_3-l_2)\delta^4(q_4+l_4+l_5)
\cr
&&\qquad \qquad \qquad \qquad \delta^4(q_3+l_6-l_5) \delta^4(l_7-l_3-l_6)\delta^4(-l_1-l_3-l_7)
\\
&=&\int \prod_{j=1}^6d^4 x_j\int \prod_{i=1}^7\frac{d^4 l_i}{(2\pi)^4}\frac{1}{l_i^2}
e^{ix_1\cdot (q_1+l_1+l_2)}e^{ix_2\cdot (q_2+l_3-l_2)}
\\
&&\qquad \qquad \qquad \qquad e^{ix_4\cdot (q_4+l_4+l_5)}
e^{ix_3\cdot (q_3+l_6-l_5)} e^{ix_5\cdot (l_7-l_3-l_6)}e^{ix_6\cdot (-l_1-l_3-l_7)}
\cr
&=&\int {\prod_{j=1}^6d^4 x_j\; \; e^{i\sum_{i=1}^4 x_i\cdot q_i}}{I_{16}\;I_{12}\;
I_{25}\;I_{35}\;I_{34}I_{46}I_{56}}
\label{DBpos}
\ee
where $I_{ij}$ are standard position space propagators\footnote{In Minkowski space the propagator has an
additional  factor of $i$.}
\be
I_{ij}=\frac{1}{(2\pi)^2(x_i-x_j)^2} \ .
\ee
Fourier-transforming \rf{DBpos} to position space simply freezes the integrals over $x_1, x_2, x_3$ and $x_4$
to the location of the operator insertions. Momentum space numerator factors become differential operators
acting on the integrand.

Repeating the derivation above for the other integrals in eq.~\rf{2loop4pt} we find that, up to derivative
and numerator factors, the integrals that can appear in their Fourier transform are
\be
Y_{123}&=&\int d^4 w I_{1w}I_{2w}I_{3w}
\\
X_{1234}&=&\int d^4 w I_{1w}I_{2w}I_{3w}I_{4w}
\\
H_{12;34}&=&\int d^4 ud^4 v I_{1u}I_{2u}\,I_{uv}I_{3v}I_{4v} \ ,
\ee
The correlation functions depend only on  $(\partial_1+\partial_2)^2H_{41;23}$ or, perhaps more symmetrically,
\be
((\partial_1+\partial_2)^2+(\partial_3+\partial_4)^2)H_{41;23} \ ;
\ee
following the ideas of \cite{Beisert:2002bb}, in this appendix we spell out some details of the reduction of the
relevant derivatives of the $H$ integral.

This integral does not require any regularization as it is finite both in the UV and in the IR. Furthermore, the
terms in which two derivatives act on the same coordinate reduce immediately to integrals of the type $Y_{ijk}$
upon use of
\be
\partial_i^2 I_{ij}=\delta^4(x_i-x_j) \ ;
\ee
We therefore need to reduce only
\be
I(x_1,x_2,x_3,x_4)=\int \prod_{i=1}^4\frac{d^4p_i}{(2\pi)^4} e^{i \sum_{i=1}^4 p_i\cdot x_i}\frac{p_1\cdot
p_2+p_3\cdot p_4}{p_1^2p_2^2p_3^2p_4^2(p_1+p_4)^2}(2\pi)^4\delta^{(4)}(\sum_{i=1}^4 p_i)
\ee

Following \cite{Beisert:2002bb}, we multiply and divide by $(x_1-x_4)^2(x_2-x_3)^2$ and rewrite the numerator
in terms of derivatives on the integrand:
\be
I&=&{}\frac{(x_1-x_4)^2(x_2-x_3)^2}{(x_1-x_4)^2(x_2-x_3)^2}I,\nonumber\\
&=&{}\frac{1}{x_{14}^2x_{23}^2}\int \prod_{i=1}^4\frac{d^4p_i}{(2\pi)^4}
\left((\partial_{p_1}-\partial_{p_4})^2e^{ip_1\cdot x_1+ip_4\cdot x_4}\right)
\left((\partial_{p_2}-\partial_{p_3})^2e^{ip_2\cdot x_2+ip_3\cdot x_3}\right)\\
&&
\quad\times\frac{p_1\cdot p_2+p_3\cdot p_4}{p_1^2p_2^2p_3^2p_4^2(p_1+p_4)^2}
(2\pi)^4\delta^{(4)}(p_1+p_2+p_3+p_4) \ .
\label{Itimesrat}
\ee

We then integrate by parts $(\partial_{p_1}-\partial_{p_4})^2$ setting aside the factor
$(p_2^2p_3^2(p_1+p_4)^2)^{-1}$ on which this combination of derivatives act trivially
and find that the integrand becomes
\be
(\partial_{p_1}-\partial_{p_4})^2\frac{p_1\cdot p_2+p_3\cdot p_4}{p_1^2p_4^2}&=&
{}-\frac{4(p_1+p_4)^2(p_1\cdot p_2+p_3\cdot p_4)}{p_1^4p_4^4}-\frac{2(p_2+p_3)^2}{p_1^2p_4^2}
\left(\frac{1}{p_4^2}+\frac{1}{p_1^2}\right)\nonumber\\
&&{}-\frac{2}{p_1^2p_4^2}\left(\frac{p_2^2}{p_4^2}-\frac{p_3^2}{p_4^2}+\frac{p_3^2}{p_1^2}
-\frac{p_2^2}{p_1^2}\right) \ .
\label{firstfactor}
\ee
The main difference between the first two and the last terms above is that in the former the denominator
factor $(p_1+p_4)^2$ cancels out. We will therefore treat these terms separately.

Plugging the last term in the original integral \rf{Itimesrat}, extracting the remaining derivatives as $x_{23}^2$
and adding to the result its image under the combined transformations $x_1\leftrightarrow x_2$ and
$x_3\leftrightarrow x_4$ it is not difficult to see that this term leads to $Y_{ijk}$-type integrals multiplied by
ratios of distances $x_{ij}^2$:
\be
\frac{1}{x^2_{14}x^2_{23}}
\left((x_{34}^2-x_{13}^2)Y_{134}-(x_{24}^2-x_{12}^2)Y_{124}
+(x_{34}^2-x_{24}^2)Y_{234}-(x_{13}^2-x_{12}^2)Y_{134}\right) \ .
\label{Ypart}
\ee
A useful relation is
\be
x_{12}^2Y_{123} &= &{}4\int d^4y\int\prod_{i=1}^3\frac{d^4p_i}{(2\pi)^4}
e^{ip_1\cdot(x_1-y)+ip_2\cdot(x_2-y)+ip_3\cdot(x_3-y)}
\frac{1}{p_1^2p_2^2p_3^2}\left(\frac{1}{p_1^2}+\frac{1}{p_2^2}\right) \\
&-&4\int \frac{d^4p_1d^4p_2}{(2\pi)^8}\frac{e^{ip_1\cdot x_{13}+ip_2\cdot x_{23}}}{p_1^4p_2^4}
-4\int\frac{d^4p_1}{(2\pi)^4}\frac{e^{ip_1\cdot x_{13}}}{p_1^4}
-4\int\frac{d^4p_2}{(2\pi)^4}\frac{e^{ip_3\cdot x_{23}}}{p_2^4}
\ .
\nonumber
\ee
which may be proven by starting with a Fourier transform representation for $Y_{123}$.

Returning to the first two terms in \rf{firstfactor} we notice that upon restoring  the factor of
$(p_2^2p_3^2(p_1+p_4)^2)^{-1}$ the mixed-momentum propagator cancels out. Acting with the
remaining derivatives, $(\partial_{p_2}-\partial_{p_3})^2$, we find
\be
&&(\partial_{p_2}-\partial_{p_3})^2\left(\frac{-4(p_1\cdot p_2+p_3\cdot p_4)}{p_1^4p_2^2p_3^2p_4^4}
-\frac{2}{p_1^2p_2^2p_3^2p_4^2}\left(\frac{1}{p_4^2}+\frac{1}{p_1^2}\right)\right)\nonumber\\
&&\qquad
={}\frac{-32}{p_1^4p_2^4p_3^4p_4^4}
\big(p_1\cdot p_2p_3\cdot p_4+p_1\cdot p_4 p_2\cdot p_3-p_1\cdot p_3p_2\cdot p_4\big).\nonumber
\ee
Integrating this expression with the measure factor following from eq.~\rf{Itimesrat} one can check that
the result may be reorganized as
\be
\frac{x_{13}^2x_{24}^2-x_{12}^2x_{34}^2-x_{14}^2x_{23}^2}{2x_{14}^2x_{23}^2}\int [d^dp]
e^{ip_1\cdot x_1+ip_2\cdot x_2+ip_3\cdot x_3+ip_4\cdot x_4}\frac{1}{p_1^2p_2^2p_3^2p_4^2}
\label{Xpart}
\ee

Combining eqs.~\rf{Xpart} and \rf{Ypart} it is easy to see that
\be
\frac{-(\partial_1\cdot\partial_2+\partial_3\cdot\partial_4)H_{41,23}}{I_{14}I_{23}}=
\frac{X_{1234}}{I_{13}I_{24}}-
\frac{X_{1234}}{I_{34}I_{12}}-
\frac{X_{1234}}{I_{23}I_{14}}+G_{3,41}-G_{2,41}+G_{4,32}-G_{1,32}\ . ~~
\label{usefulid}
\ee
where
\be
G_{3;41}=\frac{Y_{341}}{I_{31}}-\frac{Y_{341}}{I_{34}} \ .
\ee

Using similar techniques, or as a limit of \rf{usefulid} by noticing that
in the limit $x_1\rightarrow x_2$ the right-hand side of eq.~\rf{Xpart} becomes identically zero,
it is possible to show that
\be
\partial_1^2 H_{41;13} = \frac{I_{13}I_{14}}{I_{34}}Y_{134} \ .
\ee
which was originally proven in \cite{Eden:1998hh} by explicit integration.


\newpage

\begin{thebibliography}{99}

\bibitem{Alday:2010zy}
  L.~F.~Alday, B.~Eden, G.~P.~Korchemsky, J.~Maldacena and E.~Sokatchev,
  ``From correlation functions to Wilson loops,''
  JHEP {\bf 1109}, 123 (2011)
  [arXiv:1007.3243 [hep-th]].

\bibitem{Alday:2011ga}
  L.~F.~Alday, E.~I.~Buchbinder and A.~A.~Tseytlin,
  ``Correlation function of null polygonal Wilson loops with local operators,''
  JHEP {\bf 1109}, 034 (2011)
  [arXiv:1107.5702 [hep-th]].

\bibitem{Engelund:2011fg}
  O.~T.~Engelund and R.~Roiban,
  ``On correlation functions of Wilson loops, local and non-local operators,''
  arXiv:1110.0758 [hep-th].

\bibitem{Adamo:2011cd}
  T.~Adamo,
  ``Correlation functions, null polygonal Wilson loops, and local operators,''
  JHEP {\bf 1112}, 006 (2011)
  [arXiv:1110.3925 [hep-th]].

\bibitem{Eden:2010zz}
  B.~Eden, G.~P.~Korchemsky and E.~Sokatchev,
  ``From correlation functions to scattering amplitudes,''
  JHEP {\bf 1112}, 002 (2011)
  [arXiv:1007.3246 [hep-th]].

\bibitem{energy_correlators}
C. L. Basham, L. S. Brown, S. D. Ellis and S. T. Love,
ÒEnergy Correlations In Electron-Positron Annihilation In Quantum Chromodynamics:
Asymptotically Free Perturbation Theory,Ó
Phys.\ Rev.\ D {\bf 19}, 2018 (1979).
%
C. L. Basham, L. S. Brown, S. D. Ellis and S. T. Love,
ÒEnergy Correlations In Electron - Positron Annihilation: Testing QCD,Ó
Phys.\ Rev.\ Lett. {\bf 41}, 1585 (1978).
%
C. L. Basham, L. S. Brown, S. D. Ellis and S. T. Love,
Phys.\ Rev.\ D {\bf 17}, 2298 (1978).

\bibitem{Hofman:2008ar}
  D.~M.~Hofman and J.~Maldacena,
  ``Conformal collider physics: Energy and charge correlations,''
  JHEP {\bf 0805}, 012 (2008)
  [arXiv:0803.1467 [hep-th]].

\bibitem{Georgiou:2012zj}
  G.~Georgiou, V.~Gili, A.~Grossardt and J.~Plefka,
  ``Three-point functions in planar N=4 super Yang-Mills Theory for scalar operators up to length five at the one-loop order,''
  JHEP {\bf 1204}, 038 (2012)
  [arXiv:1201.0992 [hep-th]].

\bibitem{Intriligator:1998ig}
  K.~A.~Intriligator,
  ``Bonus symmetries of N=4 superYang-Mills correlation functions via AdS duality,''
  Nucl.\ Phys.\ B {\bf 551}, 575 (1999)
  [hep-th/9811047].

  \bibitem{Howe:1999hz}
  P.~S.~Howe, C.~Schubert, E.~Sokatchev and P.~C.~West,
  ``Explicit construction of nilpotent covariants in N=4 SYM,''
  Nucl.\ Phys.\ B {\bf 571}, 71 (2000)
  [hep-th/9910011].


\bibitem{Eden:2000mv}
  B.~Eden, C.~Schubert and E.~Sokatchev,
  ``Three loop four point correlator in N=4 SYM,''
  Phys.\ Lett.\ B {\bf 482}, 309 (2000)
  [hep-th/0003096].

\bibitem{Okuyama:2004bd}
  K.~Okuyama and L.~-S.~Tseng,
  ``Three-point functions in N = 4 SYM theory at one-loop,''
  JHEP {\bf 0408}, 055 (2004)
  [hep-th/0404190].

\bibitem{Roiban:2004va}
  R.~Roiban and A.~Volovich,
  ``Yang-Mills correlation functions from integrable spin chains,''
  JHEP {\bf 0409}, 032 (2004)
  [hep-th/0407140].

 \bibitem{Alday:2005nd}
  L.~F.~Alday, J.~R.~David, E.~Gava and K.~S.~Narain,
  ``Structure constants of planar N = 4 Yang Mills at one loop,''
  JHEP {\bf 0509}, 070 (2005)
  [hep-th/0502186].

  \bibitem{Escobedo:2010xs}
  J.~Escobedo, N.~Gromov, A.~Sever and P.~Vieira,
  ``Tailoring Three-Point Functions and Integrability,''
  JHEP {\bf 1109}, 028 (2011)
  [arXiv:1012.2475 [hep-th]].
%
  N.~Gromov and P.~Vieira,
  ``Quantum Integrability for Three-Point Functions,''
  arXiv:1202.4103 [hep-th].

\bibitem{Caetano:2011eb}
  J.~Caetano and J.~Escobedo,
  ``On four-point functions and integrability in N=4 SYM: from weak to strong coupling,''
  JHEP {\bf 1109}, 080 (2011)
  [arXiv:1107.5580 [hep-th]].

\bibitem{Ananth:2012tf}
  S.~Ananth, S.~Kovacs and S.~Parikh,
  ``Gauge-invariant correlation functions in light-cone superspace,''
  arXiv:1203.5376 [hep-th].

\bibitem{Eden:2011we}
  B.~Eden, P.~Heslop, G.~P.~Korchemsky and E.~Sokatchev,
  ``Hidden symmetry of four-point correlation functions and amplitudes in N=4 SYM,''
  Nucl.\ Phys.\ B {\bf 862}, 193 (2012)
  [arXiv:1108.3557 [hep-th]].

  \bibitem{Eden:2012tu}
  B.~Eden, P.~Heslop, G.~P.~Korchemsky and E.~Sokatchev,
  ``Constructing the correlation function of four stress-tensor multiplets and the four-particle amplitude in N=4 SYM,''
  Nucl.\ Phys.\ B {\bf 862}, 450 (2012)
  [arXiv:1201.5329 [hep-th]].

  \bibitem{Eden:2012fe}
  B.~Eden, P.~Heslop, G.~P.~Korchemsky, V.~A.~Smirnov and E.~Sokatchev,
  ``Five-loop Konishi in N=4 SYM,''
  Nucl.\ Phys.\ B {\bf 862}, 123 (2012)
  [arXiv:1202.5733 [hep-th]].

\bibitem{5loopKonishi}
Z.~Bajnok, A.~Hegedus, R.~A.~Janik and T.~Lukowski,
  ``Five loop Konishi from AdS/CFT,''
  Nucl.\ Phys.\ B {\bf 827}, 426 (2010)
  [arXiv:0906.4062 [hep-th]].

\bibitem{5loopKonishiTBA}
G.~Arutyunov, S.~Frolov and R.~Suzuki,
  ``Five-loop Konishi from the Mirror TBA,''
  JHEP {\bf 1004}, 069 (2010)
  [arXiv:1002.1711 [hep-th]].
%
J.~Balog and A.~Hegedus,
  ``5-loop Konishi from linearized TBA and the XXX magnet,''
  JHEP {\bf 1006}, 080 (2010)
  [arXiv:1002.4142 [hep-th]].

 \bibitem{Janik:2010gc}
  R.~A.~Janik, P.~Surowka and A.~Wereszczynski,
  ``On correlation functions of operators dual to classical spinning string states,''
  JHEP {\bf 1005}, 030 (2010)
  [arXiv:1002.4613 [hep-th]].

 \bibitem{Buchbinder:2010vw}
  E.~I.~Buchbinder and A.~A.~Tseytlin,
  ``On semiclassical approximation for correlators of closed string vertex operators in AdS/CFT,''
  JHEP {\bf 1008}, 057 (2010)
  [arXiv:1005.4516 [hep-th]].

\bibitem{Janik:2011bd}
  R.~A.~Janik and A.~Wereszczynski,
  ``Correlation functions of three heavy operators: The AdS contribution,''
  JHEP {\bf 1112}, 095 (2011)
  [arXiv:1109.6262 [hep-th]].

\bibitem{Buchbinder:2011jr}
  E.~I.~Buchbinder and A.~A.~Tseytlin,
  ``Semiclassical correlators of three states with large $S^5$ charges in string theory in $AdS_5 \times S^5$,''
  Phys.\ Rev.\ D {\bf 85}, 026001 (2012)
  [arXiv:1110.5621 [hep-th]].

\bibitem{Zarembo:2010rr}
  K.~Zarembo,
  ``Holographic three-point functions of semiclassical states,''
  JHEP {\bf 1009}, 030 (2010)
  [arXiv:1008.1059 [hep-th]].

\bibitem{Costa:2010rz}
  M.~S.~Costa, R.~Monteiro, J.~E.~Santos and D.~Zoakos,
  ``On three-point correlation functions in the gauge/gravity duality,''
  JHEP {\bf 1011}, 141 (2010)
  [arXiv:1008.1070].

\bibitem{Roiban:2010fe}
  R.~Roiban and A.~A.~Tseytlin,
  ``On semiclassical computation of 3-point functions of closed string vertex operators in $AdS_5 x S^5$,''
  Phys.\ Rev.\ D {\bf 82}, 106011 (2010)
  [arXiv:1008.4921 [hep-th]].

\bibitem{Buchbinder:2010ek}
  E.~I.~Buchbinder and A.~A.~Tseytlin,
  ``Semiclassical four-point functions in $AdS_5 \times S^5$,''
  JHEP {\bf 1102}, 072 (2011)
  [arXiv:1012.3740 [hep-th]].

\bibitem{Raju:2010by}
  S.~Raju,
  ``BCFW for Witten Diagrams,''
  Phys.\ Rev.\ Lett.\  {\bf 106}, 091601 (2011)
  [arXiv:1011.0780 [hep-th]].

\bibitem{Raju:2011mp}
  S.~Raju,
  ``Recursion Relations for AdS/CFT Correlators,''
  Phys.\ Rev.\ D {\bf 83}, 126002 (2011)
  [arXiv:1102.4724 [hep-th]].

\bibitem{Drummond:2006rz}
  J.~M.~Drummond, J.~Henn, V.~A.~Smirnov and E.~Sokatchev,
  ``Magic identities for conformal four-point integrals,''
  JHEP {\bf 0701}, 064 (2007)
  [hep-th/0607160].

\bibitem{Bern:2006ew}
  Z.~Bern, M.~Czakon, L.~J.~Dixon, D.~A.~Kosower and V.~A.~Smirnov,
``The Four-Loop Planar Amplitude and Cusp Anomalous Dimension in Maximally Supersymmetric
Yang-Mills Theory,''
  Phys.\ Rev.\ D {\bf 75}, 085010 (2007)
  [hep-th/0610248].

\bibitem{Bern:2008qj}
  Z.~Bern, J.~J.~M.~Carrasco and H.~Johansson,
  ``New Relations for Gauge-Theory Amplitudes,''
  Phys.\ Rev.\ D {\bf 78}, 085011 (2008)
  [arXiv:0805.3993 [hep-ph]].

\bibitem{OtherTreeBCJ}
S.-H.~H.~Tye and Y.~Zhang,
``Dual Identities inside the Gluon and the Graviton Scattering Amplitudes,''
JHEP {\bf 1006}, 071 (2010)
[arXiv:1003.1732 [hep-th]];
%
N.~E.~J.~Bjerrum-Bohr, P.~H.~Damgaard, T.~Sondergaard and P.~Vanhove,
``Monodromy and Jacobi-like Relations for Color-Ordered Amplitudes,''
JHEP {\bf 1006}, 003 (2010)
[arXiv:1003.2403 [hep-th]];
%

\bibitem{virtuousTrees}
J.~Broedel and J.~J.~M.~Carrasco,
``Virtuous Trees at Five and Six Points for Yang-Mills and Gravity,''
Phys.\ Rev.\  D {\bf 84}, 085009 (2011)
[arXiv:1107.4802 [hep-th]].

\bibitem{Square}
Z.~Bern, T.~Dennen, Y.-t.~Huang and M.~Kiermaier,
``Gravity as the Square of Gauge Theory,''
Phys.\ Rev.\  D {\bf 82}, 065003 (2010)
[arXiv:1004.0693 [hep-th]].

\bibitem{Oconnell}
R.~Monteiro and D.~O'Connell,
``The Kinematic Algebra From the Self-Dual Sector,''
JHEP {\bf 1107}, 007 (2011)
[arXiv:1105.2565 [hep-th]].


\bibitem{ExplicitForms}
M. Kiermaier, presented at Amplitudes 2010,\\
{\tt{http://www.strings.ph.qmul.ac.uk/$\sim$theory/Amplitudes2010}}.
%
N.~E.~J.~Bjerrum-Bohr, P.~H.~Damgaard, T.~Sondergaard and P.~Vanhove,
``The Momentum Kernel of Gauge and Gravity Theories,''
JHEP {\bf 1101}, 001 (2011)
[arXiv:1010.3933 [hep-th]];
%
C.~R.~Mafra, O.~Schlotterer and S.~Stieberger,
``Explicit BCJ Numerators from Pure Spinors,''
JHEP {\bf 1107}, 092 (2011)
[arXiv:1104.5224 [hep-th]].

\bibitem{Sondergaard:2009za}
  T.~Sondergaard,
  ``New Relations for Gauge-Theory Amplitudes with Matter,''
  Nucl.\ Phys.\ B {\bf 821}, 417 (2009)
  [arXiv:0903.5453 [hep-th]].

\bibitem{BCJLoop}
Z.~Bern, J.~J.~M.~Carrasco and H.~Johansson,
``Perturbative Quantum Gravity as a Double Copy of Gauge Theory,''
Phys.\ Rev.\ Lett.\  {\bf 105}, 061602 (2010)
[arXiv:1004.0476 [hep-th]].
%
J.~J.~M.~Carrasco and H.~Johansson,
``Five-Point Amplitudes in N=4 Super-Yang-Mills Theory and N=8
Supergravity,''
Phys.\ Rev.\  D {\bf 85}, 025006 (2012)
[arXiv:1106.4711 [hep-th]].
%
  Z.~Bern, J.~J.~M.~Carrasco, L.~J.~Dixon, H.~Johansson and R.~Roiban,
  ``Simplifying Multiloop Integrands and Ultraviolet Divergences of Gauge Theory and Gravity Amplitudes,''
  Phys.\ Rev.\ D {\bf 85}, 105014 (2012)
  [arXiv:1201.5366 [hep-th]].

\bibitem{Feng}
B.~Feng, R.~Huang and Y.~Jia,
``Gauge Amplitude Identities by On-shell Recursion Relation in S-matrix
Program,''
Phys.\ Lett.\  B {\bf 695}, 350 (2011)
[arXiv:1004.3417 [hep-th]];
%
Y.-X.~Chen, Y.-J.~Du and B.~Feng,
``A Proof of the Explicit Minimal-basis Expansion of Tree Amplitudes
in Gauge Field Theory,''
JHEP {\bf 1102}, 112 (2011)
[arXiv:1101.0009 [hep-th]];
%
Y.-J.~Du, B.~Feng and C.-H.~Fu,
``BCJ Relation of Color Scalar Theory and KLT Relation of Gauge Theory,''
JHEP {\bf 1108}, 129 (2011)
[arXiv:1105.3503 [hep-th]].


\bibitem{Bjerrum1}
N.~E.~J.~Bjerrum-Bohr, P.~H.~Damgaard and P.~Vanhove,
``Minimal Basis for Gauge Theory Amplitudes,''
Phys.\ Rev.\ Lett.\  {\bf 103}, 161602 (2009)
[arXiv:0907.1425 [hep-th]];
%
S.~Stieberger,
``Open \& Closed vs. Pure Open String Disk Amplitudes,''
arXiv:0907.2211 [hep-th].

\bibitem{BjerrumBohr:2011xe}
  N.~E.~J.~Bjerrum-Bohr, P.~H.~Damgaard, H.~Johansson and T.~Sondergaard,
  ``Monodromy--like Relations for Finite Loop Amplitudes,''
  JHEP {\bf 1105}, 039 (2011)
  [arXiv:1103.6190 [hep-th]].

\bibitem{Jin:2012mk}
Q.~Jin and R.~Roiban,
  ``On the non-planar beta-deformed N=4 super-Yang-Mills theory,''
  J.\ Phys.\ A {\bf 45}, 295401 (2012)
  [arXiv:1201.5012 [hep-th]].

\bibitem{vanNeerven:1985ja}
  W.~L.~van Neerven,
  ``Infrared Behavior Of On-shell Form-factors In A N=4 Supersymmetric Yang-mills Field Theory,''
  Z.\ Phys.\ C {\bf 30}, 595 (1986).


\bibitem{Gehrmann:2011xn}
  T.~Gehrmann, J.~M.~Henn and T.~Huber,
  ``The three-loop form factor in N=4 super Yang-Mills,''
  arXiv:1112.4524 [hep-th].

\bibitem{Bork:2012tt}
  L.~V.~Bork,
  ``On NMHV form factors in N=4 SYM theory from generalized unitarity,''
  arXiv:1203.2596 [hep-th].


\bibitem{Brandhuber:2012vm}
  A.~Brandhuber, G.~Travaglini and G.~Yang,
  ``Analytic two-loop form factors in N=4 SYM,''
  arXiv:1201.4170 [hep-th].

\bibitem{Eden:1998hh}
  B.~Eden, P.~S.~Howe, C.~Schubert, E.~Sokatchev and P.~C.~West,
  ``Four point functions in N=4 supersymmetric Yang-Mills theory at two loops,''
  Nucl.\ Phys.\ B {\bf 557}, 355 (1999)
  [hep-th/9811172].

\bibitem{Eden:1999kh}
  B.~Eden, P.~S.~Howe, C.~Schubert, E.~Sokatchev and P.~C.~West,
  ``Simplifications of four point functions in N=4 supersymmetric Yang-Mills theory at two loops,''
  Phys.\ Lett.\ B {\bf 466}, 20 (1999)
  [hep-th/9906051].

\bibitem{Plefka:2012rd}
  J.~Plefka and K.~Wiegandt,
  ``Three-Point Functions of Twist-Two Operators in N=4 SYM at One Loop,''
  arXiv:1207.4784 [hep-th].

\bibitem{Brandhuber:2011tv}
A.~Brandhuber, O.~Gurdogan, R.~Mooney, G.~Travaglini and G.~Yang,
``Harmony of Super Form Factors,''
JHEP {\bf 1110} (2011) 046
[arXiv:1107.5067 [hep-th]].

\bibitem{Bork:2010wf}
L.~V.~Bork, D.~I.~Kazakov and G.~S.~Vartanov,
``On form factors in N=4 sym,''
JHEP {\bf 1102}, 063 (2011)
[arXiv:1011.2440 [hep-th]].

\bibitem{Gehrmann:2011aa}
  T.~Gehrmann, M.~Jaquier, E.~W.~N.~Glover and A.~Koukoutsakis,
  ``Two-Loop QCD Corrections to the Helicity Amplitudes for $H \to$ 3 partons,''
  JHEP {\bf 1202}, 056 (2012)
  [arXiv:1112.3554 [hep-ph]].

\bibitem{Johansson:2012zv}
  H.~Johansson, D.~A.~Kosower and K.~J.~Larsen,
  ``Two-Loop Maximal Unitarity with External Masses,''
  arXiv:1208.1754 [hep-th].

\bibitem{Maldacena:1997re}
  J.~M.~Maldacena,
  ``The Large N limit of superconformal field theories and supergravity,''
  Adv.\ Theor.\ Math.\ Phys.\  {\bf 2}, 231 (1998)
  [Int.\ J.\ Theor.\ Phys.\  {\bf 38}, 1113 (1999)]
  [hep-th/9711200].

\bibitem{Gubser:1998bc}
  S.~S.~Gubser, I.~R.~Klebanov and A.~M.~Polyakov,
  ``Gauge theory correlators from noncritical string theory,''
  Phys.\ Lett.\ B {\bf 428}, 105 (1998)
  [hep-th/9802109].

\bibitem{Witten:1998qj}
  E.~Witten,
  ``Anti-de Sitter space and holography,''
  Adv.\ Theor.\ Math.\ Phys.\  {\bf 2}, 253 (1998)
  [hep-th/9802150].

\bibitem{Veltman:1963th}
  M.~J.~G.~Veltman,
  ``Unitarity and causality in a renormalizable field theory with unstable particles,''
  Physica {\bf 29}, 186 (1963).
%
E.~Remiddi,
  ``Dispersion Relations For Feynman Graphs,''
  Helv.\ Phys.\ Acta {\bf 54}, 364 (1982).

\bibitem{Duhr:2012fh}
  C.~Duhr,
  ``Hopf algebras, coproducts and symbols: an application to Higgs boson amplitudes,''
  JHEP {\bf 1208}, 043 (2012)
  [arXiv:1203.0454 [hep-ph]].



  \bibitem{Cachazo:2004by}
  F.~Cachazo, P.~Svrcek and E.~Witten,
  ``Gauge theory amplitudes in twistor space and holomorphic anomaly,''
  JHEP {\bf 0410}, 077 (2004)
  [hep-th/0409245].


\bibitem{UnitarityMethod}
Z.~Bern, L.~J.~Dixon, D.~C.~Dunbar and D.~A.~Kosower,
``One loop n point gauge theory amplitudes, unitarity and collinear limits,''
Nucl.\ Phys.\ B {\bf 425}, 217 (1994)
[hep-ph/9403226];
%
Z.~Bern, L.~J.~Dixon, D.~C.~Dunbar and D.~A.~Kosower,
``Fusing gauge theory tree amplitudes into loop amplitudes,''
Nucl.\ Phys.\ B {\bf 435}, 59 (1995)
[hep-ph/9409265].

\bibitem{BDDPR}
Z.~Bern, L.~J.~Dixon, D.~C.~Dunbar, M.~Perelstein and J.~S.~Rozowsky,
 ``On the relationship between Yang-Mills theory and gravity and its
implication for ultraviolet divergences,''
Nucl.\ Phys.\ B {\bf 530}, 401 (1998)
[hep-th/9802162].

\bibitem{GeneralizedUnitarity}
Z.~Bern, L.~J.~Dixon and D.~A.~Kosower,
``One-loop amplitudes for e+ e- to four partons,''
Nucl.\ Phys.\ B {\bf 513}, 3 (1998)
[hep-ph/9708239];
%
Z.~Bern, L.~J.~Dixon and D.~A.~Kosower,
``Two-loop $g \rightarrow g g$ splitting amplitudes in QCD,''
JHEP {\bf 0408}, 012 (2004)
[hep-ph/0404293].

\bibitem{BCFGeneralized}
R.~Britto, F.~Cachazo and B.~Feng,
``Generalized unitarity and one-loop amplitudes in N = 4 super-Yang-Mills,''
Nucl.\ Phys.\  B {\bf 725}, 275 (2005)
[hep-th/0412103];
%
E.~I.~Buchbinder and F.~Cachazo,
``Two-loop amplitudes of gluons and octa-cuts in N = 4 super Yang-Mills,''
JHEP {\bf 0511}, 036 (2005)
[hep-th/0506126].

\bibitem{FiveLoop}
Z.~Bern, J.~J.~M.~Carrasco, H.~Johansson and D.~A.~Kosower,
``Maximally supersymmetric planar Yang-Mills amplitudes at five loops,''
Phys.\ Rev.\  D {\bf 76}, 125020 (2007)
[0705.1864 [hep-th]].

\bibitem{CachazoSkinner}
F.~Cachazo and D.~Skinner,
``On the structure of scattering amplitudes in N=4 super Yang-Mills and N=8
supergravity,''
0801.4574 [hep-th].

\bibitem{Alday:2008yw}
  L.~F.~Alday and R.~Roiban,
  ``Scattering Amplitudes, Wilson Loops and the String/Gauge Theory Correspondence,''
  Phys.\ Rept.\  {\bf 468}, 153 (2008)
  [arXiv:0807.1889].

\bibitem{Beisert:2010jr}
  N.~Beisert, C.~Ahn, L.~F.~Alday, Z.~Bajnok, J.~M.~Drummond, L.~Freyhult, N.~Gromov and R.~A.~Janik {\it et al.},
  ``Review of AdS/CFT Integrability: An Overview,''
  Lett.\ Math.\ Phys.\  {\bf 99}, 3 (2012)
  [arXiv:1012.3982 [hep-th]].

\bibitem{JPhysA_Volume}
"Scattering amplitudes in gauge theories: progress and outlook",
R.~Roiban (ed), M.~Spradlin (ed) and A.~Volovich (ed)
J. Phys. A: Math. Theor. {\bf 44} 450301

\bibitem{Kosower:2011ty}
  D.~A.~Kosower and K.~J.~Larsen,
  ``Maximal Unitarity at Two Loops,''
  Phys.\ Rev.\ D {\bf 85}, 045017 (2012)
  [arXiv:1108.1180 [hep-th]].

\bibitem{Petkou:1999fv}
  A.~Petkou and K.~Skenderis,
  ``A Nonrenormalization theorem for conformal anomalies,''
  Nucl.\ Phys.\ B {\bf 561}, 100 (1999)
  [hep-th/9906030].

\bibitem{Osborn:1993cr}
  H.~Osborn and A.~C.~Petkou,
  ``Implications of conformal invariance in field theories for general dimensions,''
  Annals Phys.\  {\bf 231}, 311 (1994)
  [hep-th/9307010].

\bibitem{Broedel:2012rc}
  J.~Broedel and L.~J.~Dixon,
  ``Color-kinematics duality and double-copy construction for amplitudes from higher-dimension operators,''
  arXiv:1208.0876 [hep-th].

\bibitem{Staudacher:2004tk}
  M.~Staudacher,
  ``The Factorized S-matrix of CFT/AdS,''
  JHEP {\bf 0505}, 054 (2005)
  [hep-th/0412188].

\bibitem{Beisert:2005fw}
  N.~Beisert and M.~Staudacher,
  ``Long-range psu(2,2|4) Bethe Ansatze for gauge theory and strings,''
  Nucl.\ Phys.\ B {\bf 727}, 1 (2005)
  [hep-th/0504190].

\bibitem{CaronHuot:2010ek}
  S.~Caron-Huot,
  ``Notes on the scattering amplitude / Wilson loop duality,''
  JHEP {\bf 1107}, 058 (2011)
  [arXiv:1010.1167 [hep-th]].

\bibitem{Eden:2011yp}
  B.~Eden, P.~Heslop, G.~P.~Korchemsky and E.~Sokatchev,
  ``The super-correlator/super-amplitude duality: Part I,''
  arXiv:1103.3714 [hep-th].

\bibitem{Eden:2011ku}
  B.~Eden, P.~Heslop, G.~P.~Korchemsky and E.~Sokatchev,
  ``The super-correlator/super-amplitude duality: Part II,''
  arXiv:1103.4353 [hep-th].

\bibitem{Harmonic_superspace}
  A.~Galperin, E.~Ivanov, S.~Kalitsyn, V.~Ogievetsky and E.~Sokatchev,
  ``Unconstrained N=2 Matter, Yang-Mills and Supergravity Theories in Harmonic Superspace,''
  Class.\ Quant.\ Grav.\  {\bf 1}, 469 (1984).
%
  A.~S.~Galperin, E.~A.~Ivanov, V.~I.~Ogievetsky and E.~S.~Sokatchev,
  ``Harmonic superspace,''
  Cambridge, UK: Univ. Pr. (2001) 306 p

\bibitem{Bork:2011cj}
  L.~V.~Bork, D.~I.~Kazakov and G.~S.~Vartanov,
  ``On MHV Form Factors in Superspace for $\mathcal{N}=4$ SYM Theory,''
  JHEP {\bf 1110}, 133 (2011)
  [arXiv:1107.5551 [hep-th]].

\bibitem{Dixon:2004za}
L.~J.~Dixon, E.~W.~N.~Glover and V.~V.~Khoze,
``MHV Rules for Higgs Plus Multi-Gluon Amplitudes,''
JHEP {\bf 0412} (2004) 015
[arXiv:hep-th/0411092].


\bibitem{Brandhuber:2010ad}
A.~Brandhuber, B.~Spence, G.~Travaglini and G.~Yang,
``Form Factors in ${\mathcal{N}}\!=4$ Super Yang-Mills and Periodic Wilson Loops,''
JHEP {\bf 1101} (2011) 134
[arXiv:1011.1899].

\bibitem{CSW}
  F.~Cachazo, P.~Svrcek and E.~Witten,
  ``MHV vertices and tree amplitudes in gauge theory,''
  JHEP {\bf 0409}, 006 (2004)
  [hep-th/0403047].
%
  H.~Elvang, D.~Z.~Freedman and M.~Kiermaier,
  ``Proof of the MHV vertex expansion for all tree amplitudes in N=4 SYM theory,''
  JHEP {\bf 0906}, 068 (2009)
  [arXiv:0811.3624 [hep-th]].

\bibitem{Beisert:2003jj}
  N.~Beisert,
  ``The complete one loop dilatation operator of N=4 superYang-Mills theory,''
  Nucl.\ Phys.\ B {\bf 676}, 3 (2004)
  [hep-th/0307015].

\bibitem{Belitsky:2003ys}
  T.~Ohrndorf,
  ``Constraints From Conformal Covariance On The Mixing Of Operators Of Lowest Twist,''
  Nucl.\ Phys.\ B {\bf 198}, 26 (1982).
%
  Y.~.M.~Makeenko,
  ``Conformal Operators In Quantum Chromodynamics,''
  Sov.\ J.\ Nucl.\ Phys.\  {\bf 33}, 440 (1981)
  [Yad.\ Fiz.\  {\bf 33}, 842 (1981)].
%
  A.~V.~Belitsky, A.~S.~Gorsky, G.~P.~Korchemsky,
  ``Gauge/string duality for QCD conformal operators,''
  Nucl.\ Phys.\  {\bf B667}, 3-54 (2003).
  [arXiv:hep-th/0304028 [hep-th]].

\bibitem{Belitsky:2007jp}
  A.~V.~Belitsky, J.~Henn, C.~Jarczak, D.~Mueller and E.~Sokatchev,
  ``Anomalous dimensions of leading twist conformal operators,''
  Phys.\ Rev.\ D {\bf 77}, 045029 (2008)
  [arXiv:0707.2936 [hep-th]].

\bibitem{Beisert:2002bb}
  N.~Beisert, C.~Kristjansen, J.~Plefka, G.~W.~Semenoff and M.~Staudacher,
  ``BMN correlators and operator mixing in N=4 superYang-Mills theory,''
  Nucl.\ Phys.\ B {\bf 650}, 125 (2003)
  [hep-th/0208178].


\bibitem{Bern:2009xq}
  Z.~Bern, J.~J.~M.~Carrasco, H.~Ita, H.~Johansson and R.~Roiban,
  ``On the Structure of Supersymmetric Sums in Multi-Loop Unitarity Cuts,''
  Phys.\ Rev.\ D {\bf 80}, 065029 (2009)
  [arXiv:0903.5348 [hep-th]].

\bibitem{Lee:1998bxa}
  S.~Lee, S.~Minwalla, M.~Rangamani and N.~Seiberg,
  ``Three point functions of chiral operators in D = 4, N=4 SYM at large N,''
  Adv.\ Theor.\ Math.\ Phys.\  {\bf 2}, 697 (1998)
  [hep-th/9806074].

\bibitem{energy_correlators_1}
 C.~Lee and G.~F.~Sterman,
  ``Universality of nonperturbative effects in event shapes,''
  eConf C {\bf 0601121}, A001 (2006)
  [hep-ph/0603066];
  Y.~L.~Dokshitzer, G.~Marchesini and B.~R.~Webber,
  ``Nonperturbative effects in the energy energy correlation,''
  JHEP {\bf 9907}, 012 (1999)
  [hep-ph/9905339];
 J.~C.~Collins and D.~E.~Soper,
  ``The Two Particle Inclusive Cross-section In E+ E- Annihilation At Petra, Pep And Lep Energies,''
  Nucl.\ Phys.\ B {\bf 284}, 253 (1987);
  W.~Y.~Crutchfield, II, F.~R.~Ore, Jr. and G.~F.~Sterman,
  ``Quark - Gluon Correlations And Vacuum Fluctuations,''
  Phys.\ Lett.\ B {\bf 102}, 347 (1981);
  D.~de Florian and M.~Grazzini,
  ``The Back-to-back region in e+ e- energy-energy correlation,''
  Nucl.\ Phys.\ B {\bf 704}, 387 (2005)
  [hep-ph/0407241].

\bibitem{KS}
G.~P.~Korchemsky and G.~F.~Sterman,
  ``Power corrections to event shapes and factorization,''
  Nucl.\ Phys.\ B {\bf 555}, 335 (1999)
  [hep-ph/9902341];

\bibitem{BKS}
A.~V.~Belitsky, G.~P.~Korchemsky and G.~F.~Sterman,
  ``Energy flow in QCD and event shape functions,''
  Phys.\ Lett.\ B {\bf 515}, 297 (2001)
  [hep-ph/0106308];

\bibitem{stress_tensor}
  N.~A.~Sveshnikov and F.~V.~Tkachov,
  ``Jets and quantum field theory,''
  Phys.\ Lett.\ B {\bf 382}, 403 (1996)
  [hep-ph/9512370];
G.~P.~Korchemsky, G.~Oderda and G.~F.~Sterman,
  ``Power corrections and nonlocal operators,''
  hep-ph/9708346;
  M.~Testa,
  ``Exploring the light cone through semiinclusive hadronic distributions,''
  JHEP {\bf 9809}, 006 (1998)
  [hep-ph/9807204].

\bibitem{FlemingSterman}
 C.~W.~Bauer, S.~P.~Fleming, C.~Lee and G.~F.~Sterman,
  ``Factorization of e+e- Event Shape Distributions with Hadronic Final States in Soft Collinear Effective Theory,''
  Phys.\ Rev.\ D {\bf 78}, 034027 (2008)
  [arXiv:0801.4569 [hep-ph]].

\bibitem{Maldacena:2010kp}
  J.~Maldacena and A.~Zhiboedov,
  ``Form factors at strong coupling via a Y-system,''
  JHEP {\bf 1011}, 104 (2010)
  [arXiv:1009.1139 [hep-th]].

\bibitem{Klose:2012ju}
  T.~Klose and T.~McLoughlin,
  ``Worldsheet Form Factors in AdS/CFT,''
  arXiv:1208.2020 [hep-th].

\bibitem{Drummond:2008bq}
  J.~M.~Drummond, J.~Henn, G.~P.~Korchemsky and E.~Sokatchev,
  ``Generalized unitarity for N=4 super-amplitudes,''
  arXiv:0808.0491 [hep-th].

\bibitem{ArkaniHamed:2008gz}
  N.~Arkani-Hamed, F.~Cachazo and J.~Kaplan,
  ``What is the Simplest Quantum Field Theory?,''
  JHEP {\bf 1009}, 016 (2010)
  [arXiv:0808.1446 [hep-th]].

\bibitem{Huang:2011um}
  Y.~-t.~Huang,
  ``Non-Chiral S-Matrix of N=4 Super Yang-Mills,''
  arXiv:1104.2021 [hep-th].

\bibitem{Duhr:2006iq}
  C.~Duhr, S.~Hoeche and F.~Maltoni,
  ``Color-dressed recursive relations for multi-parton amplitudes,''
  JHEP {\bf 0608}, 062 (2006)
  [hep-ph/0607057].

\bibitem{DelDuca:1999rs}
  V.~Del Duca, L.~J.~Dixon and F.~Maltoni,
  ``New color decompositions for gauge amplitudes at tree and loop level,''
  Nucl.\ Phys.\ B {\bf 571}, 51 (2000)
  [hep-ph/9910563].

\end{thebibliography}
\end{document}